\documentclass[11pt]{article}
\pdfoutput=1
\usepackage{jheppub, bm}
\graphicspath{{./}}

\usepackage{amsfonts}
\usepackage{amscd}
\usepackage{amssymb}
\usepackage{amsmath,bbm}
\usepackage{graphicx}
\usepackage{epsfig}
\usepackage{framed}
\usepackage{latexsym}
\usepackage{mathtools}
\usepackage{hyperref}
\usepackage[vcentermath]{youngtab}
    
\def \be  {\begin{equation}}
\def \ee  {\end{equation}}
\def \bea {\begin{equation}\begin{aligned}}
\def \eea {\end{aligned}\end{equation}}
\def \ba  {\begin{eqnarray}}
\def \ea  {\end{eqnarray}}
\def \bb  {\begin{thebibliography}}
\def \eb  {\end{thebibliography}}
\def \lab #1 {\label{#1}}

\newcommand\cB{\mathcal{B}}

\newcommand\cH{\mathcal{H}}
\newcommand\cI{\mathcal{I}}

\newcommand\cM{\mathcal{M}}
\newcommand\cN{\mathcal{N}}

\newcommand\cT{\mathcal{T}}

\newcommand\al{\alpha}

\newcommand\C{\mathbb{C}}
\newcommand\ep{\epsilon}

\newcommand\R{\mathbb{R}}

\newcommand\la{\langle}
\newcommand\ra{\rangle}

\definecolor{cardinal}{rgb}{0.6,0,0}
\definecolor{darkgreen}{rgb}{0,0.65,0}
\definecolor{golden}{rgb}{0.92, 0.7, 0}
\definecolor{midnight}{rgb}{0, 0, 0.5}
\definecolor{darkblue}{rgb}{0.2, 0, 0.8}

\def\CR{\Cardinal}

\newcommand{\rav}{{\includegraphics[width=.4cm]{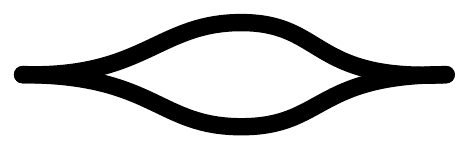}}}
\newcommand{\drav}{{\includegraphics[width=.4cm]{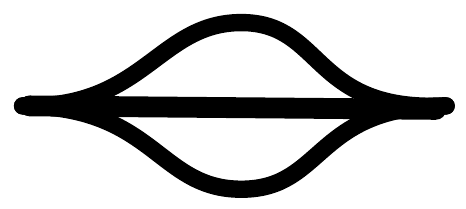}}}

\newcommand{\n}{\mathfrak n}
\newcommand{\sn}{\mathfrak n}
\newcommand{\cp}{\mathbb{CP}}

\newcommand{\Z}{{\mathbb Z}}

\newcommand{\CA}{\mathcal{A}}
\newcommand{\CB}{\mathcal{B}}

\renewcommand{\CD}{\mathcal{D}}
\newcommand{\CE}{\mathcal{E}}

\newcommand{\CG}{\mathcal{G}}
\newcommand{\CH}{\mathcal{H}}
\newcommand{\CI}{\mathcal{I}}

\newcommand{\CL}{\mathcal{L}}
\newcommand{\CM}{\mathcal{M}}
\newcommand{\CN}{\mathcal{N}}
\newcommand{\CO}{\mathcal{O}}

\renewcommand{\CR}{\mathcal{R}}
\newcommand{\CS}{\mathcal{S}}
\newcommand{\CT}{\mathcal{T}}

\newcommand{\CW}{\mathcal{W}}

\newcommand{\CY}{\mathcal{Y}}
\newcommand{\CZ}{\mathcal{Z}}

\newcommand{\Tr}{{\rm Tr \,}}

\newcommand{\pd}{\partial}

\newcommand{\wt}{\widetilde}

\newcommand{\ol}{\overline}
\newcommand{\ds}{\displaystyle}

\newcommand{\eg}{\emph{e.g.}}
\newcommand{\ie}{\emph{i.e.}}
\newcommand{\cf}{\emph{cf.}}

\title{Vortices and Vermas}
\author[1]{Mathew Bullimore}
\author[2,3]{Tudor Dimofte}
\author[2]{Davide Gaiotto}
\author[2,4,5]{Justin Hilburn}
\author[2,6,7]{Hee-Cheol Kim}
\affiliation[1]{Mathematical Institute, University of Oxford, Woodstock Road, Oxford, OX2 6GG, UK}
\affiliation[2]{Perimeter Institute for Theoretical Physics, 31 Caroline St.\,N., Waterloo, ON N2L 2Y5, Canada}
\affiliation[3]{Department of Mathematics and Center for Quantum Mathematics and Physics (QMAP), University of California, Davis, CA 95616, USA}
\affiliation[4]{Department of Mathematics, University of Oregon, Eugene, OR 97403, USA}
\affiliation[5]{Department of Mathematics, University of Pennsylvania, Philadelphia, PA 19104, USA}
\affiliation[6]{Jefferson Physical Laboratory, Harvard University, Cambridge, MA 02138, USA}
\affiliation[7]{Department of Physics, POSTECH, Pohang 790-784, Korea}

\emailAdd{mathew.bullimore@maths.ox.ac.uk}
\emailAdd{tudor@math.ucdavis.edu}
\emailAdd{dgaiotto@perimeterinstitute.ca}
\emailAdd{jhilburn@math.upenn.edu}
\emailAdd{heecheol1@gmail.com}

\abstract{In three-dimensional gauge theories, monopole operators create and destroy vortices. We explore this idea in the context of 3d $\CN=4$ gauge theories in the presence of an $\Omega$-background. In this case, monopole operators generate a non-commutative algebra that quantizes the Coulomb-branch chiral ring. The monopole operators act naturally on a Hilbert space, which is realized concretely as the equivariant cohomology of a moduli space of vortices.
The action furnishes the space with the structure of a Verma module for the Coulomb-branch algebra. This leads to a new mathematical definition of the Coulomb-branch algebra itself, related to that of Braverman-Finkelberg-Nakajima.
By introducing additional boundary conditions, we find a construction of vortex partition functions of 2d $\CN=(2,2)$ theories as overlaps of coherent states (Whittaker vectors) for Coulomb-branch algebras,
generalizing work of Braverman-Feigin-Finkelberg-Rybnikov on a finite version of the AGT correspondence. 
In the case of 3d linear quiver gauge theories, we use brane constructions to exhibit vortex moduli spaces as handsaw quiver varieties, and realize monopole operators as interfaces between handsaw-quiver quantum mechanics, generalizing work of Nakajima.}

\begin{document}
\today
\maketitle

\section{Introduction}
\label{sec:intro}

\subsection{Summary}
\begin{figure}[htp]
\centering
\includegraphics[height=3.5cm]{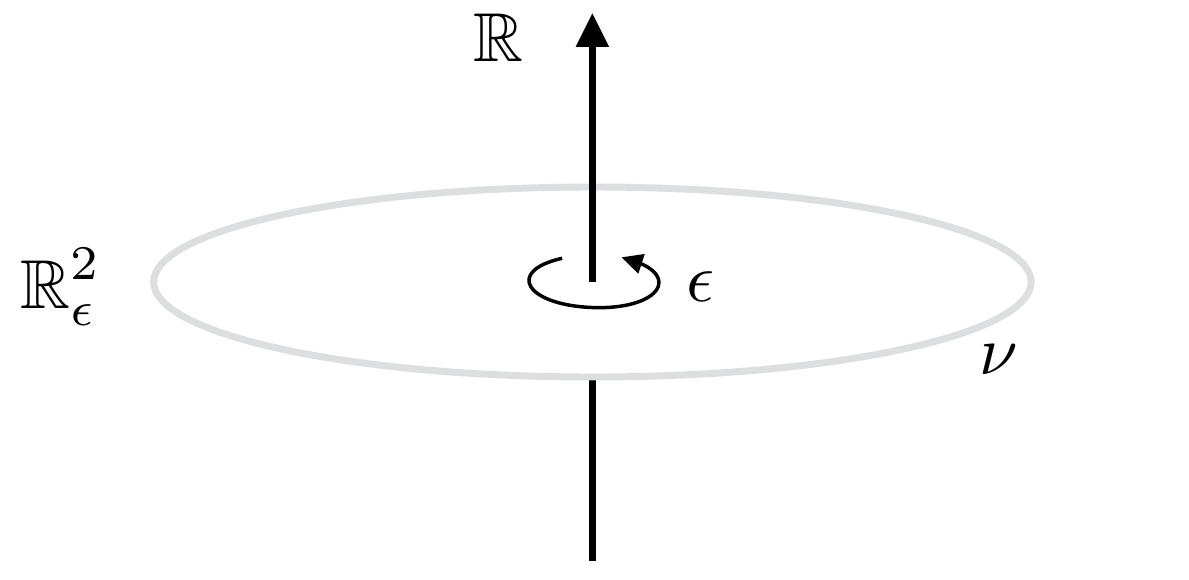}
\caption{A 3d $\CN=4$ theory in the $\Omega$-background, with a fixed vacuum $\nu$ at spatial infinity.}
\label{fig:omega}
\end{figure}

In this paper, we study various setups involving a three-dimensional gauge theory $\CT$ with $\CN=4$ supersymmetry placed in an $\Omega$-background $\R^2_\epsilon\times \R$ (Figure \ref{fig:omega}). Such a theory is labelled by a compact gauge group $G$ and a quaternionic representation $\CR$ describing the hypermultiplet content. We will require that the theory has only isolated massive vacua when generic mass and FI parameters are turned on, and place the system in such a vacuum $\nu$ at infinity in the plane of the $\Omega$-background.

The vacuum and $\Omega$-background effectively compactify this system to one-dimensional $\CN=4$ supersymmetric quantum mechanics at the origin of $\mathbb{R}^2_\ep$, with a Hilbert space $\CH_\nu$ of supersymmetric ground states. By analyzing solutions of the BPS equations that are independent of the coordinate along $\mathbb{R}$, we find the following description of the Hilbert space:
\begin{itemize}
\item The half-BPS particles of the three-dimensional gauge theory that preserve the same supersymmetry as the $\Omega$-background are vortices localized at the origin of $\R^2_\epsilon$. They are characterized by a vortex number $\n$: the flux of the abelian part of the gauge field through $\mathbb{R}^2_\ep$. The Hilbert space
\be \label{H-intro}
\CH_\nu = \bigoplus_\n H^*_{G_\nu} \big(\CM_\nu^{\n}\big)\, ,
\ee
is the direct sum of the equivariant cohomology of vortex moduli spaces $\CM_\nu^{\n}$ with respect to the symmetries $G_\nu$ preserved by the vacuum $\nu$. 
\end{itemize}

We also provide a mathematical description of the vortex moduli space $\cM_\nu^\n$ as the moduli space of based holomorphic maps from $\mathbb{CP}^1$ into the Higgs branch $\cM_H$ of the theory $\cT$ where the vortex number $\n$ corresponds to the degree of the map. More precisely, it is the moduli space of such maps into a Higgs-branch ``stack.'' We expect that this description is more general and holds even when the theory $\cT$ does not have vortex solutions in the standard sense, for example when $\CT$ is a pure gauge theory.

The theory $\CT$ has monopole operators labelled by cocharacters $A$ of the gauge group $G$, which create or destroy vortices. Together with vectormultiplet scalar fields, the monopole operators generate a Coulomb-branch chiral ring, which is the coordinate ring $\C[\CM_C]$ of the Coulomb branch in a given complex structure. The Coulomb-branch chiral ring is quantized to a noncommutative algebra $\C_\ep[\CM_C]$ in the presence of the $\Omega$ background. A systematic construction of this ring and its quantization was the topic of \cite{BDG-Coulomb,Nak-Coulomb,BFN-II}. One motivation for the present paper is to provide a new construction of $\C_\ep[\CM_C]$ from its action on vortices.

We will compute the action of monopole operators on $\CH_\nu$ by analyzing solutions of three-dimensional BPS equations in $\mathbb{R}^2_\ep \times \mathbb{R}$. Schematically, a monopole operator labelled by a cocharacter $A$ takes a vortex with charge $\n$ to one with charge $\n+A$. The following statement is one of the main results of this paper:
\begin{framed}
The action of monopole operators on $\CH_\nu$ endows it with the structure of a Verma module for the quantized Coulomb branch algebra $\C_\ep[\CM_C]$.
\end{framed}
\noindent Intuitively, this corresponds to the statement that the entire Hilbert space is generated from the vacuum state by acting with monopole operators of positive charge. We will demonstrate it explicitly for various theories with unitary gauge groups, and prove it given some assumptions on the structure of the Coulomb branch.

\begin{figure}[htp]
\centering
\includegraphics[height=2.3cm]{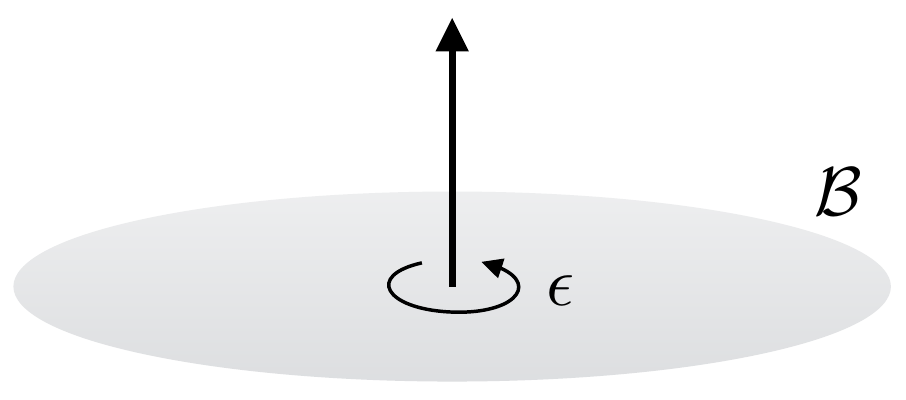}
\caption{A 3d $\CN=4$ theory in the $\Omega$-background, with a boundary condition.}
\label{fig:boundary}
\end{figure}

We can now enrich the setup of Figure \ref{fig:omega} by adding a boundary condition $\CB$ at some point in the $\R$ direction and filling $\R^2_\epsilon$, as in Figure \ref{fig:boundary}. We will consider boundary conditions that preserve 2d $\CN=(2,2)$ supersymmetry on the boundary and are therefore compatible with the $\Omega$ background. Such a boundary condition defines a state in the supersymmetric quantum mechanics:
\be
\mathrm{Boundary\; Condition} \; \CB \quad  \longrightarrow \quad \mathrm{State} \; | \CB \ra \in \CH_\nu\,.
\ee 
The state is characterized by the additional relations obeyed by operators in $\C_\ep[\CM_C]$ when acting on it. In physical terms, the state is characterized by the behavior of monopole operators brought to the boundary.

\begin{figure}[htp]
\centering
\includegraphics[height=3cm]{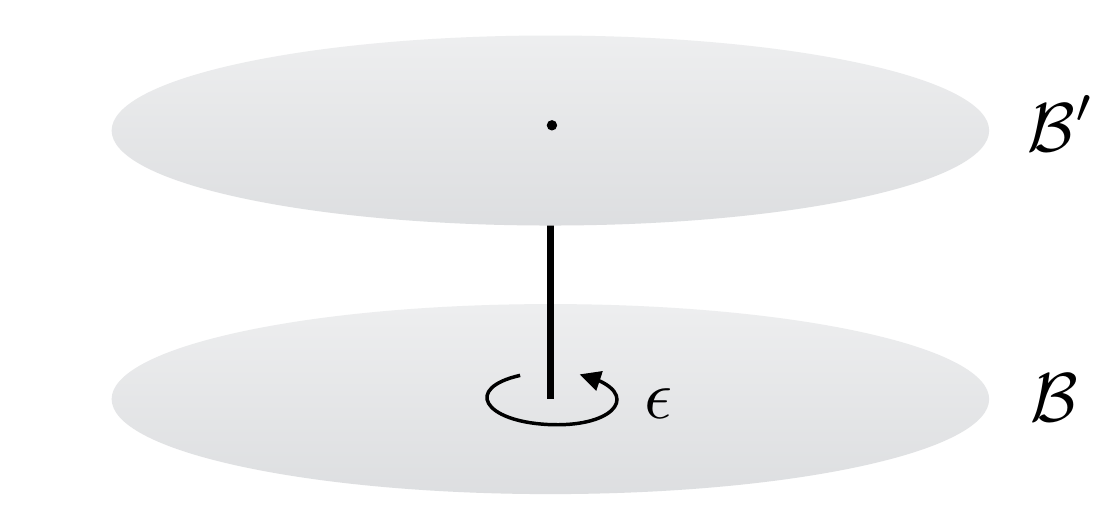}
\caption{A 3d $\CN=4$ theory in the $\Omega$-background, sandwiched between half-BPS boundary conditions.}
\label{fig:interval}
\end{figure}

We will also consider the setup shown in Figure \ref{fig:interval}, with boundary conditions $\cB$ and $\cB'$ at either end of an interval $\R^2_\epsilon\times \CI$.  At low energies, this system has an effective description as a 2d $\CN=(2,2)$ theory $\CT_{\CB,\CB'}$ in the $\Omega$-background. The partition function of this system admits two equivalent descriptions: directly as the partition function of the two-dimensional theory, or as an inner product
\be \CZ_{\CB,\CB'} = \langle \CB |\CB '\rangle \ee
in the Hilbert space $\CH_\nu$ of the three-dimensional theory $\CT$.

It is particularly interesting to consider boundary conditions $\CB$ and $\CB'$ that preserve the gauge symmetry $G$. Such `Neumann' boundary conditions were studied extensively in~\cite{BDGH}. They depend on a choice of $G$-invariant Lagrangian splitting $\CR=L\oplus L^*$ of the hypermultiplet  representation, and on complex boundary FI parameters $\xi=e^{t_{2d}}$. The states $|\CN_{L,\xi}\rangle \in \CH_\nu$ created by these boundary conditions have an explicit description as an equivariant cohomology class in the vortex moduli space, or rather a sum of classes for all vortex numbers. They turn out to be coherent states in the supersymmetric quantum mechanics, satisfying an equation of the form
\be V_A\, \big| \CN_{L,\xi} \big\rangle \;\sim\; \xi^A \big| \CN_{L,\xi}\big\rangle\,. \label{Whit-intro} \ee
Mathematically, these conditions identify $\big| \CN_{L,\xi} \big\rangle$ as a generalized ``Whittaker vector''  for the Coulomb-branch algebra $\C_\epsilon[\CM_C]$.

If we now consider an interval with Neumann boundary conditions $\CN_{L,\xi}$ and $\CN_{L',\xi'}$ at either end, we will find at low energies a 2d $\CN=(2,2)$ gauge theory $\cT_{L,L'}$ with gauge group $G$, chiral matter content transforming in the representation $L \cap (L')^*$, and FI parameter $q=\xi/\xi'$. Its partition function $\CZ_{L,L'}$ acquires two equivalent descriptions:
\begin{itemize}
\item The partition function is a standard vortex partition function \cite{Shadchin-2d, DGH} of the two-dimensional theory $\CT_{L,L'}$. This is an equivariant integral
\be \label{Zvx-intro}
\CZ_{L,L'} = \sum_\n q^\n \int_{\CM_{L,L';\nu}^{\n}} 1\hspace{-.15cm}1
\ee
of a fundamental class over the moduli space of $\n$ vortices in the two-dimensional gauge theory $\cT_{L,L'}$. 
\item The partition function is an inner product
\be
\CZ_{L,L'} = \big\langle \CN_{L',\xi'}\,\big|\, \CN_{L,\xi} \big\rangle
\ee
of states defined by the boundary conditions $\CN_{L,\xi},\,\CN_{L',\xi'}$ in the Hilbert space $\CH_\nu$ of the three-dimensional theory $\CT$ on $\R^2_\epsilon \times \R$.
\end{itemize}
The equivalence of these two descriptions means that
\begin{framed}
The vortex partition function of a 2d $\cN=(2,2)$ theory $\cT_{L,L'}$ arising from a 3d $\cN=4$ theory $\cT$ on an interval with Neumann boundary conditions $\cN_L$ and $\cN_{L'}$ is equal to an overlap of generalized Whittaker vectors for the quantized Coulomb branch algebra $\mathbb{C}_\ep[\cM_C]$ in $\CH_\nu$.
\end{framed}

As we explain in Section \ref{sec:AGT} below, this can be seen as a ``finite'' version of the AGT correspondence. Indeed, in very particular examples the Coulomb-branch algebra $\mathbb{C}_\ep[\cM_C]$ is known to be a finite W-algebra, motivating the name.
One consequence of writing the partition function $\CZ_{L,L'}$ as an inner product of vectors that satisfy the Whittaker-like condition \eqref{Whit-intro} is that the vortex partition function itself must satisfy differential equations in the parameters $q$ that quantize the twisted chiral ring of $\cT_{L,L'}$.

Throughout the paper, we find it useful to describe the physics of half-BPS vortex particles via an $\CN=4$ supersymmetric quantum mechanics on their worldlines. For each vacuum $\nu$ and vortex number $\n$, there is an $\cN=4$ supersymmetric quantum mechanics $Q(\nu,\n)$ whose Higgs branch is the moduli space of vortices $\CM_\n^{\mathrm{vortex}}$. Its space of supersymmetric ground states coincides with subspace of the 3d Hilbert space \eqref{H-intro} of vortex number $\n$,
\be \cH_{Q(\nu,\n)} = H^*_{G_\nu}(\CM_\nu^\n)\,. \label{QMH-intro} \ee
Both the complex mass parameters of $\cT$ and the $\Omega$-background deformation parameter $\epsilon$ are twisted masses in the supersymmetric quantum mechanics; they are the equivariant parameters for the symmetry $G_\nu$ preserved by the vacuum $\nu$.

The quantum mechanics $Q(\nu,\n)$ can be given a simple description as a 1d gauge theory (with finite-dimensional gauge group) when $\CT$ itself is a type-A quiver gauge theory. Then $\CT$ can be engineered on a system of intersecting D3 and NS5 branes \cite{HananyWitten} and a vortex of charge $\n$ corresponds to adding $|\n|$ finite-length D1 branes to this geometry in appropriate positions \cite{HananyHori, HananyTong-branes}. From the brane construction one reads off $Q(\nu,\n)$ as a quiver quantum mechanics whose moduli space is precisely $\CM_\nu^\n$.

The monopole operators of $\CT$ change vortex number and so should correspond to interfaces between different supersymmetric quantum mechanics. Very schematically, a monopole operator $V_A$ is represented as an interface between the quantum mechanics $Q(\nu,\n)$ and $Q(\nu,\n+A)$. It defines a correspondence between the moduli spaces; roughly speaking, this is a map
\be \CL_A \to \CM_\nu^\n \times \CM_\nu^{\n+A} \label{corresp-intro} \ee
from a monopole moduli space $\CL_A$ to the product of vortex moduli spaces. Upon taking cohomology, this induces a map of Hilbert spaces \eqref{QMH-intro}. We will construct such correspondences for general theories $\CT$, and explain how they reproduce the Coulomb-branch algebra.

When $\CT$ is an $A$-type quiver gauge theory, we find an explicit description of these interfaces by coupling the supersymmetric quantum mechanics (as a 1d gauge theory) to matrix-model degrees of freedom at the interface. This provides a physical setup for a construction of Nakajima \cite{Nakajima-handsaw} (see Section \ref{sec:quiver-intro} below) and extends it to more general $A$-type quivers.

\subsection{Relation to other work}

There are numerous connections between this paper and previous work and ideas. We briefly mention a few  prominent ones.

\subsubsection{Vortices, J-functions, and differential equations}
\label{sec:J-intro}

BPS vortices have a very long history in both mathematics and physics. They were initially discovered in abelian Higgs models, \ie\ $U(1)$ gauge theories with scalar matter \cite{Abrikosov, NielsenOlesen}. Vortex moduli spaces were later studied by mathematicians, \eg\ \cite{Taubes-LG, JaffeTaubes}, who established an equivalence between vortices and holomorphic maps. See \cite{Tong-TASI} for a review with further references.

Vortices in 2d $\CN=(2,2)$ theories played a central role in early work on mirror symmetry \cite{MorrisonPlesser, Vafa-MS, Givental-MS, HoriVafa}. In mathematics, vortex partition functions such as \eqref{Zvx-intro} (and its K-theory lift) arose in Gromov-Witten theory, and are sometimes known as equivariant J-functions, \cf\ \cite{GiventalKim, Givental-homgeom, GiventalLee, CoatesGivental} and references therein.

From these early works it became clear that vortex partition functions should be solutions to certain differential equations --- interpreted either as Picard-Fuchs equations or more intrinsically as ``quantizations'' of twisted-chiral rings of 2d $\CN=(2,2)$ theories (in the spirit of \cite{CV-tt*}). Such differential equations have shown up over and over again in various guises, from (\eg) topological string theory \cite{IntegrableHierarchies} to the AGT correspondence with surface operators \cite{AGGTV} and the 3d-3d correspondence \cite{DGG}. We re-derive them here using the construction of vortex partition functions as overlaps of Whittaker vectors.

\subsubsection{$\Omega$-background}
\label{sec:Omega-intro}

The $\Omega$-background was originally introduced for four-dimensional gauge theories with $\cN=2$ supersymmetry in~\cite{Nek-SW}, building on the previous work~\cite{LNS-SW,Moore:1998et,Moore:1997dj}. The idea that an $\Omega$-background is related to quantization of moduli spaces goes back to the work of Nekrasov-Shatashvili \cite{NShatashvili} and related works such as \cite{DG-RMQ, AGGTV, DGOT, GMNIII}.

As explained in~\cite{BDG-Coulomb, BDGH}, a 3d $\CN=4$ gauge theory admits two distinct families of $\Omega$-backgrounds that provide a quantization of either the Higgs branch and Coulomb branch in a given complex structure. These $\Omega$-backgrounds may be viewed as deformations of the two distinct families of Rozansky-Witten twists introduced in~\cite{RW,BT-twists}. The former class was studied recently in~\cite{Yagi-quantization} in the context of a sigma model onto the Higgs branch. In this paper, we study the latter: the $\Omega$-background that quantizes the Coulomb branch. This is a dimensional reduction of the usual four-dimensional $\Omega$-background in the Nekrasov-Shatashvili limit where the deformation is confined to a single plane~\cite{NShatashvili}.

In Section~\ref{sec:Hilbert}, we demonstrate that the Hilbert space $\CH_\nu$ is given by the equivariant cohomology of a moduli spaces of vortices. This observation is not unexpected: a 2d theory with at least $\CN=(2,2)$ supersymmetry localizes to BPS vortex configurations in the presence of an $\Omega$-background \cite{Shadchin-2d}. It is therefore natural to find a Hilbert space populated by BPS vortex particles in three dimensions.

\subsubsection{Finite AGT correspondence}
\label{sec:AGT}

One of our main results is that the Hilbert space $\CH_\nu$ of a 3d theory in the $\Omega$-background is a Verma module for the quantized Coulomb branch algebra, and that 2d vortex partition functions arise as overlaps of Whittaker-like vectors in $\CH_\nu$.
Special cases of these statements were discovered in mathematics by Braverman and Braverman-Feigin-Finkelberg-Rybnikov \cite{Brav-W,BFFR-W}. Much earlier, Kostant \cite{Kostant-Whittaker} introduced overlaps of Whittaker vectors to construct eigenfunctions of the Toda integrable system, which happen to be examples of 2d vortex partition functions.

The physical setup for these references is the 3d $\CN=4$ theory $T[G]$ and its generalization $T_\rho[G]$, introduced by Gaiotto and Witten as an S-duality interface in 4d gauge theory \cite{GW-Sduality}. The Higgs branch of $T_\rho[G]$ is the cotangent bundle of a partial flag variety $T^*(G_\C/P_\rho)$ for $G$, and its quantized Coulomb-branch algebra is a finite W-algebra $W_\rho[\mathfrak g^\vee]$ for the Langlands dual algebra.%
\footnote{Finite W-algebras originated in \cite{Tjin-W,dBT-W} and thereafter explored extensively in mathematics, as summarized in the review~\cite{Losev-W}.} 
Notice that holomorphic maps to $T^*(G_\C/P_\rho)$ are all supported on the base $G_\C/P_\rho$. By studying the Hilbert space of $T_\rho[G]$ in an $\Omega$-background, one therefore expects to find (roughly) that the equivariant cohomology
\be H^*(\text{based maps $\cp^1\to G_\C/P_\rho$}) \label{qmap-intro} \ee
is a Verma module for $W_\rho[\mathfrak g^\vee]$. Moreover, one expects that the vortex partition function for a 2d sigma-model with target $G_\C/P_\rho$ is an overlap of Whittaker vectors for $W_\rho[\mathfrak g^\vee]$. These are precisely the claims made by \cite{Brav-W,BFFR-W} (after modifying \eqref{qmap-intro} by partially compactifying the space of based maps and passing to intersection cohomology to to account for the fact that this compactification is not necessarily smooth).

When $G=U(n)$ is of type $A$, the theory $T_\rho[U(n)]$ is a linear-quiver gauge theory. Moreover, in the presence of generic mass and FI deformations, it has isolated massive vacua.
It is thus amenable to the gauge-theory methods of the current paper, and we will discuss it in many examples.
We also generalize to theories $T_\rho^{\rho^\vee}[U(n)]$ whose Higgs branches are intersections of nilpotent orbits and Slodowy slices in $\mathfrak{sl}_n\C$.

One of the main goals of \cite{Brav-W, BFFR-W} was to develop and prove a `finite' analogue of the AGT conjecture. 
To relate to AGT, recall that the AGT conjecture \cite{AGT} states that instanton partition functions of 4d $\CN=2$ theories of class $\CS$ are conformal blocks for a W-algebra.
In mathematics (see for instance~\cite{BFN-instW,SchiffmannVasserot, MaulikOkounkov}), this conjecture has been viewed as a consequence of two more fundamental statements: 1) that a W-algebra acts on the equivariant cohomology of instanton moduli spaces; and 2) that the instanton partition function is an inner product of Whittaker vectors for the W-algebra. Together, (1) and (2) imply that the instanton partition function satisfies conformal Ward identities that ensure it is W-algebra conformal block.
By analogy, the statement that 2d vortex partition functions arise as inner products of Whittaker vectors \eqref{Whit-intro} for finite W-algebras can be viewed as a finite version AGT.

We expect it should be possible to understand the full AGT conjecture using a higher-dimensional analogue of the setup in this paper, as outlined in~\cite{Tachikawa-instanton} (\cf\ \cite{Gaiotto-states, MMM-AGT, Taki-AGT}. Specifically, one would like to consider a 5d $\CN=2$ theory in an $\Omega$ background $\R^4_{\epsilon_1,\epsilon_2}\times \R$, with instanton operators generating a W-algebra or generalization thereof. Compatifying on an interval $\R^4_{\epsilon_1,\epsilon_2}\times I$ with half-BPS Neumann boundary conditions would lead to a 4d $\CN=2$ theory, whose instanton partition function naturally becomes interpreted as an overlap of Whittaker vectors. This geometry could be further enriched with codimension-two defects along $\R^2_{\epsilon_1}\times \R$ or $\R^2_{\epsilon_2}\times \R$, leading to similar statements about `ramified' instanton partition functions and affine Lie algebras~\cite{Alday:2010vg,KT-chainsaw}.
This would be very interesting to explore.

\subsubsection{Handsaw quivers and interfaces}
\label{sec:quiver-intro}

In Section \ref{sec-adhm}, we employ a description of BPS vortex-particles using $\cN=4$ supersymmetric quantum mechanics. For type-A quiver gauge theories $\CT$ whose Higgs branches are cotangent bundles of partial flag varieties, the supersymmetric quantum mechanics describing vortex particles are precisely the ``handsaw'' quivers that appeared in work of Nakajima \cite{Nakajima-handsaw}.  The infrared images of the interfaces that represent the action of monopole operators were defined in \cite{Nakajima-handsaw} as correspondences between pairs of vortex moduli spaces, as in \eqref{corresp-intro}. Here we develop gauge-theory definitions of these interfaces and extend the discussion to more general type-A quiver gauge theories $\CT$. The interfaces are closely analogous to those found in \cite{GaiottoKim} for five-dimensional gauge theories.

\subsubsection{Symplectic duality}

There are many relations known between geometric structures assigned to Higgs and Coulomb branches of 3d $\CN=4$ gauge theories, often referred to collectively as ``symplectic duality'' \cite{BPW-I, BLPW-II}. This includes an equivalence of categories of modules associated to the Higgs and Coulomb branches, whose physical origin was studied in \cite{BDGH}. The relation proposed in this paper might also be included in the symplectic duality canon. It is somewhat different in character from the equivalence of categories discussed in \cite{BDGH}, most notably in its asymmetric treatment of the Higgs and Coulomb branches. The $\Omega$-background that quantizes the Higgs branch (related to the one studied here by mirror symmetry) should lead to a relation between quasi-maps to the Coulomb branch and Verma modules for Higgs-branch algebras.

\subsection{Outline of the paper}

We begin in Section \ref{sec:setup} by reviewing the basic structure of 3d $\CN=4$ theories, their BPS operators and excitations, and the $\Omega$-background. In Section \ref{sec:Hilbert} we describe the Hilbert space $\CH_\nu$, and give it a mathematical definition in terms of holomorphic maps to a Higgs-branch stack. In Section \ref{sec-mono} we then derive the action of monopole operators (more generally, the Coulomb-branch algebra) on $\CH_\nu$. We construct this action mathematically in terms of correspondences, leading to a new ``definition'' of the Coulomb-branch algebra complementary to that of Braverman-Finkelberg-Nakajima. In Section \ref{sec:bound} we introduce half-BPS boundary conditions and 2d vortex partition functions as overlaps of Whittaker vectors. Finally, in Section \ref{sec-adhm} we use D-branes to derive quiver-quantum-mechanics descriptions of the 1d theories on the worldlines of vortices, and describe the matrix-model interfaces corresponding to monopole operators.
In Section \ref{sec:ab-quiv} we demonstrate our various constructions in the case of a simple 3d abelian quiver gauge theory, whose Higgs branch is a resolved $\C^2/\Z_N$ singularity and whose Coulomb-branch algebra is a central quotient of $\mathfrak{sl}_N$.

\section{Basic setup}
\label{sec:setup}

We begin with a review of 3d $\CN=4$ theories, their symmetries, and their moduli spaces, setting up some basic notation. We then describe various half-BPS excitations and operators in 3d $\CN=4$ theories. Notably, half-BPS monopoles, vortices, and boundary conditions can be aligned so as to preserve two common supercharges. The BPS equations for this pair of supercharges will feature throughout the paper. In Section \ref{sec:SQM} we rewrite the 3d theory on $\C\times \R$ as a 1d $\CN=4$ quantum mechanics along $\R$ (with infinite-dimensional gauge group and target space). In terms of the quantum mechanics, vortices are simply identified as supersymmetric ground states, and monopoles as half-BPS operators or interfaces. The quantum-mechanics perspective also gives us an easy way to describe the $\Omega$-background, as an ordinary twisted-mass deformation.

\subsection{3d $\CN=4$ theories}
\label{sec:N=4}

We consider a 3d $\cN=4$ supersymmetric gauge theory with compact gauge group $G$ and hypermultiplets transforming in the representation $R \oplus \bar R$ where $R$ is a unitary representation of $G$.

Recall that this theory has an R-symmetry $SU(2)_C\times SU(2)_H$, where the two factors rotate vectormultiplet and hypermultiplet scalars, respectively. (Alternatively, these are metric isometries that rotate the $\cp^1$'s of complex structures on the Coulomb and Higgs branches.) The theory also has flavor symmetry $G_C\times G_H$, acting via tri-Hamiltonian isometries of the Coulomb and Higgs branches. Explicitly, $G_C$ is the Pontryagin dual
\be G_C = \text{Hom}(\pi_1(G),U(1)) \approx U(1)^{\text{$\#$ $U(1)$ factors in $G$}}\,. \ee
In the infrared, $G_C$ may be enhanced to a nonabelian group. This Higgs-branch symmetry $G_H$ is the group of unitary symmetries of $R$ acting independently of $G$; it fits into the exact sequence
\be  G \to U(R) \to G_H \to 1\,. \label{GGH} \ee

Momentarily, we will fix a choice of complex structures on the Coulomb and Higgs branches, left invariant by a $U(1)_C\times U(1)_H$ subgroup of the R-symmetry. All choices are equivalent. In the fixed complex structures, we denote the holomorphic hypermultiplet scalars as $(X,Y) \in R\oplus \bar R$, with $U(1)_H$ charges $(+\frac12,+\frac12)$; the vectormultiplet scalars split into a holomorphic field $\varphi\in \mathfrak g_\C$ of $U(1)_C$ charge $+1$, and a real $\sigma\in \mathfrak g$ that enters the construction of holomorphic monopole operators.

The Higgs branch can be described either as a hyperk\"ahler quotient or an algebraic symplectic quotient
\be \CM_H  = (R\oplus \bar R)/\!/\!/G= \{\mu_\R = \mu_\C=0\}/G = \{\mu_\C=0\}/{G_\C}\,, \label{MHred}\ee
where $\mu_\R,\mu_\C$ are the real and complex moment maps for the action of $G$ on the representation $R\oplus\bar R$. The moment maps are given by
\be
\mu_\R = \bar XTX-YT\bar Y \qquad \mu_\C = YTX
\ee
where $T$ are the generators of $G$. The Coulomb branch was constructed in full generality in \cite{BDG-Coulomb,Nak-Coulomb,BFN-II}. It takes the form of a holomorphic Lagrangian fibration
\be \CM_C \longrightarrow \mathfrak t_\C/W\,, \label{MCint} \ee
where the base is parameterized by $G$-invariant polynomials in $\varphi$, and generic fibers are `dual complex tori' $T_\C^\vee\simeq (\C^*)^{\text{rank G}}$. The fibers are parameterized by expectation values of monopole operators, which we will return to later.

The theory admits canonical mass and FI deformations that preserve 3d $\CN=4$ supersymmetry. Masses are constant, background expectation values of vectormultiplet scalars associated to the $G_H$ flavor symmetry, and thus take values in the Cartan subalgebra of $G_H$,
\be m_\R \in \mathfrak t^{(H)}\,,\qquad  m_\C \in \mathfrak t_\C^{(H)}\,.\ee
By combining the masses with the dynamical vectormultiplet scalars, we can lift them to elements in the Cartan of the full $U(R)$ symmetry of the hypermultiplets, schematically denoted $\sigma+m_\R$ and $\varphi+m_\C$.
One can think of $m_\C$ as generating an infinitesimal complexified $\mathfrak u(1)_m$ action on the Higgs branch, and $\varphi+m_\C$ as generating a corresponding $\mathfrak u(1)_{\varphi+m}$ action on the hypermultiplets. We shall mostly be interested in complex masses, which deform the ring of holomorphic functions on the Coulomb branch (the Coulomb-branch chiral ring).

Similarly, FI parameters are constant, background values of \emph{twisted} vectormultiplet scalars associated to the $G_C$ Coulomb-branch symmetry,
\be t_\R \in \mathfrak t^{(C)}\,,\qquad t_\C \in \mathfrak t_\C^{(C)}\,. \label{FIt} \ee
These transform as a triplet of $SU(2)_H$ rather than the usual $SU(2)_C$. We shall mostly be interested in real FI parameters $t_\R$, which resolve the Higgs branch,
\be
\CM_H = \{\mu_\R +t_\R=0=\mu_\C\}/G\,.
\ee
Algebraically, we also have
\be \CM_H \simeq \{\mu_\C=0\}^{\rm stab}/G_\C\,, \label{MHstab} \ee
where the stable locus is a certain open subset of $\{\mu_\C=0\}$ determined by the choice of $t_\R$.

We make a major simplifying assumption: that for generic $m_\C$ and $t_\R$ the theory is fully massive, with a finite set $\{\nu\}$ of isolated massive vacua. Geometrically, this means that the Higgs branch is fully resolved and the $\mathfrak u(1)_m$ action on the Higgs branch has isolated fixed points; or equivalently that the Coulomb branch is fully deformed to a smooth space on which the $\mathfrak u(1)_t$ action has isolated fixed points. In either description, the fixed points correspond to the massive vacua $\{\nu\}$.

\subsection{The half-BPS zoo}

We are interested in the interactions of half-BPS monopole operators, vortices, and boundary conditions in a 3d $\CN=4$ theory. Each of these objects preserves a different half-dimensional subalgebra of the 3d $\CN=4$ algebra, which we summarize in Table \ref{tab:SUSY}.

Here and throughout the paper the Euclidean spacetime coordinates are denoted $x^1,x^2,x^3$, or
\be z = x^1+ix^2\,,\qquad \bar z = x^1-ix^2\,,\qquad t = x^3\,. \label{zzt} \ee
The 3d supercharges are denoted $Q_\alpha^{a\dot a}$, where $\alpha$ is an $SU(2)_E$ Lorentz index, and $a,\dot a$ are $SU(2)_H$, $SU(2)_C$ R-symmetry indices. There is a distinguished $U(1)_E\subset SU(2)_E$ that preserves the complex $z$-plane in spacetime, rotating $z$ with charge one. Similarly, there are distinguished $U(1)_H\times U(1)_C\subset SU(2)_H\times SU(2)_C$ subgroups of the R-symmetry that preserve a fixed choice of complex structures on the Higgs and Coulomb branches. We index the supercharges so that they transform with definite charge under $U(1)_E\times U(1)_H\times U(1)_C$, namely
\be \label{QU1}
 \begin{array}
{c|c|c|c|c|c|c|c|c}
 & Q_-^{1\dot 1} & Q_-^{1\dot 2} & Q_-^{2\dot 1} & Q_-^{2\dot 2} &Q_+^{1\dot 1} & Q_+^{1\dot 2} & Q_+^{2\dot 1} & Q_+^{2\dot 2} \\\hline
U(1)_E & -  & - & - & - & + & + & + & +  \\
U(1)_H & + & + & - & - & + & + & - & - \\
U(1)_C & + & - & + & - & + & - & + & - 
\end{array}
\ee
where $+,-$ denote charges $+\frac12$, $-\frac12$. The superalgebra then takes the form
\be \{Q_\alpha^{a\dot a},Q_\beta^{b\dot b}\} = -2\epsilon^{ab}\epsilon^{\dot a\dot b}\sigma_{\alpha\beta}^\mu P_\mu + 2\epsilon_{\alpha\beta}(\epsilon^{ab}Z^{\dot a \dot b}+\epsilon^{\dot a\dot b}Z^{a b})\,, \ee
where $\sigma^\alpha{}_\beta$ are the standard Pauli matrices, and the central charges act as infinitesimal gauge or flavor transformations with parameters
\be Z^{11} = (Z^{22})^\dagger \sim t_\C\,,\quad Z^{12}\sim i t_\R\,;\qquad Z^{\dot 1\dot 1} = (Z^{\dot 2\dot 2})^\dagger \sim \varphi+m_\C\,,\quad Z^{\dot 1\dot 2}\sim i(\sigma+m_\R)\,.\ee

We can partially align the half-BPS subalgebras preserved by various objects by requiring that the subalgebras all have a common $U(1)_H\times U(1)_C$ R-symmetry. This fixes the algebras to the form in Table \ref{tab:SUSY}.
Although we are mainly interested in Coulomb-branch chiral ring operators, vortices, and $\CN=(2,2)$ boundary conditions, it is instructive to include a few other half-BPS objects as well.

\begin{table}[htb]
\be \notag
\begin{array}{l|c|c|c|c|c|c|c|c}
& Q_-^{1\dot 1} & Q_-^{1\dot 2} & Q_-^{2\dot 1} & Q_-^{2\dot 2} &Q_+^{1\dot 1} & Q_+^{1\dot 2} & Q_+^{2\dot 1} & Q_+^{2\dot 2} \\\hline
 \text{3d $\CN=4$ SUSY vacua} & \bullet & \bullet & \bullet & \bullet & \bullet & \bullet & \bullet & \bullet \\
 \text{2d $\CN=(2,2)$ b.c. filling $z$-plane} & \bullet &&& \bullet && \bullet & \bullet & \\
 \text{particles \& Wilson lines along $t$} & \bullet  && \bullet &&& \bullet && \bullet \\
 \text{vortices \& vortex lines along $t$} & \bullet & \bullet &&&&& \bullet & \bullet \\
 \text{$\CM_H$ chiral ring operators} & \bullet & \bullet &&& \bullet & \bullet &&\\
 \text{$\CM_C$ chiral ring operators} & \bullet && \bullet && \bullet && \bullet \\
\end{array}
\ee
\caption{Supercharges preserved by various half-BPS boundary conditions, line operators, and local operators in 3d $\CN=4$ theory.}
\label{tab:SUSY}
\end{table}

Some brief comments are in order:
\begin{itemize}
\item There exist half-BPS boundary conditions preserving any 2d $\CN=(p,q)$ subalgebra with $p+q=4$. The 2d $\CN=(2,2)$ b.c. shown here are rather special in that this subalgebra is preserved under 3d mirror symmetry, which swaps dotted and undotted R-symmetry indices on the $Q$'s. Such b.c. were studied in \cite{BDGH}.
\item The half-BPS particles come in two varieties, related by mirror symmetry. In a gauge theory they can be identified as ordinary ``electric'' particles and vortices. Each preserve a particular 1d $\CN=4$ subalgebra.
 Similarly, a 3d $\CN=4$ theory has two types of half-BPS line operators (Wilson lines and vortex lines), discussed in \cite{GomisAssel}, which preserve the same 1d $\CN=4$ subalgebras as the BPS particles.
\item There are two half-BPS chiral rings. They only contain bosonic operators, whose expectation values are holomorphic functions on either the Higgs or Coulomb branches. Two supercharges ($Q_-^{1\dot1}$ and $Q_+^{1\dot 1}$) are preserved by both types of operators; these are the supercharges that would define the chiral ring of a 3d $\CN=2$ theory, which has no distinction between Higgs and Coulomb branches.
\end{itemize}

Most importantly for us, there is a pair of supercharges $Q_-^{1\dot 1}$ and $Q_+^{2\dot 1}$ preserved by all three of the objects we want to study: boundary conditions, vortices, and Coulomb-branch chiral ring operators. We will denote these as
\be Q := Q_-^{1\dot 1}\,,\qquad Q':=Q_+^{2\dot 1}  \label{QQ'}\ee
in the remainder of the paper. Their sum is the twisted Rozansky-Witten supercharge $\wt Q_{RW} = Q+Q'$. They do not quite commute with each other, but rather have
\be (\wt Q_{RW})^2  =  2\{Q,Q'\} = 4 Z^{\dot 1\dot 1} \sim \varphi+m_\C\cdot \label{QQ'comm} \ee
In other words, their commutator in a gauge theory is a combined gauge and flavor rotation, with parameters $\varphi$, $m_\C$. This is good enough for many purposes. In particular, if we consider a path integral with operator insertions and boundary conditions all of which preserve $Q$ and $Q'$ (and thus are invariant under $\varphi+m_\C$), the path integral will localize to field configurations that are invariant under both $Q$ and $Q'$.%
\footnote{The localization can be understood as a two-step procedure. First, one localizes with respect to the twisted RW supercharge $\wt Q_{RW}$. Its fixed locus is invariant under $\varphi+m_\C$, and thus has an action of $Q$. Then one can localize with respect to $Q$.}

\subsection{The quarter-BPS equations}
\label{sec:BPS}

The field configurations in a 3d $\CN=4$ gauge theory preserved by both $Q$ and $Q'$ from \eqref{QQ'} satisfy an interesting set of equations. They can easily be derived by considering the action of $Q$ and $Q'$ on the various fields of the 3d theory; however, a more conceptual derivation follows from the quantum-mechanics perspective of Section \ref{sec:SQM}.

To describe the equations, we introduce the complexified covariant derivatives%
\footnote{Throughout the paper we will assume the gauge field $A_\mu$ (and the scalar $\sigma$) to be Hermitian, so $D=d-iA$ and $F = i[D,D]$.}
\be \begin{array}{ll}2D_z &= D_1-iD_2\,,\\ 2D_{\bar z} &= D_1+iD_2\,,\\ \CD_t &= D_t-(\sigma+m_\R)\,.\end{array} \label{Dzt}\ee
The equations state that the chiral scalars in a hypermultiplet are holomorphic in the $z$-plane and constant in ``time'' with respect to the modified $\CD_t$ derivative
\be D_{\bar z} X = D_{\bar z} Y = 0\,,\qquad \CD_t X=\CD_t Y = 0\,. \label{BPS-eq-XY}\ee
In addition, the $D_{\bar z}$ derivative is constant in time, and real and complex moment-map constraints are imposed as 
\be \label{BPS-eq-D}  \begin{array}{l}
[\CD_t,D_{\bar z}]=0 \,, \cr
 4[D_z,D_{\bar z}] + [\CD_t,\CD_t^\dagger] = \mu_\R+t_\R\,, \cr
 \mu_\C+t_\C=0\,. \end{array} \ee
Finally, the vectormultiplet scalars obey
\be
[D_z, \varphi]=[D_{\bar z},\varphi]=[D_t,\varphi] = 0\,, \qquad [ \sigma , \varphi ] = 0\,, \qquad [ \varphi,  \varphi^\dagger] = 0 \, ,
\label{nondyn-BPS-1}
\ee
and
\be
(\varphi+m_\C)\cdot X = 0\,, \qquad (\varphi+m_\C)  \cdot  Y  = 0 \, .
\label{nondyn-BPS-2}
\ee
As usual, we write $(\varphi + m_\C)\cdot \Phi$ or $(\sigma + m_\R)\cdot \Phi$
to mean the action of a combined gauge and flavor transformation on $\Phi$ in the appropriate representation 
 --- say $R$ for $\Phi=X$ or $\bar R$ for $\Phi = Y$.
Most of the equations in \eqref{nondyn-BPS-1}--\eqref{nondyn-BPS-2} can be understood as requiring that the anticommutator $\{Q,Q'\}\sim \varphi+m_\C$ vanish when acting on any field.
The final equation in~\eqref{nondyn-BPS-1} requires that the complex scalar $\varphi$ lie in a Cartan subalgebra $\mathfrak t\subset \mathfrak{g}$.

These equations have several specializations, corresponding to the fact that $Q$ and $Q'$ are simultaneously preserved by 3d SUSY vacua, vortices, and Coulomb-branch operators (Table \ref{tab:SUSY}).

\subsubsection{Supersymmetric vacua}
\label{sec:vacua}

The classical supersymmetric vacua of the 3d $\CN=4$ gauge theory correspond to solutions of the BPS equations that are independent of $t$ and $z,\bar z$, and have vanishing gauge field:
\bea
\mu_\C + t_\C & = 0    \qquad & \mu_\R + t_\R = 0 \\
[\varphi ,\varphi^\dagger] & = 0 \qquad & [\varphi,\sigma] = 0 \\
(\varphi+m_\C) \cdot X & = 0 \qquad  & (\sigma+m_\R) \cdot X = 0 \\
(\varphi+m_\C) \cdot Y & = 0 \qquad  &(\sigma+m_\R) \cdot Y = 0 
\label{susy-vacua}
\eea

We are interested in situations where $m_\R$ and $t_\C$ vanish, but $m_\C$ and $t_\R$ are generic. We require that for generic $m_\C$ and $t_\R$ these equations have a finite number of isolated solutions $\{\nu\}$, \ie\ that the theory is fully massive. As mentioned at the end of Section~\ref{sec:N=4}, these solutions can be identified as fixed points on the Higgs branch of a complexified $\C^*$ flavor symmetry generated by $m_\C$.

\subsubsection{Vortices}
\label{sec:BPSvortices}

Next, let us consider time-independent solutions of the BPS equations. In temporal gauge $A_t=0$, equations \eqref{BPS-eq-D} and \eqref{BPS-eq-XY} imply that
\be \label{vortex-eq}
 \begin{array}{l} D_{\bar z}X=D_{\bar z}Y = 0\,, \cr
 \mu_\C +t_\C= 0\,, \cr
 4[D_z,D_{\bar z}] = \mu_\R+t_\R\,.
\end{array}
\ee
These are generalized vortex equations, which describe half-BPS vortex excitations of the 3d $\CN=4$ gauge theory. They generally only have solutions when $t_\C=0$. Quantizing the moduli space of solutions to these equations is the main goal of Section~\ref{sec:Hilbert}.

The generalized vortex equations should be supplemented by the additional constraints $D_{\bar z}\varphi=D_z\varphi = D_{\bar z}\sigma=D_z\sigma=0$ and 
\be \label{vortex-m} \begin{array}{rll}
(\varphi+m_\C)\cdot X & = 0 \qquad  & (\sigma+m_\R) \cdot X = 0 \\
(\varphi+m_\C)\cdot   Y & = 0 \qquad  & (\sigma+m_\R) \cdot Y = 0  \end{array}
\ee
from \eqref{BPS-eq-XY} and \eqref{nondyn-BPS-2}.
When $m_\R=m_\C=0$, we can simply set $\varphi=\sigma=0$ to satisfy these constraints. In this case, the time-independent BPS equations are fully equivalent to \eqref{vortex-eq}. As $m_\C$ is turned on, the additional constraints \eqref{vortex-m} have the effect of restricting the moduli space of solutions of \eqref{vortex-eq} to fixed points of a combined gauge and flavor rotation. This will lead to the use of equivariant cohomology when quantizing the moduli space of vortices.

\subsubsection{Monopoles}
\label{sec:BPSmon}

Finally, if we turn off the FI parameters $t_\R=t_\C=0$ and set the hypermultiplets to zero, $X=Y=0$, equations \eqref{BPS-eq-D} become the monopole equations
\be F = *D \sigma\,. \label{bps-mono} \ee
Together with $D\varphi = [\sigma,\varphi]=[\varphi,\varphi^\dagger]=0$ from \eqref{nondyn-BPS-1}, these describe half-BPS monopole solutions of the 3d $\CN=4$ gauge theory.

We recall that near the center of a monopole the field $\sigma$ has a profile
\be \sigma = \frac{A}{2r}\,, \label{mono-profile} \ee
where $r$ is Euclidean distance from the center and $A\in \Lambda_{\rm cochar} = \text{Hom}(U(1),G) \subset \mathfrak g$ is an element of the cocharacter lattice of $G$ that specifies the magnetic charge. (Charges $A,A'$ related by an element of the Weyl group are equivalent.)
In the quantum theory, one defines a corresponding monopole operator $V_A$ by requiring that fields have a singularity of the form \eqref{mono-profile} near a given point. The Coulomb-branch chiral ring $\C[\CM_C]$ is then generated by such monopole operator and by guage-invariant polynomials in $\varphi$.

Altogether, the full set of BPS equations can be understood intuitively as describing vortices in the $z$-plane that propagate in time $t$, and that can be created or destroyed at the location of monopole operators.
Of the four supercharges preserved by BPS vortices, only two are preserved by monopoles.

\subsection{3d $\CN=4$ as 1d $\CN=4$ quantum mechanics}
\label{sec:SQM}

When describing the interactions of vortices, monopoles, and boundary conditions, an extremely useful perspective is to view the 3d $\CN=4$ theory as a one-dimensional $\CN=4$ quantum mechanics, whose supersymmetry algebra involves the same four supercharges preserved by vortices in Table \ref{tab:SUSY}.%
\footnote{This sort of gauged $\CN=4$ quantum mechanics played a prominent role in \cite{Witten-path}. Many of the basic results there are directly applicable here.} 
Then vortices can be understood as supersymmetric ground states in the quantum mechanics. Similarly, boundary conditions that fill the $z$-plane become half-BPS b.c. in the quantum mechanics (preserving $Q$ and $Q'$); and monopoles become half-BPS operators (also preserving $Q$ and $Q'$).

We can give a rather explicit description of the $\CN=4$ quantum mechanics --- though the details will not be relevant for most of this paper. We use the language of 2d $\CN=(2,2)$ superfields and superspace, reduced to one dimension. The quantum mechanics is a gauge theory, whose gauge group
\be \CG = \text{Hom}(\C_z,G) \label{G-QM} \ee
is the group of gauge transformations in the $z$-plane. Its fields are valued in functions (or sections of various bundles) on the $z$-plane.

The fields $X,Y$ in a 3d hypermultiplet become chiral fields in the quantum mechanics, as does the $\bar z$-component of the gauge connection $A_{\bar z}$. A more gauge-covariant way of saying that is that the covariant derivative $D_{\bar z}$ should be treated as a chiral field. There is a natural superpotential
\be W = \int |dz|^2 \, Y D_{\bar z} X  \label{W-SQM}
\ee
that contains the $z,\bar z$ kinetic terms for $X$ and $Y$.
The 1d vectormultiplet contains all the 3d scalars $\varphi,\sigma$ as well as the gauge field $A_t$; they fit in the the vector superfield
\be V \,\sim\, \theta^+\bar\theta^- \varphi + \theta^-\bar\theta ^+\varphi^\dagger + \theta^+\bar \theta^+(A_t+i\sigma) + \theta^-\bar \theta^-(A_t-i\sigma)+...+\theta^4D\,.\ee
The field $\varphi$ is a twisted chiral, the leading component of the twisted-chiral superfield
\be \Sigma = D_+\ol D_- V \,\sim\,   \varphi + \theta^+\bar\lambda_++\bar\theta^-\lambda+_-  + \theta^+\bar\theta^-(D+i[\CD_t,\CD_t^\dagger])+...\,. \label{Sigma}\ee
The natural K\"ahler potential then takes the form
\be K = \int |dz|^2\, \Big( \Tr\,\Sigma\Sigma^\dagger + \big|e^{\frac V2}X\big|^2+ \big|e^{-\frac V2}Y\big|^2+ \Tr\big|e^{\frac V2}D_{\bar z}e^{-\frac V2}\big|^2\big)\,. \label{K-SQM} \ee
where $e^{\frac V2}X$ schematically denotes the exponentiated action of $V\in \mathfrak g$ on $X$, and similarly for $Y$.

To include masses $m_\R,m_\C$, we may introduce a background vectormultiplet for the Higgs-branch flavor symmetry. The complex masses $m_\C$ become background values of twisted-chiral fields. Similarly, a real FI parameter $t_\R$ enters in a standard twisted superpotential $\wt W =  \int |dz|^2 \langle t_\R,\Sigma\rangle$.

The vortex equations of Section \ref{sec:BPSvortices} are easily derived as equations for supersymmetric vacua in this $\CN=4$ quantum mechanics. Namely, $D_{\bar z}X=D_{\bar z}Y = \mu_\C=0$ arise as F-term equations for the superpotential \eqref{W-SQM}, the real equation $[D_z,D_{\bar z}]=\mu_\R+t_\R$ is the D-term, and the supplemental constraints $D_{\bar z}\varphi=(\varphi+m_\C)\cdot X=(\varphi+m_\C)\cdot Y=0$ (etc.) arise as twisted-mass terms from the K\"ahler potential.

The quarter-BPS equations for $Q$ and $Q'$ in Section \ref{sec:BPS} can be derived from the $\CN=4$ quantum mechanics in a similar way. In particular, equations \eqref{BPS-eq-XY}--\eqref{BPS-eq-D} are a combination of F-terms and Morse flow
\be dW = 0\,,\qquad D_t \Phi = g^{\Phi\Phi'}\frac{\delta h}{\delta\Phi'} \label{Morse} \ee
with respect to a Morse function
\be h = \int |dz|^2 \langle \sigma,\mu_\R(D_{\bar z},X,Y)\rangle = \int |dz|^2 \langle\sigma,\mu_\R(X,Y)-4[D_z,D_{\bar z}]\rangle  \,.\ee
Here $\mu_\R(D_{\bar z},X,Y)$ denotes the moment map for gauge group \eqref{G-QM} of the quantum mechanics, which contains a contribution from the chiral field $D_{\bar z}$ and its conjugate.

Such Morse flows may be more familiar in $\CN=2$ quantum mechanics, where instantons that preserve a single supercharge appear as Morse flow for a single real function $\hat h$ \cite{Witten-Morse}. In the present case, our $\CN=4$ quantum mechanics has many $\CN=2$ subalgebras embedded inside. Each subalgebra is labelled by a phase $\zeta$, and contains the two supercharges $\CR_\zeta= \zeta^{-\frac12}Q_-^{1\dot 1}+\zeta^{\frac12} Q_+^{2\dot 1}=\zeta^{-\frac12}Q+\zeta^{\frac12} Q'$ and $\wt \CR_\zeta = \zeta^{-\frac12}Q_-^{1\dot 2}-\zeta^{\frac12}Q_+^{2\dot 2}$, which obey $\{\CR_\zeta,\wt \CR_\zeta\} = -2i\CD_t$ for any $\zeta$. The instantons that preserve $\CR_\zeta$ take the form of Morse flow with respect to
\be \hat h_\zeta = h + \text{Re}(W/\zeta) \label{hzeta} \ee
The instantons that preserve both $Q$ and $Q'$ individually must be Morse flows for \eqref{hzeta} for all $\zeta$, and therefore obey \eqref{Morse}.

\subsection{$\Omega$-background}
\label{sec:omega}

We would also like introduce an $\Omega$-deformation associated to the vector field
\be V = x^1\pd_2-x^2\pd_1 = \tfrac i2 (\bar z\pd_{\bar z}-z\pd_z) \ee
that rotates the $z$-plane, with a complex parameter $\epsilon$. There are many equivalent ways to understand this deformation. A standard approach (analogous to the $\Omega$-background in 4d $\CN=2$ theory \cite{Nek-SW}, see Section \ref{sec:Omega-intro}) is to work in the twisted-Rozansky-Witten topological twist, and to deform the supercharge and the Lagrangian in such a way that $(\wt Q_{RW})^2 \sim \varphi+m_\C -i \epsilon \CL_V$. Alternatively, one may view the $\Omega$-background as a twisted-mass deformation of the 1d $\CN=4$ quantum mechanics of Section~\ref{sec:SQM}. This latter approach, which we describe here, makes several important properties manifest.

The four supercharges of the quantum mechanics (the ``vortex'' row of Table \ref{tab:SUSY}) are all left invariant by a simultaneous $U(1)_E$ rotation in the $z$-plane and a $U(1)_H$ R-symmetry rotation. Let us call this diagonal subgroup
\be U(1)_\epsilon  \,\subset\, U(1)_E\times U(1)_H\,. \label{def-U1e} \ee
It is an ordinary flavor symmetry of the 1d quantum mechanics, and thus we can introduce a background vectormultiplet for it, with a nonzero complex-scalar field $\epsilon$ (analogous to $\varphi$ in \eqref{Sigma}). Thus $\epsilon$ becomes a twisted-mass deformation in the quantum mechanics.

Formulated this way, it is clear that the $\Omega$-background preserves all four supercharges of the quantum mechanics.%
\footnote{This conclusion was also reached from a different viewpoint in \cite[Sec. 5]{ClossetCremonesi}, which constructed the $\Omega$-background by coupling to supergravity.} %
Moreover, it is easy to see how it will deform the quarter-BPS equations of Section \ref{sec:BPS}: any appearance of $\varphi + m_\C$ should be replaced by
\be \begin{array}{rl} \varphi+m_\C \;\to\;& 
 \varphi+m_\C -i\epsilon \CL_V + \epsilon\,r_H\,,
  \end{array}
\label{Omega-phi} \ee
representing a simultaneous $G\times G_H \times U(1)_\epsilon$ transformation with parameters $(\varphi,m_\C,\epsilon)$. (Here `$r_H$' is the generator of $U(1)_H$.)

Notably, this means that the nondynamical constraints \eqref{nondyn-BPS-2} in the quarter-BPS equations, or \eqref{vortex-m} in the vortex equations, are deformed to
\be \big(\varphi+m_\C + \epsilon\big( z D_z+\tfrac12\big)\big)\cdot X = 0\,,\qquad  \big(\varphi+m_\C + \epsilon\big( z D_z+\tfrac12\big)\big)\cdot Y = 0\,. \label{vx-def} \ee
We have used here the fact that $D_{\bar z}X=D_{\bar z}Y=0$ to replace $\CL_V$ with $zD_z$.
Since $X$ and $Y$ transform in conjugate representations of $G$ and $G_H$, the parameters $\varphi$ and $m_\C$ (viewed as actual complex numbers) will typically appear with opposite signs in these two equations. On the other hand, $X$ and $Y$ both have R-charge $+\frac12$ under $U(1)_H$, leading to the extra $+\tfrac\epsilon2$ term in each equation.


\section{Hilbert space}
\label{sec:Hilbert}

In this section, we analyze in some detail the Hilbert space $\CH_\nu$ of a 4d $\CN=4$ theory in the $\Omega$-background, with a fixed massive vacuum $\nu$ at spatial infinity.

From the perspective of $\CN=4$ supersymmetric quantum mechanics (Section \ref{sec:SQM}), $\CH_\nu$ is a space of supersymmetric ground states. By standard arguments \cite{Witten-Morse}, we expect that $\CH_\nu$ should be realized as the cohomology
\be \CH_\nu  \,=\, H^*(\CM_\nu,\C) \label{HH-1} \ee
of a classical moduli space $\CM_\nu$. The moduli space $\CM_\nu$ is the space of time-independent solutions to the BPS equations of Section \ref{sec:BPSvortices}. As discussed there, it is a particular generalization of a vortex moduli space. We will describe some general features of $\CM_\nu$ in Section \ref{sec:general}, and related it to a space of holomorphic maps to the Higgs-branch stack in Section \ref{sec:stack}.

In the presence of complex masses and the $\Omega$-background, $H^*(\CM_\nu)$ should be replaced by an equivariant cohomology group
\be \CH_\nu  \,=\, H_{G_\nu}^*(\CM_\nu,\C)\,, \label{HH-2} \ee
where $G_\nu$ is an appropriate group of symmetries acting on $\CM_\nu$. We will only consider theories where the action of $G_\nu$ has isolated fixed points. Then, by virtue of the localization theorem in equivariant cohomology \cite{AtiyahBott}, $\CH_\nu$ acquires a distinguished basis labelled by the fixed points. We will describe this abstractly in Section \ref{sec:fixed}. Then, in Sections \ref{sec:abel-moduli}--\ref{sec:nonab-moduli}, we will analyze $\CM_\nu$ and $\CH_\nu$ very explicitly for families of abelian and non-abelian theories, including SQED, SQCD, and triangular-quiver gauge theories.

Here and throughout the rest of the paper we set $t_\C=0$, to allow nontrivial vortex configurations. We leave $t_\R$ generic, so that the Higgs branch is fully resolved. We also usually set $m_\R=0$ for simplicity, as this parameter does not affect the BPS sector that we are considering.

\subsection{General structure}
\label{sec:general}

We begin by studying time-independent solutions to the BPS equations in the absence of $\Omega$-background $(\epsilon=0)$ and with mass parameters set to zero $(m_\C=0)$. We can then set $\varphi=\sigma=0$, and reduce the BPS equations to the generalized vortex equations given in~\eqref{vortex-eq}.

Suppose $\nu$ is a vacuum that survives mass deformations, and becomes fully massive in the presence of generic $m_\C$. This can be thought of as a point on the resolved Higgs branch where the gauge symmetry is fully broken, but a maximal torus $T_H\subset G_H$ of the flavor symmetry and the R-symmetry $U(1)_H$ are preserved. Let $G\cdot \nu$ denote the $G$-orbit of $\nu$ in the space $R\oplus \bar R$ of hypermultiplet scalars.

We are interested in the moduli space of solutions to the time-independent BPS equations that tend to the vacuum $\nu$ at spatial infinity,
\be \label{vortex-eq-2}
\CM_\nu  = \left\{ \begin{array}{l} \CD_{\bar z}X=\CD_{\bar z}Y = 0 \cr
 \mu_\C = 0 \cr
 [\CD_z,\CD_{\bar z}] = \mu_\R+t_\R
\end{array} \;\text{ \quad with \quad $X,Y\overset{|z|\to\infty} \longrightarrow G\cdot\nu$} \right\}/ \, \CG\, , 
\ee
where $\CG$ is the infinite-dimensional group of gauge transformations on $\R^2$ that are constant at infinity. The last condition ensures that gauge transformations preserve the orbit $G \cdot \nu$ at infinity. We call this the moduli space of `generalized vortices'.

If we compactify the $z$-plane to $\mathbb{CP}^1$, a point in this moduli space may be equivalently described by the following data:
\begin{enumerate}
\item A $G$-bundle on $\mathbb{CP}^1$, trivialized near $\infty$.
\item Holomorphic sections $(X,Y)$ of an associated bundle in the representation $R\oplus \bar R$ tending to $\nu$ at infinity and satisfying $\mu_\C=0$ and $ [\CD_z,\CD_{\bar z}]=\mu_\R+t_\R$. 
\end{enumerate}

The moduli space will split into components labelled by a `vortex number' $\n \in \pi_1(G)$. This labels topological type of the $G$-bundle on $\cp^1$, 
\be \n = \frac{1}{2\pi}\int_{\cp^1} \Tr(F) \label{defn} \ee
This number can also be defined as the winding number of a gauge transformation $g:S^1\to G$ on the equator of $\cp^1$ that relates trivializations of the bundle on the northern and southern hemispheres. This makes it clear that $ \n \in \pi_1(G)$. We will mainly be interested in cases where $\pi_1(G)$ is a free abelian group, namely, $G=U(N)$ with $\pi_1(G)\simeq\Z$ and products thereof. It is only in such cases that solutions of \eqref{vortex-eq-2} are `vortices' in the traditional sense. Nevertheless, we expect our construction to valid more generally and continue to use the term `vortex number' for $\n \in \pi_1(G)$.

The moduli space of solutions splits into disconnected components
\be \CM_\nu = \bigcup_\sn \CM_\nu^\sn\,, \ee \label{dvort}
where
labelled by the vortex number $\n$. In Section \ref{sec:stack}, we will see that not all vortex numbers are realized: whether or not the component $\CM_\nu^\sn$ is empty depends on the precise choice of vacuum $\nu$.

The components $\CM_\nu^\sn$ of the moduli space are K\"ahler manifolds with rather large abelian symmetry groups
\be G_\nu =  T_H \times U(1)_\epsilon\,. \label{symM} \ee
Here $T_H$ is the maximal torus of the Higgs-branch flavor symmetry preserved by the vacuum $\nu$; and $U(1)_\epsilon$ is the combination \eqref{def-U1e} of Higgs-branch R-symmetry and rotation in the $z$-plane that acts as a flavor symmetry of $\CN=4$ quantum-mechanics. We will work equivariantly with respect to $T_H$ and $U(1)_\epsilon$ when turning on $m_\C$ and $\epsilon$, respectively.

The assumption that $\nu$ is an isolated fixed point of $T_H$ on a smooth Higgs branch ensures that $\CM_\nu$ has isolated fixed points under the combined symmetry $T_H\times U(1)_\epsilon$. We will describe them in Section \ref{sec:fixed}. Similarly, the fact that $G$ symmetry is fully broken at the vacuum $\nu$ ensures that $\CG$ is fully broken in a neighborhood of each fixed point on $\CM_\nu$, and therefore that a neighborhood of each fixed point is smooth. More generally, we expect that the $\CG$ action in \eqref{vortex-eq-2} is free and that the whole space $\CM_\nu$ is smooth, but we will not prove this here.

\subsection{Algebraic description}
\label{sec:stack}

We expect the moduli space of generalized vortices to have a complex-algebraic description as well. This is obtained by dropping the real moment-map equation and dividing by complex gauge transformations,
\be \label{vortex-eq-alg}
\CM_\nu  \simeq \left\{ \begin{array}{l} \CD_{\bar z}X=\CD_{\bar z}Y = 0\,, \cr
 \mu_\C = 0\,
\end{array} \quad \text{with} \quad X,Y\overset{|z|\to\infty}\longrightarrow G_\C\cdot\nu  \right\}/ \; \CG_\C\,.
\ee
Usually, a stability condition must be imposed in the algebraic quotient. However, any solution that tends to a massive vacuum at infinity is automatically stable, so no further conditions are necessary in \eqref{vortex-eq-alg}. This construction makes manifest that the moduli space $\CM_\nu$ is K\"ahler. The equivalence of descriptions \eqref{vortex-eq-2} and \eqref{vortex-eq-alg} is a version of the Hitchin-Kobayashi correspondence for the generalized vortex equations, which we will not attempt to prove here. (Algebraically, \eqref{vortex-eq-alg} could be taken as a definition of $\CM_\nu$.)

From the algebraic point of view, a point in $\CM_\nu$ is specified by
\begin{enumerate}
\item A choice of $G_\C$-bundle $E$ on $\cp^1$, trivialized near $\infty$. 
\item Holomorphic sections $X,Y$ of an associated bundle in the representation $R\oplus \bar R$, satisfying $\mu_\C=0$ and sitting inside the orbit $G_\C\cdot\nu$ at infinity. 
\end{enumerate}

Once we allow for complex gauge transformations, we may pass to a `holomorphic frame' where $\CD_{\bar z}=2 \pd_{\bar z}$. The holomorphic sections can then be described concretely as polynomial matrices $X(z),Y(z)$ in the affine coordinate $z$. We must still quotient by holomorphic gauge transformations that preserve the choice of gauge. These are polynomial valued group elements $g(z)$. The resulting description of the moduli space is familiar in the physics literature, for example in the work of Morrison and Plesser \cite{MorrisonPlesser} and in the `moduli matrix' construction of~\cite{Eto-matrix,Eto-moduli}. 

Mathematically, we have described what are based maps from $\cp^1$ into the \emph{stack} $[\CM_H]:=[\mu_\C^{-1}(0)/G_\C]$. Recall from \eqref{MHstab} that that the actual Higgs branch $\CM_H = (\mu_\C^{-1}(0))^{\rm stab}/G_\C$ involves a stability condition that depends on the real FI parameter $t_\R$. The stability condition prevents certain combinations of the hypermultiplet fields from vanishing. Provided $R$ is a faithful representation of $G$, maps from $\cp^1$ into the stack $[\CM_H]$ only differ from ordinary holomorphic maps into the Higgs branch in that they may violate the stability condition at various points $z\in \cp^1$. Since $\infty \in \cp^1$ must map to the vacuum $\nu\in \CM_H$, which is a point on the actual Higgs branch, holomorphicity ensures that the points $z\in \cp^1$ where stability is violated are isolated and finite.

Thus we can simply say that 
\be \CM_\nu \,\simeq\, \{f:\cp^1\to [\CM_H] \quad \text{such that} \quad f(\infty) = \nu\,. \} \label{Mstack}\ee
In this picture the decomposition \ref{dvort} comes from looking at the fibers of the map
\be
\CM_\nu \to \text{Bun}_{G_\C}(\cp^1) \to \pi_0(\text{Bun}_{G_\C}(\cp^1)) \cong \pi_1(G_{\C}) \label{MBunG}
\ee
and the vortex number is often called the degree because it constrains the degrees of the polynomial matrices $X(z)$, $Y(z)$.

\subsection{Fixed points and the Hilbert space}
\label{sec:fixed}

As discussed around \eqref{HH-1}, the perspective of supersymmetric quantum mechanics suggests that
 the Hilbert space $\CH_\nu$ should be identified with the de Rham cohomology of the classical moduli space $\CM_\nu$ of generalized vortices.
Care must be taken to properly interpret this cohomology, because $\CM_\nu$ is noncompact.%
\footnote{Supersymmetric quantum mechanics suggests that the Hilbert space actually consists of $L^2$ harmonic forms on $\CM_\nu$.} %
Such subtleties disappear, however, once complex masses $m_\C$ and the $\Omega$-deformation parameter $\epsilon$ are turned on. Physically, their effect is to make the quantum mechanics fully massive. Both $m_\C$ and $\epsilon$ play the role of twisted masses (scalar fields in background vectormultiplets) associated to the symmetries \eqref{symM} of the space $\CM_\nu$. Namely, $m_\C \in \mathfrak t_\C^{(H)}$ generates an infinitesimal $T_H$ rotation and $\epsilon \in \C$ generates a $U(1)_\ep$ rotation.  The resulting massive vacua of the supersymmetric quantum mechanics are identified as the fixed points of these symmetries on $\CM_\nu$.

Mathematically, in the presence of twisted masses $m_\C$ and $\epsilon$ the Hilbert space is identified as the  equivariant cohomology of the moduli space of generalized vortices,
\be \CH_\nu = H^*_{T_H\times U(1)_\epsilon}(\CM_\nu)\,. \ee
The equivariant cohomology has a distinguished basis, whose elements $|p\rangle$ correspond to cohomology classes supported on each fixed point $p$ of $T_H\times U(1)_\epsilon$.%
\footnote{Ordinarily in mathematics (\cf\ \cite{AtiyahBott}) the equivariant cohomology of a point $H^*_G(p)$ is an infinite-dimensional space, generated by invariant polynomials in the Lie algebra $\mathfrak g_\C$. In our case, this would mean polynomials in $m_\C$ and $\epsilon$. However, because $m_\C$ and $\epsilon$ are fixed parameters rather than dynamical fields, such polynomials are just complex numbers and do not correspond to new states. For example, $m_\C|p\rangle$ is just a rescaling of the state $|p\rangle$. In contrast, had we \emph{gauged} the $T_H$ symmetry (say), $m_\C$ would be promoted to a dynamical field and we would have found the usual infinity of states.\label{foot:gauge}} %
Thus
\be \label{Hilb-fixed} \CH_\nu = \bigoplus_{\text{fixed $p$}} \C\,|p\rangle\,.\ee

There is a slight ambiguity in the normalization of fixed-point states. One natural option is to take $|p\rangle$ to denote the Poincar\'e dual of the fundamental class of the fixed point $p\in \CM_\nu$, \ie\ an equivariant delta-function $\bm\delta_p$ supported at $p$. In terms of the inclusion map $i:\{p\}\hookrightarrow \CM_\nu$, one would say that $|p\rangle$ is the push-forward of the fundamental class of the point,
\begin{subequations}\label{normp}
\be |p\rangle = \bm\delta_p =  i_*(1\!\!1_p)\,.\ee
Alternatively, we could normalize $|p\rangle$ by the Euler class of the normal bundle to $p$ in~$\CM_\nu$,
\be |p\rangle = \frac{1}{e(N_p)}\bm\delta_p = \frac{1}{\omega_p(m_\C,\epsilon)}\bm\delta_p\,, \ee
\end{subequations}
where $\omega_p(m_\C,\epsilon)$ denotes the equivariant weight of the normal bundle. This normalization is dual to (\ref{normp}a), in the sense that the pull-back $i^*(|p\rangle)=1\!\!1_p$ is the fundamental class of the point. Notice that the combined operation $i^*i_* 1\!\!1_p = e(N_p)1\!\!1_p$ is multiplication by the Euler class.

From a physical perspective, neither normalization is especially preferred, but a choice must be made. Almost exclusively throughout this paper we will use (\ref{normp}b).

The Hilbert space \eqref{Hilb-fixed} has a natural inner product coming from the supersymmetric quantum mechanics: the overlap of states $\langle p'| p\rangle$ is given by computing the path integral of the supersymmetric quantum mechanics with a state $|p\rangle$ at $t\to -\infty$ and a state $\langle p'|$ at $t\to\infty$.
In terms of equivariant cohomology, $\langle p'| p\rangle$ is given by the equivariant integral $\int_{\CM_\nu}$ of the product of classes representing $\langle p'|$ and $|p\rangle$.
If we use the convention (\ref{normp}b) for both $\langle p'|$ and $|p\rangle$, then the inner product is
\be \langle p' | p\rangle =   \frac{\delta_{p,p'}}{\omega_p(m_\C,\ep)}\,,\ee
where as before $\omega_p = e(N_p)$ is the equivariant weight of the normal bundle to $p$ under the combined $T_H\times U(1)_\epsilon$ action.
\footnote{Notice that by describing the space of ground states as the cohomology of a supercharge, we lost track of unitary. Thus our $\CH$ is not a Hilbert space in the formal sense, but we will continue using this terminology. }

In the remainder of this section, we give some explicit descriptions of $\CM_\nu$ and its fixed points for abelian and some basic nonabelian theories.

\subsection{Example: SQED}
\label{sec:abel-moduli}

Let us consider $G=U(1)$ with $N\geq 1$ hypermultiplets $(X_i,Y_i)$ of charges $(+1,-1)$ and introduce a negative real FI parameter $t_\R<0$. The Higgs branch is a hyper-K\"ahler quotient given by imposing the moment map constraints
\be
\mu_\C := \sum_iX_i Y_i = 0,\,\qquad \mu_\R :=  \sum_i |X_i|^2-|Y_i|^2 = -t_\R 
\ee
and dividing by the $U(1)$ gauge symmetry. This gives a description of the Higgs branch as the cotangent bundle $\cM_H=T^*\cp^{N-1}$ with the compact base parameterized by the coordinates $X_i$ at $Y_i=0$. Algebraically, we can impose the complex moment map constraint $\mu_\C=0$ together with the stability condition $X\neq 0$, and divide by $G_\C =GL(1,\C) = \C^*$. 

The Higgs-branch flavor symmetry is $PSU(N)$, and we choose a maximal torus $T_H = \big[\prod_i U(1)_i\big]/U(1)$ such that $(X_i,Y_i)$ have charges $(+1,-1)$ under the $U(1)_i$, and zero under all other $U(1)_j$. Correspondingly, we introduce complex masses $(m_1,...,m_N)\in \mathfrak t^{(H)}_\C$. The vacuum equations require the hypermultiplets to be invariant under a simultaneous gauge and flavor transformation,
\be  (\varphi+m_{i})X_i = 0  \qquad -(\varphi+m_{i})Y_i = 0 \,.\ee
When $t_\R<0$ and the masses are generic, there are $N$ massive vacua
\be \nu_j:\qquad X_i=\sqrt{-t_\R}\delta_{ij} \qquad Y_i=0 \qquad \varphi=-m_j \,, 
\ee
which are the isolated fixed points of $T_H$. In the algebraic description of the Higgs branch, they correspond to the coordinate hyperplanes in the base.

Let us consider vortices that tend to a vacuum, say $\nu_1$, at spatial infinity.
Following the algebraic description of Section \ref{sec:stack}, we first choose a $GL(1,\C)$ bundle on $\cp^1$, which is classified by a vortex number $\n \in \pi_1(U(1))\simeq \Z$. The fields $X_i,Y_i$ become sections of an associated bundle in the representation $R\oplus \bar R$, namely $\CO(\n)^N\oplus \CO(-\n)^N$, and therefore $X_i(z)$ and $Y_i(z)$ are polynomials of degrees at most $\n$ and $-\n$, respectively. The moduli space $\CM_{\nu_1}^\n$ is the space of such polynomials satisfying the complex moment map constraint $\sum_i X_i Y_i=0$ and hitting the vacuum $\nu_1$ as $z\to \infty$. There are several options:
\begin{itemize}
\item If $\n > 0$ then only the $X_i(z)$ can be nonzero. Hitting the vacuum $\nu_1$ requires the leading coefficients of $X_{i}(z)$ with $i\neq 1$ to vanish while the leading coefficient of $X_1(z)$ is nonvanishing. A constant complex gauge transformation sets
\be \label{QEDpoly}
X_1(z) = z^\sn + \sum_{l=0}^{\sn-1}x_{1,l} z^l \, , \qquad
 X_i(z) = \sum_{l=0}^{\sn-1}x_{i,l} z^l \quad (i \neq 1)\,.
\ee
The coefficients $x_{i,l}$ are unconstrained and parameterize $\CM_{\nu_1}^\sn \simeq \C^{N\sn}$.

\item If $\n=0$, both $X_i$ and $Y_i$ are sections of $\CO(0)$, and hence may be nonzero constants. However, the requirement that they hit the vacuum $\nu_1$ at infinity sets them equal to their vacuum values $X_i \sim \delta_{i1}$, $Y_i=0$. Thus $\CM_{\nu_1}^0$ is a point.

\item If $\n< 0$ then only the $Y_i(z)$ can be nonzero. This is incompatible with the vacuum $\nu_1$, so $\CM_{\nu_1}^\sn$ is empty.

\end{itemize}
If instead $t_\R >0$, the vacuum $\nu_1$ would have $Y_i\sim \delta_{i1}$ and $X_i=0$, and the component $\CM_{\nu_1}^\sn$ would be empty for positive $\n$. In general,
\be \text{$\CM_{\nu_j}^\sn$ nonempty\; $\Leftrightarrow$\; $t_\R\cdot \n \leq 0$\,.} \label{QEDpos} \ee

In order to determine the Hilbert space $\CH$ in the presence of complex masses and the $\Omega$-background, we must find the fixed points of the $T_H\times U(1)_\epsilon$ action on $\CM_{\nu_1}^\n$. We analyze this action by considering combined $G\times T_H\times U(1)_\epsilon$ transformations of the fields $X_i(z)$, and compensating for $T_H\times U(1)_\epsilon$ rotations with gauge transformations. An infinitesimal transformation with parameters $\varphi,m,\epsilon$ (respectively) sends
\be X_i(z) \mapsto (\varphi+m_i + \tfrac\epsilon2 + \epsilon z\pd_z) X_i(z)\,. \label{QED-action} \ee
For $\n\geq 0$, there is a unique fixed point $X_1(z) = z^\n$ and $X_{i}(z)=0$ for $i\neq 1$, with compensating gauge transformation $\varphi = -m_1 - (\n+\frac12)\epsilon$. The fixed point is therefore just the origin of the space $\CM_{\nu_1}^\n = \C^{N\n}$. Denoting the corresponding state in the quantum mechanics as $|\n\rangle$, we therefore find that
\be \CH = \bigoplus_{\n\geq 0} \C\,|\n\rangle\,. \ee

The tangent space to the origin in $\CM_{\nu_1}^\n$ is parameterized by the remaining coefficients $x_{i,l}$ in \eqref{QEDpoly}, which transform as
\be \begin{array}{rl} x_{i,l} &\,\mapsto\, (\varphi+m_i+(l+\tfrac12)\epsilon)x_{i,l} \\ &\, = \,(m_i-m_1+(l-\n)\epsilon)x_{i,l} \end{array} \ee
under an infinitesimal $T_H\times U(1)_\epsilon$ rotation. Therefore, the inner product on the Hilbert space is given by
\be \langle \n'| \n\rangle = \delta_{\n',\n} \prod_{i=1}^N\prod_{l=0}^{\n-1} \frac{1}{m_i-m_1+(l-\n)\epsilon}\,. \ee
It is convenient to introduce a characteristic polynomial for the flavor symmetry, $P(x) = \prod_{i=1}^N (x+m_i)$, and write this as
\be \langle \n'| \n\rangle = \delta_{\n',\n} \prod_{l=0}^{\n-1} \frac{1}{P(\varphi+(l+\tfrac12)\epsilon)}\,,\ee
with $\varphi = -m_1 - (\n+\frac12)\epsilon$ given by its value at the fixed point.

\subsection{Nonabelian theories}
\label{sec:nonab-moduli}

The analysis of non-abelian gauge theories is more intricate as there is a rich space of polynomial gauge transformations preserving holomorphic gauge. 

A straightforward approach is to simply fix this additional freedom completely, as in the `moduli matrix' approach of \cite{Eto-matrix,Eto-moduli}. Here, the idea is to work on the complex plane $\C$ rather than $\cp^1$, so that the $G_\C$-bundle can be fully trivialized. The polynomial matrices $X(z)$, $Y(z)$ are parameterized in such a way that no residual gauge symmetries remain; and as $z\to\infty$ these matrices are required to approach a fixed, chosen lift of the vacuum $\nu$. In the case of $G=U(K)$, this leads to a cell decomposition of the vortex moduli space
\be \label{M-cells} \CM_\nu^\n \;= \bigcup_{k} \CM_\nu^{\n,k}\,, \ee
where each cell $\CM_\nu^{\n,k}$ is labelled by a cocharacter $k=(k_1,...,k_N) \in \Z^K$ such that $\n = \sum_{i=1}^N k_i$. We call $k$ the \emph{abelianized} vortex number. Each cell has a unique fixed point of the $T_H\times U(1)_\epsilon$ symmetry.  We consider $U(K)$ with $N$ hypermultiplets in more detail below. This approach can also be extended to quiver gauge theories with unitary gauge groups.

An alternative approach proceeds by decomposing $\CM_\nu$ as a union of fibers of the map
\be
U: \CM_\nu \to \text{Bun}_{G_\C}(\cp^1)
\ee
from \eqref{MBunG}. As we will see the points of $\text{Bun}_{G_\C}(\cp^1)$ and hence the fibers of $U$ are not necessarily closed so we will get a stratification of $\CM_\nu$.

To be more explicit, we first need to understand a few basic facts about $G_{\C}$-bundles on $\cp^1$. Let $T_{\C}$ be a maximal torus of $G_{\C}$, let $\Lambda_{\rm cochar}$ be the lattice of cocharacters of $T_{\C}$, and let $W$ be the Weyl group of $G_{\C}$. A result of Grothendieck \cite{Groth-class} states that any $G_\C$-bundle $E$ on $\cp^1$ admits a reduction of structure group to $T_{\C}$ and hence the set of isomorphism classes of $G_{\C}$-bundles on $\cp^1$ is in bijection with $\Lambda_{\rm cochar}/W$. More concretely, a reduction of $E$ to $T_{\C}$ consists of trivializations of $E$ on the two hemispheres such that the gauge transformation relating them is valued in $T_{\C}$ and hence defines an \emph{abelianized vortex number} $k \in \Lambda_{\rm cochar}$. By composing $k$ with the embedding $T_\C\subset G_\C$ one can compute the topological vortex number $\n(k) \in \pi_1(G)$.\footnote{Alternatively, one may recall that $\pi_1(G)$ is isomorphic to a quotient of the cocharacter and coroot lattices of $G$, $\pi_1(G) \simeq \Lambda_{\rm cochar}/\Lambda_{\rm coroot}$. The quotient induces the map $k\mapsto \n(k)$.} For example, if $G=U(K)$, the abelianized vortex number takes values $k=(k_1,...,k_K) \in \Z^K$ and $\n(k) = \sum_{i=1}^N k_i$.

Thus we have a decomposition
\be \label{M-strata} \CM_\nu^\n \;= \bigcup_{\text{$[k] \in \Lambda_{\rm cochar}/W$ such that $\n(k)=\n$}} \CM_\nu^{\n,[k]}\,,\ee
where $\CM_\nu^{\n,[k]} := U^{-1}([k])$. As mentioned earlier a point $[k] \in \text{Bun}_{G_\C}(\cp^1)$ and hence the fiber $\CM_\nu^{\n,[k]}$ is only locally closed in general. In particular, each Weyl orbit has a unique dominant representative and a $G_{\C}$-bundle with abelianized vortex number $[k]$ can deform to a bundle with abelianized vortex number $[k']$ if and only if $k'$ is greater than $k$ in the standard order on dominant cocharacters.  

To understand the fibers $\CM_\nu^{\n,[k]}$, notice that once we reduce the structure group of a $G_\C$ bundle $E$ to $T_\C$, the associated bundle in the representation $R\oplus \bar R$ splits as a direct sum of line bundles,
\be \label{nonab-RR*} \bigoplus_{\lambda_i} \CO(\langle k,\lambda_i\rangle) \oplus \CO(-\langle k,\lambda_i\rangle)\,,   \ee
where $\{\lambda_i\}$ are the weights of $R$ with respect to $T_\C$, and $\langle k,\lambda_i\rangle \in \Z$ is the natural pairing between $k\in \text{Hom}(\C^*,T_\C)$ and $\lambda_i\in \text{Hom}(T_\C,\C^*)$. Similar to the abelian case, we now consider the space of polynomials $X_i(z)$, $Y_i(z)$ of degrees $\leq \langle k,\lambda_i\rangle$ and $\leq -\langle k,\lambda_i\rangle$, respectively, such that
\begin{itemize}
\item[a)] the complex moment map vanishes, $\mu_\C=0$
\item[b)] at $z=\infty$, the sections $X_i(z)$, $Y_i(z)$ lie in the orbit $G_\C \cdot \nu$ of a vacuum $\nu$.
\end{itemize}

Quite unlike the abelian case, there typically remains a large group of unbroken gauge transformations that must still be accounted for. These come from automorphisms of $E$. If $z^k\in T_\C[z,z^{-1}]$ is the $T_\C$-valued gauge transformation on the equator of $\cp^1$ coming from the reduction of structure the gauge transformations are elements of
\be P_k =  G_\C[z] \,\cap\, z^k G_\C[z^{-1}]z^{-k}\,, \label{triang-gauge} \ee
where $G_\C[z]$ denotes group elements with polynomial entries. These transformations preserve the degrees of the $X_i,Y_i$ polynomials. The fiber $\CM_\nu^{\n,[k]}$ is precisely the space of polynomials satisfying the conditions above, modulo $P_k$. Each fiber contains a number of fixed points for the $T_H\times U(1)_\epsilon$ symmetry, which we will describe more explicitly in examples below.

Either the cell decomposition \eqref{M-cells} or the decomposition into strata \eqref{M-strata} can be used to analyze fixed points and to construct the Hilbert space \eqref{Hilb-fixed}. However, we warn readers that the two are not globally compatible --- the cells of \eqref{M-cells} usually cut across multiple strata of \eqref{M-strata}.

\subsubsection{SQCD via moduli matrix}
\label{sec:SQCD-moduli}

As an example, we consider $G=U(K)$ with $N\geq K$ fundamental hypermultiplets $(X^a{}_i,Y^i{}_a)$, where $1\leq i\leq N$, $1\leq a\leq K$. Introducing a negative FI parameter $t_\R<0$, the Higgs branch is a hyper-K\"ahler quotient describing the cotangent bundle $T^*Gr(K,N)$. The real moment-map constraint requires $X$ to have maximal rank; this provides the stability condition in the algebraic description of the Higgs branch, and we have
\be \CM_H \simeq
\{YX = 0\,,\; \text{rank}(X)=K\}/GL(K,\C) \, . \ee

The flavor symmetry is $G_H = PSU(N)$ and we may introduce complex masses $(m_1,...,m_N)$ valued in a Cartan subalgebra. The classical vacuum equations require that $\varphi = \text{diag}(\varphi_1,...,\varphi_K)$ is diagonal and that the hypermultiplets are invariant under a combined gauge and flavor transformation:
\be \label{sigmaphi-QCD}  (\varphi_a+m_{i})X^a{}_i = 0 \qquad (\varphi_a+m_{i})Y^i{}_a = 0\,. \ee
In order to satisfy both the stability condition and equation \eqref{sigmaphi-QCD} in the presence of generic mass parameters, exactly $K$ entries of the matrix $X$ in distinct rows and columns must be nonzero.
The possible choices of nonvanishing entries are labelled by subsets $I = \{ i_1,\ldots, i_K\} \subset \{ 1,\ldots, N\}$ of size $K$. Therefore, there are ${N \choose K }$ distinct massive vacua, of the form
\be \nu_I \quad : \quad  \varphi_a=-m_{i_a} \qquad X^a{}_i = \delta_{i,i_a}\,,\qquad Y^i{}_a = 0 \, . \ee
These vacua are the fixed points of a maximal torus $T_H = U(1)^N/U(1)$ of the flavor symmetry acting on the Higgs branch. 

The connected components $\CM_{\nu_I}^\n$ of the vortex moduli space associated to vacuum $\nu_I$ are labelled by an integer $\n \in \pi_1(U(K))= \Z$. When $t_\R<0$, the only nonempty moduli spaces are $\CM_\nu^{\n}$ for $\n\geq 0$. In each component, the vortices are parameterized by the polynomial-valued matrix $X(z)$, with $Y(z)$ set to zero, modulo gauge transformations. Let us denote the $K \times K$ minor formed from the columns $J=\{ j_1,\ldots,j_k\}$ by $X_J(z)=\text{det}|\!|X^a{}_j|\!|^{a\in\{1,...,K\}}_{j\in J}$. Then we have the vacuum condition 
\be  \text{deg}\left( X_J(z) \right) = \begin{cases} \n  & \mathrm{if} \quad J = I \\
\leq  \n-1\, & \mathrm{if} \quad J \neq I  \end{cases}
\label{QCD-vac-cond}
\, .\ee
Due to Pl\"ucker identities, it is only necessary to impose the second condition for minors involving just one column outside of $I$.

For simplicity, let us consider the vacuum $\nu_I$ with $I = \{1,\ldots,K\}$. Then polynomial gauge transformations can be used to bring any such matrix into a canonical `triangular' gauge-fixed form with square part
\be \label{QCD-X-mm}
 X^a{}_i(z) = \begin{cases}   0 & i <a  \\
 z^{k_a} + \sum_{l=0}^{k_a-1} x^a{}_{i,l} z^l & i=a  \\
   \sum_{l=0}^{k_i-1} x^a{}_{i,l} z^l & a<i\leq K  \\
\end{cases}
\ee
for some non-negative integers $(k_1,...,k_K)$ satisfying $\sum_a k_a=\n$. The remaining columns with $i =K+1,\ldots,N$ are fixed by the second condition in~\eqref{QCD-vac-cond}. 
For example, if $K=2$ and $N=3$, the canonical form looks like
\be  X =  \left( \begin{array}{c@{\quad}c@{\quad}c} z^{k_1}+... & a z^{k_2-1}+...  & * \\
  0 & z^{k_2}+... & *  \end{array} \right)\,, \ee
where `$...$' indicates lower-order terms. The unconstrained coefficients in the matrix $X^a{}_i(z)$ parameterize a cell $\CM_\nu^{\n,k}$ in the vortex moduli space.

Every cell $\CM_\nu^{\n,k}$ has a unique fixed point. To see this, we note that the combined action of $T_H\times U(1)_\epsilon$ and the maximal torus $T\subset G$ of the gauge group sends
\be X^a{}_i(z) \;\mapsto\; (\varphi_a + m_i + (z\pd_z+\tfrac12)\epsilon)\,X^a{}_i(z)\,. \label{QCD-transX}\ee
The origin $X^i{}_a(z) = \delta_{i,i_a} z^{k_a}$ is the unique fixed point of \eqref{QCD-transX}, with a compensating gauge transformation
\be  \varphi_a =  - m_a - (k_a+\tfrac12)\epsilon  \qquad a = 1,\ldots, K \,. \label{QCD-detphi} \ee
The weights of the tangent space at the fixed point can be computed by observing that small deformations away from the fixed point are parameterized as
\be \label{QCD-X-def}
 X^a{}_i(z) = \begin{cases} z^{k_a} + \sum_{l=0}^{k_a-1} \delta x^a{}_{i,l} z^l & i=a\,, \\
   \sum_{l=0}^{k_i-1} \delta x^a{}_{i,l} z^l & i \in \{1,\ldots,K\} -\{a\} \,, \\
   \sum_{l=0}^{k_a-1} \delta x^a{}_{i,l} z^l & i \in  \{ K+1,\ldots,N \} \,. \end{cases}
\ee
This matrix only obeys the vacuum condition to linear order in the small deformations, which is adequate to describe the tangent space at the fixed point.
Now multiplying the weights of the coordinates in \eqref{QCD-X-def} we find, after a small calculation, that the equivariant weight of the tangent space is
\begin{align}
 \omega_{\n,k} &=  \prod_{a<b}(-1)^{k_a-k_b} \frac{\varphi_a-\varphi_b+(k_a-k_b)\ep}{\varphi_a-\varphi_b}  \prod_{a=1}^K \prod_{l=0}^{k_a-1}   P(\varphi_a+(l+\tfrac12)\epsilon) \notag \\
  &= (-1)^{\n K} \prod_{a<b} \frac{m_b-m_a}{\varphi_a-\varphi_b}  \prod_{a=1}^K \prod_{l=0}^{k_a-1}   P(\varphi_a+(l+\tfrac12)\epsilon)
 \label{QCD-weights}
\end{align}
with $P(x) = \prod_{i=1}^N(x+m_i)$ and $\varphi_a$ evaluated as in \eqref{QCD-detphi}. 

Now each fixed point contributes a state $|\n,k\rangle$ to the Hilbert space of the supersymmetric quantum mechanics, labelled by an integer $\n\geq 0$ and non-negative integers $k_a$ such that $\sum_a k_a=\n$. The component of the Hilbert space with fixed $\n$ thus has dimension ${\n + K-1\choose K-1}$. The inner product of states is given by the inverse of the equivariant weight,
\be \langle \n,k|\n',k'\rangle = \delta_{\n,\n'}\delta_{k,k'} \frac{1}{\omega_{\n,k}}\,.
\label{QCD-innerprod}
\ee

\subsubsection{SQCD via strata}
\label{sec:SQCD-strata}

We can reproduce the same result as an example of the more sophisticated approach that we expect applies more broadly. As explained above, the space $\CM_\nu^\n$ admits a decomposition into fibers labelled by a reduction of the structure group of the gauge bundle. For $G = U(K)$, this corresponds to a set of integers $(k_1,...,k_K)\in \Z^K$ modulo permutations that satisfy $\sum_a k_a = \n$. We can denote the equivalence class under permutations by $[k]$ and write
\be \CM_\nu^\n \; = \bigcup_{[k]} \CM_\nu^{\n,[k]}\,. \label{QCD-strata}\ee

In the stratum corresponding to a given $[k]$, the hypermultiplets $X,Y$ are matrices of polynomials whose entries $X^a{}_i(z)$ and $Y^i{}_a(z)$ have degrees $\leq k_a$ and $\leq -k_a$, respectively. In order for $X,Y$ to lie in the $G_\C$-orbit of the vacuum $\nu$ at infinity, we must have $k_a\geq 0$ for all $a$. This implies that
\be \text{$\CM_\nu^\n$  nonempty} \quad \Leftrightarrow\quad \n\geq 0\,,\ee
and that for each $\n\geq 0$ there are finitely many nonempty strata $\CM_\nu^{\n,[k]}$. These strata are labelled by partitions of $\n$, \ie\ by Young diagrams $k$ of size $\n$. (For a general FI parameter we would have $\CM_\nu^\n$ nonempty if and only if $t_\R\cdot\n<0$.) 

Since $\deg Y^i{}_a\leq -k_a$ and $k_a$ are nonnegative, we also find that the the $Y$'s must be constant, possibly zero. The condition that $Y$ lies in the $G_\C$ orbit of $\nu$ as $z\to \infty$ then implies that $Y$ vanishes identically. The complex moment-map constraint $YX=0$ is satisfied automatically.

As before, we concentrate on the vacuum $\nu_I$ with $I = \{1,\ldots,K\}$. The complex orbit $G_\C \cdot \nu_I$ consists of matrices $X$ with nonvanishing leading minor $X_I(z) \neq 0$ and $X^a{}_i=0$ for $i>K$.
Therefore, we find that the stratum $\CM_\nu^{\n,[k]}$ is the space of polynomial matrices $X^a{}_i(z)$ with
\be \label{stratum-QCD}
\begin{cases}\text{deg}\,X^a{}_i \leq k_a & i\leq a \\
 \text{deg}\,X^a{}_i < k_a & i>a \end{cases}\quad \text{and}\quad
 \text{deg}\left( X_I(z) \right)  = \n\,,
\ee
modulo residual gauge transformations. The residual gauge transformations as in \eqref{triang-gauge} are polynomial matrices $g^a{}_b(z)$ such that $\text{deg}\,g^a{}_b \leq k_a-k_b$. These are the transformations that preserve the degrees in \eqref{stratum-QCD}. 

For example, if $K=2$, $N=3$, and $k= (0,2)$, we can parametrize the stratum by matrices of the form
\be \label{QCD-X-eg}
X(z) = \left( \begin{array}{c@{\quad}c@{\quad}c} a & a'  & 0 \\
  c z^2+dz+e & c' z^2+d'z+e' & d''z+e''  \end{array} \right)
\ee
such that $ac'-a'c\neq 0$, modulo gauge transformations of the form
\be g(z) = \left( \begin{array}{c@{\quad}c} \alpha & 0 \\ \beta z^2+\gamma z+\eta & \delta  \end{array} \right) \,.\ee
The stratum can be covered by two coordinate charts, corresponding to $a\neq 0$ or $a'\neq 0$ in \eqref{QCD-X-eg}. If $a\neq 0$ then gauge transformations can be fixed by bringing $X$ to a canonical form
\begin{subequations}\label{QCD-X-fix}
\be X(z) = \left( \begin{array}{c@{\quad}c@{\quad}c} 1 & a'  & 0 \\
  0 & z^2+d'z+e' & d''z+e''  \end{array} \right)\,. \ee
On the other hand, if $a'\neq 0$ then $X$ can be brought to the form
\be X(z) = \left( \begin{array}{c@{\quad}c@{\quad}c} \tilde a & 1  & 0 \\
   z^2+\tilde dz+\tilde e & 0 & \tilde d''z+\tilde e''  \end{array} \right)\,. \ee
\end{subequations}
Globally, the stratum is the total space of the line bundle $\CO(0)^{\oplus 2}\oplus \CO(1)^{\oplus 2}\to\cp^1$, where $[a:a']$ are homogeneous coordinates on the base, $d',e'$ are sections of $\CO(0)$, and $d'',e''$ are sections of~$\CO(1)$.

Let us now reconsider the fixed points and the equivariant weights of their tangent spaces. This does not require a understanding the global structure of the strata. The combined action of $T_H\times U(1)_\epsilon$ and $T\subset G$ is given in~\eqref{QCD-transX}. First, a fixed point of $U(1)_\epsilon$ requires the entries of $X(z)$ to be monomials. Moreover, in order for $X(z)$ to be a fixed point of $T_H$ (modulo the $T\subset G$ action), at most $K$ entries of the leading $K\times K$ block of $X$ can be nonzero, one in each row and each column. In order to be compatible with the vacuum $\nu$, exactly $K$ such entries must be nonzero, and must be monomials of maximal degree. Thus, the fixed points are
\be X^a{}_i = \delta_{i,i_a} z^{k_a}  \, ,\ee
where $i_a$ is any permutation of $1,...,K$.

This would suggest that there are $K!$ fixed points per stratum. However, if some of the abelian vortex numbers are equal (\eg\ $k_a=k_b$) then residual gauge transformations contain elements of the Weyl group that identify corresponding fixed points (\eg\ swapping $i_a\leftrightarrow i_b$). Altogether, the distinct fixed points in $\CM_\nu^{\n,[k]}$ end up in 1-1 correspondence with points in the Weyl orbit of $k$. Taking a union over all strata, we find that the fixed points in $\CM_\nu^\n$ are labelled by all sets of non-negative integers $(k_1,\ldots,k_K)$ such that $\sum_a k_a=\n$. Given a cocharacter $k$, complex gauge transformations can be used to bring the corresponding fixed point to a canonical form
\be |\n,k\rangle \quad\leftrightarrow\quad \begin{cases} X^a{}_i = \delta^a{}_i z^{k_a} & i\leq k\,, \\ X^a{}_i = 0 & i>k\,,\end{cases}\qquad Y=0\,. \label{QCD-fix-can} \ee
as found previously.

Returning to our previous example with $K=2$ and $N=3$, the moduli space of vortices with $\n=2$ has three fixed points, corresponding to $k=(2,0)$, $(1,1)$ and $(0,2)$. The fixed points $(2,0)$ and $(0,2)$ lie in the stratum $\CM_\nu^{2,[0,2]}$ and are given by the points
\be (2,0)\,:\quad X(z) = \begin{pmatrix} 0 & 1 & 0 \\ z^2 & 0 & 0 \end{pmatrix}\,,\qquad (0,2)\,:\quad X(z) = \begin{pmatrix} 1 & 0 & 0 \\ 0 & z^2 & 0 \end{pmatrix}\,,\ee
in the coordinate charts (\ref{QCD-X-fix}b) and (\ref{QCD-X-fix}a), respectively. The remaining fixed point
\be (1,1)\,:\quad X(z) =\begin{pmatrix} z & 0 & 0 \\ 0 & z & 0 \end{pmatrix} \ee
lies in the stratum $\CM_\nu^{2,[1,1]}$.

Computation of the corresponding equivariant weights is performed in three steps:
\begin{enumerate}
\item  Lift the fixed point to the space of polynomials in \eqref{stratum-QCD}, and compute the weights of its tangent space there. 
\item Remove weights of the residual polynomial gauge transformations to compute the weight of the tangent space in the stratum $\CM_\nu^{\n,[k]}$.
\item Add weights corresponding to deformations of the $G_\C$-bundle  that parameterize the normal bundle of the stratum $\CM_\nu^{\n,[k]}$ inside $\CM_\nu^\n$.
\end{enumerate}
This gives the weight of the tangent space at the fixed point in $\CM_\nu^\n$.

To be explicit, consider the fixed point $|\n,k\rangle$, presented in the canonical form \eqref{QCD-fix-can}. As always, we concentrate on the vacuum with $I=\{1,\ldots,K\}$. A neighborhood of the fixed point in the space of polynomials~\eqref{stratum-QCD} is parameterized by
\be \begin{cases} X^a{}_i = \sum_{l=0}^{k_a}x^a{}_{i,l} z^l & i \leq K\,, \\ X^a{}_i  = \sum_{l=0}^{k_a-1} x^a{}_{i,l} z^l & i>K\,. \end{cases}\ee
from which we obtain the equivariant weight
\be \begin{array}{l} \ds \prod_{i=1}^K \prod_{a=1}^K \prod_{l=0}^{k_a} (\varphi_a+m_i+(l+\tfrac12)\epsilon) \prod_{i=K+1}^N \prod_{a=1}^K \prod_{l=0}^{k_a-1} (\varphi_a+m_i+(l+\tfrac12)\epsilon) \\[.1cm]
  \hspace{1in} \ds  = \prod_{a=1}^K \prod_{i=1}^K (\varphi_a+m_i+(k_a+\tfrac12)\epsilon) \prod_{a=1}^K \prod_{l=0}^{k_a-1}   P(\varphi_a+(l+\tfrac12)\epsilon)\,, \end{array} \label{QCD-w1} \ee
where $P(x):=\prod_{i=1}^N (x+m_i)$ is again the characteristic polynomial for the flavor symmetry. 

The residual gauge transformations contain polynomials of degree $\leq k_a-k_b$, and transform in the adjoint representation of the gauge group. Thus, they remove the following weights
\be  \prod_{a,b=1}^K \prod_{l=0}^{k_a-k_b} \frac{1}{\varphi_a-\varphi_b+l\epsilon}\,. \label{QCD-w2} \ee
Combining these two contributions gives the equivariant weight of the tangent space to the fixed point lying inside the stratum $\CM_\nu^{\n,[k]}$.

Finally, $G_\C$-bundle deformations are adjoint-valued 1-forms that can be added to the holomorphic connection $A_z$. The matrix elements of a deformation are holomorphic one-forms valued in the bundles $\CO(k_a-k_b)$ on $\cp^1$. From Serre duality, this is equivalent to global sections of $\CO(k_b-k_a-2)$ and therefore they contribute additional weights
\be \prod_{a,b=1}^K \prod_{l=0}^{k_b-k_a-2} (\varphi_a-\varphi_b-(l+1)\epsilon) \label{QCD-w3} \ee
Putting together the three contributions \eqref{QCD-w1}, \eqref{QCD-w2}, and \eqref{QCD-w3}, there are many cancellations and we finally arrive at the equivariant weight \eqref{QCD-weights} obtained via the moduli matrix description.

\subsubsection{Triangular Quivers}
\label{sec-triang-hilb}

\begin{figure}[htp]
\centering
\includegraphics[height=3cm]{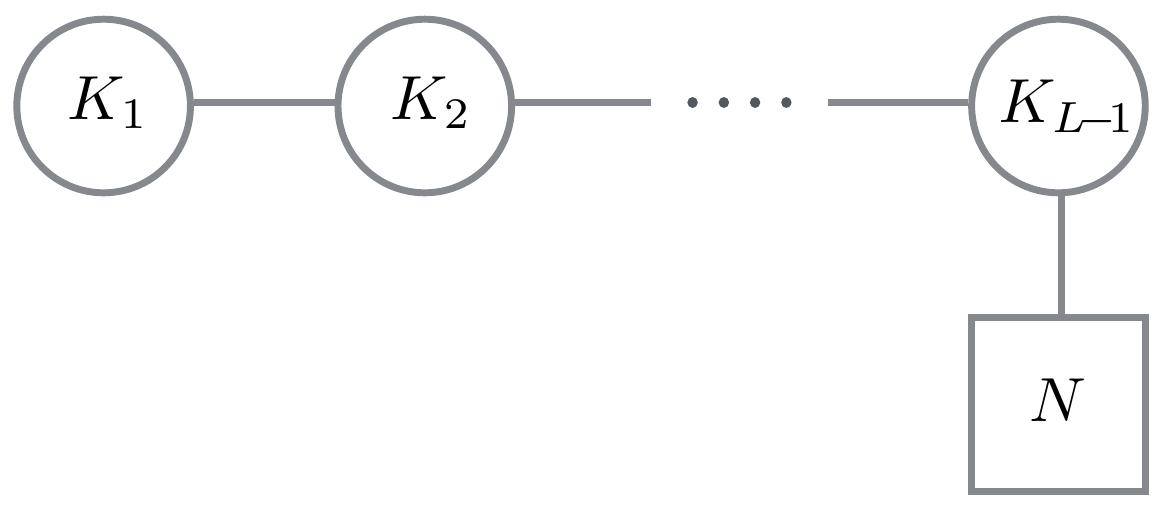}
\caption{A triangular, linear quiver.}
\label{fig:triangularquiver}
\end{figure}

The computation above can be extended to a `triangular' linear quiver with gauge group $\otimes_{\al=1}^{L-1} U(K_\al)$, hypermultiplets in the bifundamental representation of $U(K_\al) \times U(K_{\al+1})$ for $\al = 1\ldots,L-2$, and $K_L=N$ hypermultiplets in the fundamental representation of $U(K_{L-1})$. We assume that $K_1 < \ldots < K_N$. This quiver is illustrated in figure~\ref{fig:triangularquiver}. Here we are much more schematic: we only summarize the results.
 
The data of a triangular quiver can be repackaged as a partition $\rho = [\rho_1,\ldots,\rho_L]$ of $N$ with $\rho_\al = K_\al-K_{\al-1}$ and this theory is sometimes known as $T_\rho(SU(N))$. The Higgs branch flavor symmetry is $G_H = PSU(N)$. The Coulomb branch flavor symmetry is enhanced in the infrared to $G_C = S(\otimes_{j}U(\ell_j) )$ where $\ell_j$ is the number of times $j$ appears in the partition $\rho$.
 
We turn on real FI parameters $\{ t_1,\ldots,t_{L-1} \}$ such that $t_j > t_{j+1}$ and complex masses $\{m_1,\ldots,m_N\}$. The massive vacua are labelled by nested subsets
\be \label{nested-subsets}
\cI_1 \subset \cI_2 \subset \cdots \subset \cI_{L-1} \subset \cI_L =  \{ 1,\ldots,N\} \qquad 
\ee
with $| \, \cI_\al \,  | = K_\al$. We can label the elements of these subsets by $\cI_\al = \{ i_{\al,1}, \ldots, i_{\al,K_\al}\}$. The number of such vacua is given in terms of the partition $\rho$ by $N! / \rho_1! \ldots,\rho_L! $.

Solutions of the vortex equations are labelled by a vortex number for each node $\vec \n  = \{ \n_1,\ldots,\n_{L-1} \}$ and the fixed points by decompositions $\vec k  = \{ k_{\al,a} \}$ with $\sum_{a}k_{\al,a} = \n_\al$. The corresponding equivariant weights are
\be
\omega_{\vec n,\vec k} = \prod_{\al = 1}^{L-1}  \left[ \; \prod_{a<b}^{K_\al} (-1)^{k_{\al,a}-k_{\al,b}} \frac{\varphi_{\al,a}-\varphi_{\al,b}+(k_{\al,a}-k_{\al,b})\ep}{\varphi_{\al,a}-\varphi_{\al,b} }  \prod_{a=1}^{K_\al} \prod_{l=0}^{k_{\al,a}-1}   \frac{ P_{\al+1}(\varphi_{\alpha,a}+(l+\tfrac12)\epsilon) }{Q_{\al-1} (\varphi_{\alpha,a}+(l+1)\epsilon) } \;  \right]
\label{triangular-eqweight}
\ee
where 
\be
	\varphi_{\al,a} = -m_{i_{\al,a}} - (k_{\al,a}+\tfrac12)\epsilon \, ,
\ee
and in particular $\varphi_{N,i} = -m_i - \tfrac{\ep}{2}$ for the flavor node. As before, we introduce generating functions for gauge invariant operators at each node $Q_\al(z) := \prod_{a=1}^{K_\al} (z-\varphi_{\al,a})$, with $Q_0(z)=1$ by definition. In addition, we introduce polynomials $P_\alpha(z) :=\prod_{a=1}^{K_\al}(x+m_{i_{\al,a}})$ with $P_L(z) = P(z) = \prod_{a=1}^N(z+m_a)$.  As above this defines the inner product on the Hilbert space.

\section{The action of monopole operators}
\label{sec-mono}

We are now ready to explain the action of Coulomb branch operators on the Hilbert space of a 3d $\CN=4$ gauge theory in $\Omega$-background. From the perspective of supersymmetric quantum mechanics, these are half-BPS operators that preserve the Hilbert space of supersymmetric ground states. Classically, they correspond to singular solutions of the BPS equations from Section \ref{sec:BPS}.

In Section \ref{sec:Hilbert}, we showed that the Hilbert space is the equivariant cohomology of the moduli space of solutions to the time-independent BPS equations. We studied this moduli space of generalized vortices by complexifying the gauge group, removing real moment-map constraints, and fixing a holomorphic gauge $A_{\bar z} =0$. Then a vortex configuration could be described as an algebraic $G_\C$-bundle $E$ on the $z$-plane $\C_z$, together with holomorphic sections $X,Y$ of an associated $R\oplus \bar R$ bundle, such that $\mu_\C(X,Y)=0$ and a vacuum boundary condition at $|z|\to\infty$. We refer to the bundle and sections
\be \CE = (E,X,Y) \label{CE} \ee
collectively as the ``holomorphic data.''

In this section, we examine how the holomorphic data evolve in ``time'' $t = x^3$ when we impose the complete time-dependent BPS equations from Section \ref{sec:BPS}. The equation that controls their evolution is
\be
[\CD_t,\CD_{\bar z}]=0 \, .
\ee
This ensures that the holomorphic data are generically constant in time. More precisely, if we denote the holomorphic data at time $t$ by $\CE_t=(E_t,X_t,Y_t)$, we generically find that at two nearby times $t$ and $t'$, $\CE_t$ and $\CE_{t'}$ are related by a globally invertible, holomorphic gauge transformation $g(z;t,t')\in G_\C[z]$.

\begin{figure}[htb]
\centering
\includegraphics[width=2.7in]{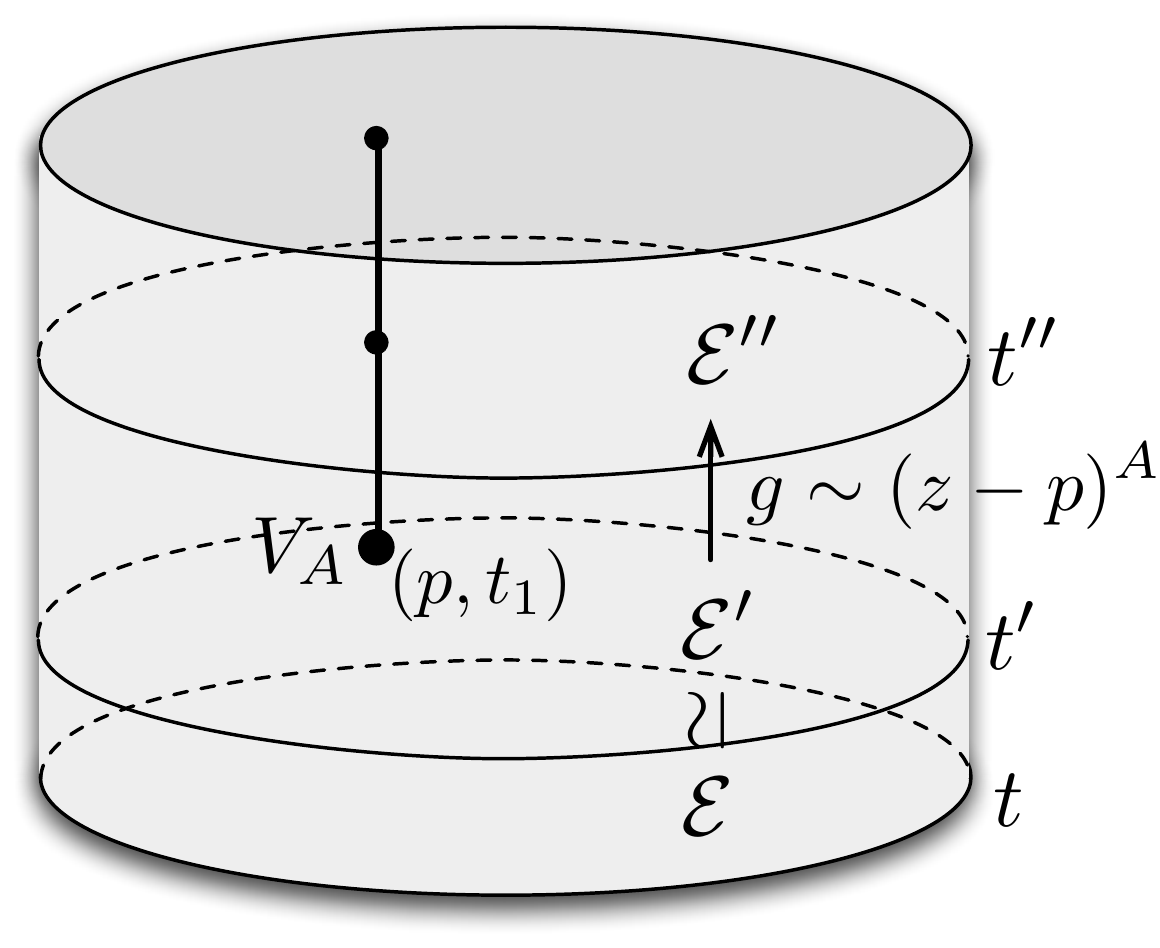}
\caption{Modification of the holomorphic data at $z=p$ and $t=t_1$.}
\label{fig:slices}
\end{figure}

At a collection of times $\{t_i\}$, however, the $G_\C$-bundle may develop a singularity and the holomorphic data can jump, as illustrated in Figure \ref{fig:slices}. This means that at nearby times $t<t_i$ and $t'>t_i$, the data $\CE_t$ and $\CE_{t'}$ are related by a gauge transformation $g(z;t,t')$ that is only invertible in the complement of some point $z=p$. For example, if the group is $G_\C = GL(N,\C)$, we might find that $\det g(z;t,t')\sim (z-p)^A$ has a zero or pole at $z=p$. One usually calls this a singular gauge transformation. In mathematics, it is known as a Hecke modification. Such modifications were analyzed by Kapustin and Witten \cite{Kapustin-Witten} in a four-dimensional lift of our current setup, with sections $X,Y$ in the adjoint representation.

In terms of the ambient 3d $\CN=4$ theory, a singular gauge transformation corresponds to the insertion of a monopole operator $V_A$ at the point $(p,t_i)$.
The monopole operator is labelled by some dominant cocharacter $A$ of $G$ (its magnetic charge), and the $G$-bundle on a small $S^2$ surrounding the monopole operator has nonzero Chern class (magnetic flux)
\be \n(A) = \frac{1}{2\pi} \int_{S^2_{p,t_i}} \Tr(F) \,\in\, \pi_1(G)\,.\ee
In the $\Omega$-background the monopole operator must be inserted at the origin $p=0$ of $\C_z$ in order to preserve $U(1)_\ep$. We then expect the monopole operator acts on vortex states in the Hilbert space as
\be V_A |\n,k\rangle \sim \sum_{w\,\in\, \text{rep}(A)} c_{A,w;\n,k} |\n+\n(A),k+w\rangle\,, \label{mon-genaction}\ee
where $w$ are weights of the finite-dimensional representation $\text{rep}(A)$ of the Langlands-dual group $G^\vee$ with highest weight $A$.

Our task in the remainder of this section is to make equation \eqref{mon-genaction} precise, determining the coefficients $c_{A,\n,k}$. We will begin in Section \ref{sec:SQED-mono} with a simple abelian example. We will then give a very general (if somewhat formal) description of the action \eqref{mon-genaction} in Section \ref{sec:corresp},
drawing on methods from topological quantum field theory.  In particular, we will find that the action of monopole operators on vortices is induced from classical \emph{correspondences} between vortex moduli spaces.

The correspondences themselves have the structure of a convolution algebra, discussed in Section \ref{sec:conv}. This leads to a new mathematical definition of the Coulomb-branch algebra $\C_\epsilon[\CM_C]$, complementary to that of Braverman-Finkelberg-Nakajima \cite{BFN-II}.

Finally, in Section \ref{sec:Verma} we explain that action of monopoles on vortices identifies each Hilbert space $\CH_\nu$ as a module for the Coulomb-branch algebra of a very special type, namely a highest-weight Verma module.

\subsection{Example: SQED}
\label{sec:SQED-mono}

A simple way to illustrate the action of monopoles on vortices and its many properties is by looking at the elementary example of $G=U(1)$ gauge theory with $N$ fundamental hypermultiplets. As in Section \ref{sec:abel-moduli}, we choose a vacuum $\nu$ in which $X_1\neq 0$ and all other hypermultiplet fields vanishing. We found there that the vortex moduli space was $\CM_\nu^\n =\C^{\n N}$ for $\n\geq 0$,
parameterized by the coefficients of
\be X_i(z) = \delta_{i,1}z^\n + \sum_{l=0}^{\n-1}x_{i,l}z^l  \,. \ee
The states in the Hilbert space $\CH_\nu = \oplus_\n \CH_\nu^\n = \oplus_\n\, \C\,|\n\rangle$ were equivariant cohomology classes $| \n \rangle$ corresponding to the fixed points at the origin of each $\CM_\nu^\n$.

Now consider the insertion of a monopole operator $v_A$ of charge $A\in \Z$ at the origin of the $z$-plane and at some time $t_*$. On a small sphere surrounding this operator we have
\be \frac1{2\pi}\int_{S^2} F = A\,, \ee
so by topological considerations alone, the operator $v_A$ must act on the basis $|\n\rangle$ by
\be
v_A | \, \n \,\ra = \begin{cases} c_{A,\n} | \, \n+A\,\ra & \mathrm{if} \; \n+A \geq 0 \\ 0 & \mathrm{if} \; \n+A <0
\end{cases} \, .
\ee
We would like to determine the non-zero coefficients $c_{A,\n}$.

As explained in general terms above, the presence of the monopole operator induces a Hecke modification of the holomorphic data. We can represent this modification as a gauge transformation
\be v_A\,:\qquad g(z) = z^A \qquad \ee
that is invertible away from the origin in the $z$-plane. The transformation must preserve the fact that $X$ and $Y$ are holomorphic sections. Since $Y=0$, we can just focus on~$X$. The effect of transformation is then summarized as follows:
\begin{itemize}
\item If $A\geq0$, the gauge transformation sends $X_i(z)\mapsto z^A X_i(z)$. This creates $A$ vortices at the origin of the $z$-plane.
\item If $A<0$, the transformation sends $X_i(z) \mapsto z^{-|A|} X_i(z)$. Regularity of this modification requires that $X_i(z)$ have a zero of order $A$ at $z=0$. In other words, there must exist $A$ vortices at the origin of the $z$-plane to be destroyed by the monopole operator.
\end{itemize}
We emphasize that not all Hecke modifications of the holomorphic bundle $E$ are allowed Hecke modifications of the full data $(E,X,Y)$.

To determine the coefficients $c_{A,\n}$ we examine the action of the singular gauge transformation in the neighborhood of the fixed points of $\CM_\nu^{\n}$ and $\CM_\nu^{\n+A}$. Note that if $A>0$ then the gauge transformation is simply a composition of singular gauge transformations of unit charge, $g(z)=z$. In terms of monopole operators, $v_A = (v_+)^A$, where $v_+$ has unit charge. Similarly, if $A<0$ then the singular gauge transformation is a composition of fundamental transformations $g(z)=z^{-1}$, hence $v_A = (v_-)^{|A|}$. Thus it suffices to determine the action of $v_+$ and $v_-$.

Thus, let us act with $g(z)=z$ on the state $|\n-1\rangle$. A vortex configuration in the neighborhood of the origin of $\CM_\nu^{\n-1}$ looks like
\be X_i(z) = z^{\n-1}\delta_{i,1} + \sum_{l=0}^{\n-2} x_{i,l+1} z^l\,,\ee
and is mapped to
\be X_i'(z) = z^\n \delta_{i,1} + \sum_{l=1}^{\n-1} x_{i,l} z^l\,.\ee
Thus the image of $g(z)$ is the subspace of $\CM_\nu^\n$ where $x_{i,0}=0$ for all $i$. In terms of equivariant cohomology, this means that the fixed-point class $|\n-1\rangle$ is mapped to the fixed-point class $|\n\rangle$, times an `equivariant delta function' that imposes the constraints $x_{i,0}=0$, and accounts for the additional directions in the tangent tangent space to the origin in $\CM_\nu^\n$. We find
\be v_+|\n-1\rangle = P(\varphi+\tfrac12\epsilon)|\n\rangle\,,\ee
where $\varphi = -m_1-(\n+\tfrac12)\epsilon$ is the value of $\varphi$ at the fixed point $|\n\rangle$.

On the other hand, acting with $g(z)^{-1}=z^{-1}$, we find that a subspace of $\CM_\nu^\n$ where $x_{i,0}=0$ maps isomorphically onto $\CM_\nu^{\n-1}$. Therefore, we expect that $v_-|\n\rangle = |\n-1\rangle$ for $\n>0$, and $v_-|0\rangle =0$.

More formally, we may observe that acting with $g(z)=z$ embeds each moduli space $\CM_\nu^\n$ as a subspace of the moduli space $\CM_\nu^{\n+1}$\,:
\be \label{SQED-inj}
 \begin{array}{ccccccccc}
   \CM^0_\nu &\overset{g}\hookrightarrow& \CM^1_\nu  &\overset{g}\hookrightarrow&\CM^2_\nu  &\overset{g}\hookrightarrow& \CM^3_\nu &\overset{g}\hookrightarrow & \cdots \\
   \rotatebox{90}{$=$} &&  \rotatebox{90}{$=$} &&   \rotatebox{90}{$=$} &&   \rotatebox{90}{$=$} \\
   \{pt\} && \C^N && \C^{2N} && \C^{3N} &&\cdots
   \end{array}
\ee
These embeddings induce natural push-forward and pull-back maps on equivariant cohomology. Setting
\be v_+ = g_*\,,\qquad v_- = g^*\,,\ee
we obtain $H^*(\CM_\nu^\n) \overset{v_+}{\underset{v_-}\rightleftharpoons}
H^*(\CM_\nu^{\n+1})$, or
\be \CH_\nu^0 \,\overset{v_+}{\underset{v_-}\rightleftharpoons}\,
\CH_\nu^1 \,\overset{v_+}{\underset{v_-}\rightleftharpoons}\,
\CH_\nu^2 \,\overset{v_+}{\underset{v_-}\rightleftharpoons}\,
\CH_\nu^3 \,\overset{v_+}{\underset{v_-}\rightleftharpoons} \cdots\,. \ee

We can summarize the action on vortices as
\be \label{SQED-module} \begin{array}{rl}
\varphi |\n\rangle &=  (-m_i-(\n+\tfrac12)\epsilon)|\n\rangle \\[.1cm]
 v_+|\n\rangle &= P\big(\varphi+\tfrac12\epsilon)\big |\n+1\rangle \\[.1cm]
 v_-|\n\rangle &= |\n-1\rangle\,.
\end{array} \ee
A short computation shows that the monopole operators obey the algebra
\be \label{SQED-algebra} \begin{array}{c} v_+v_- = P(\varphi+\tfrac12\epsilon)\,,\qquad v_-v_+ = P(\varphi-\tfrac12\epsilon)\,, \\[.2cm]
 [\varphi,v_\pm] = \mp\epsilon v_\pm\,. \end{array} \ee
For example, the relation $v_-v_+ = P(\varphi+\tfrac12\epsilon)$ captures the fact that in equivariant cohomology $g^*g_*$ equals the Euler class of the normal bundle to $\CM_\nu^\n$ in $\CM_\nu^{\n+1}$. Relations \eqref{SQED-algebra} precisely describe the quantum Coulomb-branch algebra $\C_\epsilon[\CM_C]$ for SQED derived in \cite{BDG-Coulomb}.
\footnote{To compare directly with formulas of \cite{BDG-Coulomb} and \cite{BDGH}, one should reverse the sign of $\epsilon$.} 
In the limit $\epsilon\to 0$, we recover a commutative ring with the relation $v_+v_- = P(\varphi)$. This is the expected Coulomb-branch chiral ring: it is the coordinate ring of $\C^2/\Z_N$, deformed by complex masses.

We may recall from \cite{BDG-Coulomb, BDGH} that the Coulomb-branch algebra is graded by the topological $G_C\simeq U(1)$ symmetry under which monopole operators are charged. In particular $\varphi$ has weight zero, and the weight of any monopole operator $v_A$ is the product $t_\R A$ of the magnetic charge and the real FI parameter. The Hilbert space is a \emph{highest-weight module} for the Coulomb-branch algebra with respect to this grading. This means that:
\begin{itemize}
\item the `Cartan' generator $\varphi$ is diagonalized on weight spaces $|\n\rangle$,
\item if we act repeatedly on any weight space $|\n\rangle$ with an operator $v_A$ of positive grading $t_\R A>0$, we will eventually get zero.
\end{itemize}
More so, as long as the $m_i$ are generic (so that the prefactors $P(\varphi+\tfrac12\epsilon)$ never vanish), every state $|\n\rangle$ can be obtained by acting freely on $|0\rangle$ with operators $v_A$ of negative grading. This identifies the Hilbert space as a Verma module.

For general $N$, the algebra \eqref{SQED-algebra} is known as a spherical rational Cherednik algebra.
For $N=2$, it is simply isomorphic the universal enveloping algebra of $\mathfrak{sl}_2$, with the quadratic Casimir fixed in terms of the complex masses $m_i$. Namely, defining $h=2\varphi$, $e=-v_-$, $f=v_+$, we find
\be
[h,e]=2 \ep e\,, \qquad [h,f]=-2\ep f\,, \qquad [e,f] = \ep h\,,
\ee
and
\be
C_2 = \frac{1}{2}h^2+ef+fe = \frac{1}{2}((m_1-m_2)^2-\ep^2)\,.
\ee
This algebra admits two different Verma modules, corresponding to the two possible vacua $\nu_1,\nu_2$ that we could have chosen in defining the Hilbert space $\CH_\nu$.

\subsection{Algebraic formulation}
\label{sec:corresp}

The structure we found in the preceding example can be readily formalized and generalized, by adapting the quantum-mechanics approach that we used to construct Hilbert spaces in Section \ref{sec:Hilbert}.

Physically, the action of monopole operators in the Hilbert space should be computed by performing the path integral on $\C_z\times \R_t$ with particular boundary conditions:
\begin{itemize}
\item a fixed vacuum $\nu$ at $|z|\to\infty$,
\item fixed vortex states $|\n,k\rangle$ at $t\to-\infty$ and $|\n',k'\rangle$ at $t\to \infty$, and
\item a monopole operator $V_A$ inserted at the origin $(z,t)=(0,0)$.
\end{itemize}
The insertion of the monopole operator $V_A$ amounts to removing a three-ball neighborhood of the origin, and specifying a particular state in the radially-quantized Hilbert space $\CH(S^2)$ there.
From topological considerations, we know that the amplitude is nonzero if and only if $\n'-\n=\n(A)$ and $k'-k\in \text{rep}(A)$.

\begin{figure}[htb]
\centering
\includegraphics[width=2.2in]{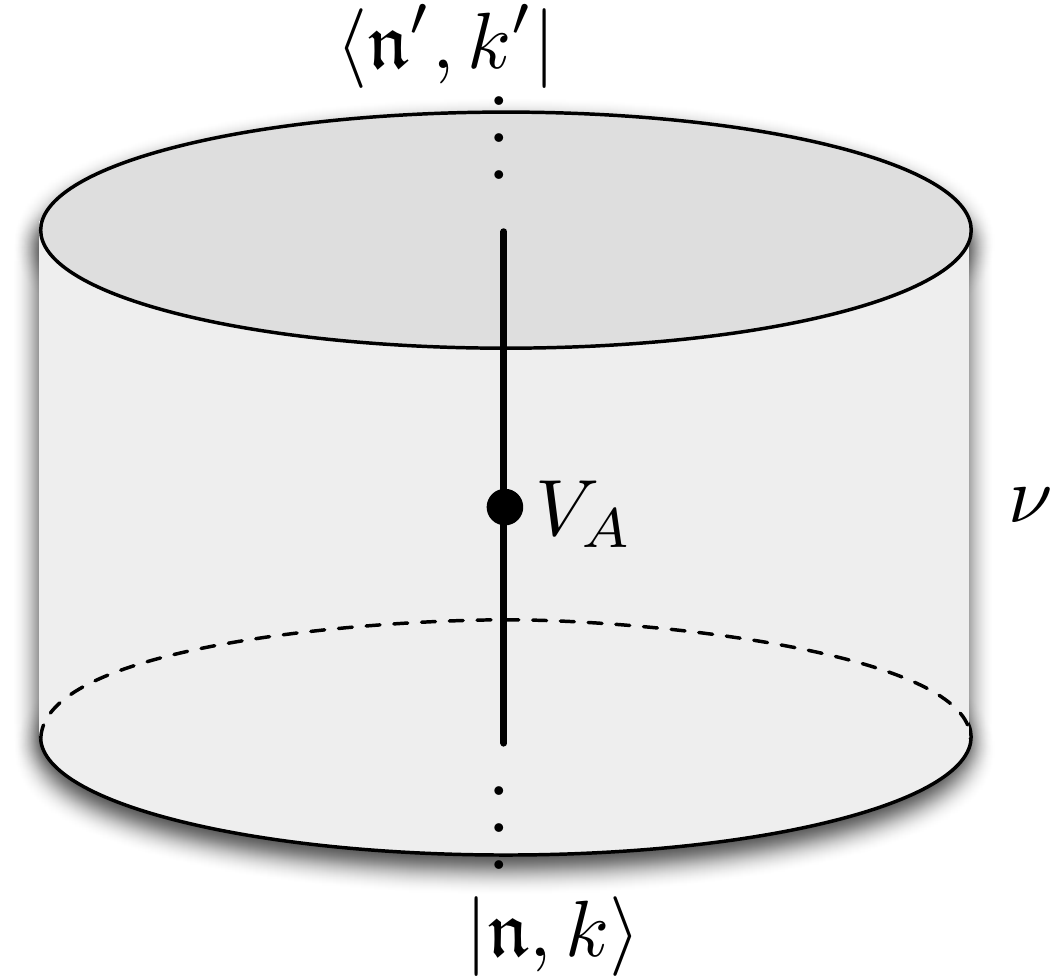}
\caption{Configuration of boundary conditions that computes the matrix element of a monopole operator.}
\label{fig:monpath}
\end{figure}

Since all the boundary conditions preserve two common supercharges, the path integral will localize on solutions of the quarter-BPS equations from Section \ref{sec:BPS}. Moreover, after complexifying the gauge group and passing to a holomorphic gauge, the equation
\be
[\CD_t,\CD_{\bar z}]=0 \, .
\ee
ensures time-evolution of the holomorphic data is trivial away from the insertion of the monopole operator. We may therefore collapse $\big(\C_z\times \R_t - (0,0))\big)$ to a `UFO' or `raviolo' curve\footnote{We thank D. Ben-Zvi and J. Kamnitzer for introducing us to these respective descriptors.}
\be \includegraphics[width=.8cm]{raviolo}\;=\;\C\cup_{\C^*}\!\C\,, \label{def-rav} \ee
consisting of two copies of the spatial plane $\C_z$, identified everywhere except for the origin (the position of the monopole operator). The path integral now reduces to an integral over the space of solutions to the BPS equations on $\includegraphics[width=.8cm]{raviolo}$, with appropriate boundary conditions.

To be more concrete, recall the notation $\CE=(E,X,Y)$ from \eqref{CE} for an algebraic $G_\C$-bundle $E$ on $\C_z$ together with sections $X,Y$ of associated $R\oplus \bar R$ bundles satisfying the complex moment-map constraint $\mu_\C(X,Y)=0$ and landing on the orbit $G_\C \cdot \nu$ at $z\to \infty$. Since the massive vacuum $\nu$ trivializes the bundles at $z=\infty$, we can always compactify the $z$-plane to $\cp^1$. 

In an algebraic formulation, the space of solutions to BPS equations on the raviolo is given by a pair $\CE$, $\CE'$, together with an identification by a gauge transformation $g$ away from the origin,
\be \CM_{\rav\,\nu} \,=\, \big\{\text{$(\CE,\CE';g)$ : $\CE \overset{g}{\overset{\sim}{\to}} \CE'$ on $\C_z-\{0\}$}\big\}/\CG\times\CG'\,. \label{Mravnu} \ee
We quotient by isomorphisms of the data $\CE$ and $\CE'$ \ie\ by holomorphic gauge transformations $\CG\times \CG'$. 
This moduli space has natural maps to two copies of the vortex moduli space $\CM_\nu$, simply obtained by forgetting either $\CE'$ and $g$, or $\CE$ and $g$,
\be \raisebox{-.5in}{$\includegraphics[width=1.5in]{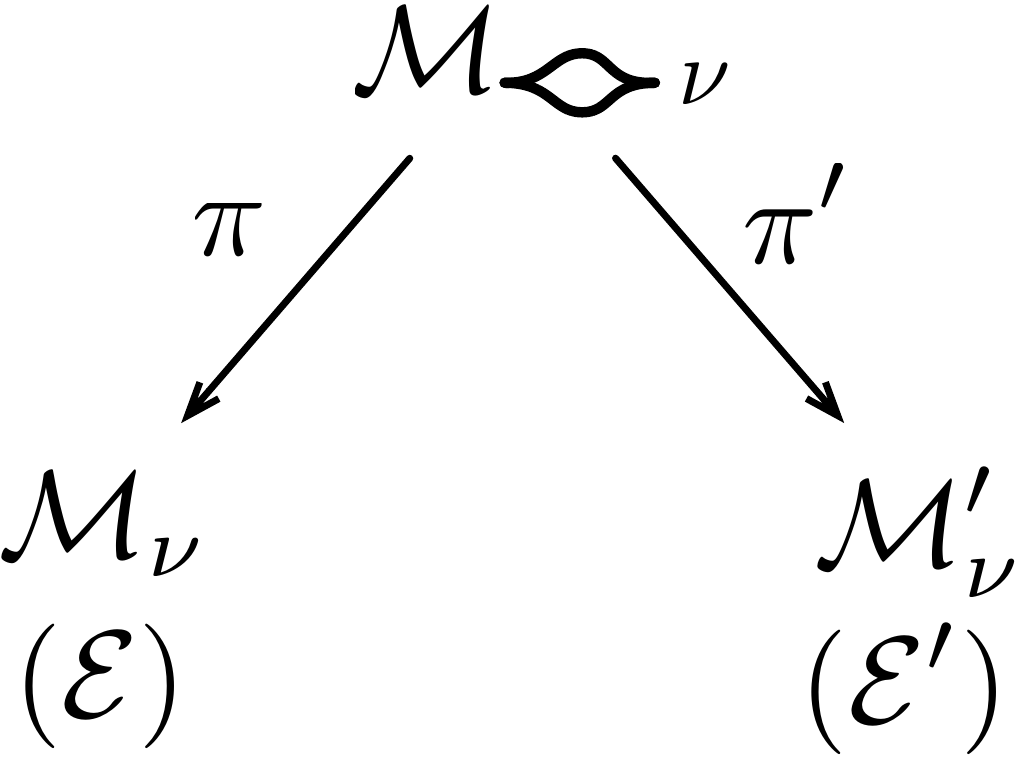}$} \,\; .\label{Vmaps} \ee
This is called a \emph{correspondence}.

We saw that quantum vortex states correspond to equivariant cohomology classes $|\n,k\rangle\in H^*(\CM_\nu)$. (We will suppress the equivariant $T_H\times U(1)_\epsilon$ action in order to simplify notation.) In a similar way, the insertion of any monopole operator $V_A$ defines an equivariant cohomology class
\be V_A \,\in\, H^*(\CM_{\rav\,\nu})\,. \ee
We will describe these classes explicitly in a moment. The action of a monopole operator on a vortex state translates to a `push-pull' action on cohomology, induced by the correspondence \eqref{Vmaps}. Namely, we use $\pi$ to pull-back the class $|\n,k\rangle$ to $H^*(\CM_{\rav\,\nu})$, take the cup-product with the class $V_A$, and use $\pi'$ to push-forward to $H^*(\CM_\nu')$,
\be V_A\,|\n,k\rangle\, =\, \pi'_*\big(V_A\,\cdot\, \pi^*(|\n,k\rangle) \big)\,. \label{Maction} \ee
The push-forward $\pi'_*$ is an equivariant integration along the fibers of the map $\pi'$, and encapsulates the integration over the moduli space of solutions to the BPS equations in the localized path integral.

\subsubsection{Components and monopole operators}
\label{sec:mon-hom}

Just as the vortex moduli space splits into components labelled by vortex number
\be \CM_\nu  \,= \bigsqcup_{\n\in \pi_1(G)} \CM_\nu^\n\,,\ee
the raviolo moduli space also has connected components labelled by pairs of vortex numbers, describing the topological type of the bundles $E,E'$ on the two copies of $\cp^1$
\be \CM_{\rav\,\nu} \,= \bigsqcup_{\n,\n'\in \pi^1(G)} \CM_{\rav\,\nu}^{\n,\n'}\,. \ee
Thus, the correspondence \eqref{Vmaps} splits into components
\be \raisebox{-.3in}{$\includegraphics[width=1.3in]{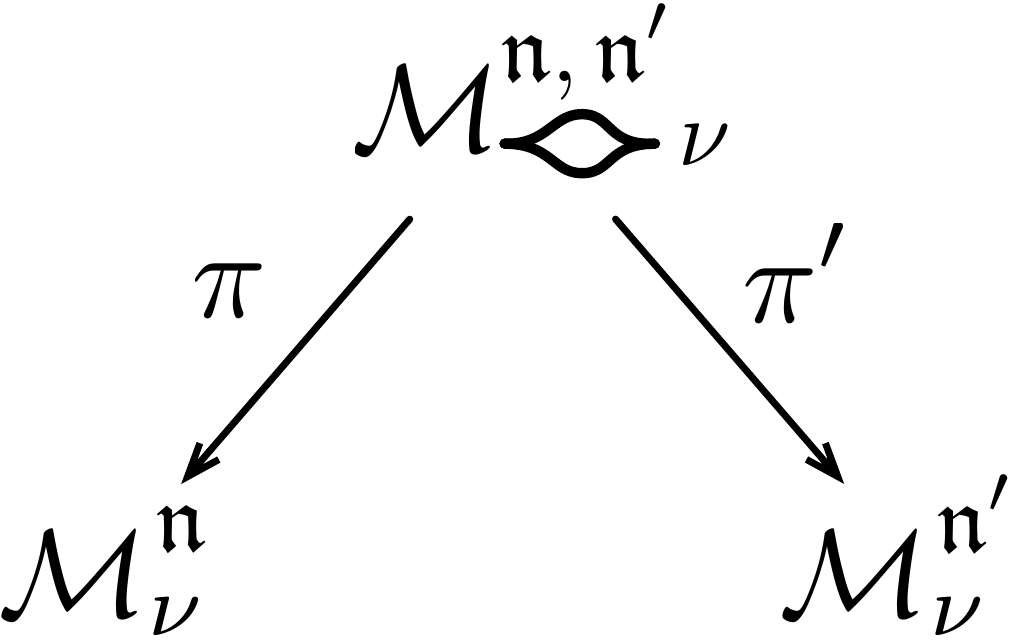}$} \label{Vmaps-n} \ee

In addition, each raviolo space $\CM_{\rav\,\nu}^{\n,\n'}$ has a further decomposition (in fact, a stratification) by the magnetic charge of monopole operators. Since the gauge transformation $g$ in \eqref{Mravnu} is regular away from the origin, it must lie in the $\CG\times \CG'$ orbit of
\be \qquad g(z) = z^A\,,\qquad A\in \text{Hom}(U(1),G) \label{gA} \ee
for some cocharacter $A$. Here we think of $z^A$ as an element in the maximal torus of the gauge group, with Laurent-polynomial entries. (See \eqref{gA-SQCD} below.)
Two cocharacters related by an element of the Weyl group lead to the same $\CG\times \CG'$ orbit, so we may assume that $A$ is a dominant cocharacter. Then
\be \CM_{\rav\,\nu}^{\n,\n'} \;= \bigcup_{A\,\in\, \Lambda_{\rm cochar}^{\rm dom}} \CM_{\rav\,\nu}^{\n,\n';A}\,. \ee
Of course, a particular singular gauge transformation changes the vortex number by a fixed amount $\n(A)$. Thus, $\CM_{\rav\,\nu}^{\n,\n';A}$ is actually empty unless $\n(A) = \n'-\n$.

It is natural to identify each basic monopole operator $V_A$ with the equivariant fundamental class of the closure $\ol{\CM_{\rav\,\nu}^{\n,\n';A}}$ of  $\CM_{\rav\,\nu}^{\n,\n';A}$,
\be V_A\;\leftrightarrow\; 1\!\!1_{\ol{\CM_{\rav\,\nu}^{\n,\n';A}}} \;\in\; H^*(\CM_{\rav\,\nu}^{\n,\n'})\,. \label{VAclass} \ee
This is the concrete way in which monopole operators enter the push-pull action \eqref{VAclass}.
More generally, the Coulomb-branch chiral ring may contain ``dressed'' monopole operators, defined by superposing polynomials in the $\varphi$ fields on top of a monopole singularity. Dressed monopole operators are represented as characteristic classes of various bundles on $\ol{\CM_{\rav\,\nu}^{\n,\n';A}}$.

Note that the `strata' $\CM_{\rav\,\nu}^{\n,\n';A}$ are not closed unless the cocharacter $A$ is miniscule. The closure of a particular stratum contains other strata, and has interesting topology related to the physics of monopole bubbling, discussed \eg\ in \cite[Sec. 10]{Kapustin-Witten}.

\subsubsection{SQED revisited}
\label{sec:SQED-alg}

We now reproduce the action of monopole operators on vortices in SQED, in terms of the canonical correspondences \eqref{Vmaps}. 

Let us consider the simplest possible correspondence $\CM_{\rav\,\nu}^{0,1}$. It consists of a pair $\CE$, $\CE'$, and a gauge transformation $g$ identifying them away from the origin. Since $\n=0$, the bundle in $\CE$ is trivial and we can use the complexified gauge group $\CG$ to set $X=(1,0,...,0)$ and $Y=0$. The gauge transformation $g$ can be any element of the form $g = az$ for nonzero $a$; it acts on $\CE$ to produce a bundle in $\CE'$ of degree one and
\be X' = (az,0,...,0)\,,\qquad Y'=0\,. \ee
The gauge group $\CG'$ can now be used to fix $a=1$. Therefore, $\CM^{0,1}_{\rav,\nu}$ is simply a point. By forgetting $\CE'$, it maps isomorphically to $\CM_\nu^0$, which is also a point. On the other hand, by forgetting $\CE$, it maps to the origin of $\CM_\nu^1 = \C^N$.

More generally, whenever $\n<\n'$ we can use up the gauge freedom $\CG\times \CG'$ to write every point of $\CM_{\rav\,\nu}^{\n,\n'}$ uniquely as
\be X_i=\delta_{i,1}z^\n+\sum_{l=1}^{\n-1}x_{i,l}z^l\,,\qquad g=z^{\n'-\n}\,,\qquad
 X_i'=\delta_{i,1}z^{\n'} +\sum_{l=\n'-\n}^{\n'-1}x_{i,l}z^{l}\,,\ee
with $Y=Y'=0$. These points are fully determined by the form of $X$, which is unconstrained; therefore, by forgetting $g$ and $X'$ we get an isomorphism with $\CM_\nu^\n$. On the other hand, by forgetting $X$ and $g$ we get a map into $\CM_\nu^{\n'}$. This is an injection, because $X'$ is constrained so that coefficients of $z^d$ vanish when $d<\n'-\n$. Thus
\be \label{corresp-SQED} \CM_\nu^{\n} \;\underset{\pi}{\overset{\sim}\leftarrow}\; \CM_{\rav\,\nu}^{\n,\n'} \;\underset{\pi'}{\hookrightarrow}\; \CM_\nu^{\n'} \,.\ee
Similarly, for $\n>\n'$, the isomorphism and injection are reversed.

Let us now compute the action $v_A|\n\rangle$ for $A>0$ using the correspondence $\CM_{\rav\,\nu}^{\n,\n+A}$. Recall that $|\n\rangle$ corresponds to the fundamental class in $\CM_\nu^\n$. We use the isomorphism $\pi$ in \eqref{corresp-SQED} to pull it back to the fundamental class $\pi^*(|\n\rangle)=1\!\!1$ in $H^*(\CM_{\rav\,\nu}^{\n,\n+A})$. The monopole operator $v_A$ also corresponds to the fundamental class in $H^*(\CM_{\rav\,\nu}^{\n,\n+A})$, so $v_A\wedge \pi^*(|\n\rangle) = \pi^*(|\n\rangle)$. Finally, we use the injection $\pi'$ to push forward $\pi^*(|\n\rangle)$ to $H^*(\CM_\nu^{\n+A})$, obtaining the vortex class $|\n+1\rangle$ times an equivariant delta function corresponding to the normal directions of $N^*(\pi'(\CM_{\rav\,\nu}^{\n,\n+a}))\subset T^*\CM_\nu^{\n+A}$. The delta function is easily computed to give
\be v_A |\n\rangle = \prod_{l=0}^{A-1}P(\varphi+(l+\tfrac12)\epsilon)\, |\n+A\rangle \qquad A > 0 \ee
in agreement with Section \ref{sec:SQED-mono}.

Similarly, if $A<0$ is a negative integer, then we compute $v_A|\n\rangle$ by using the injection $\pi:\CM_{\rav\,\nu}^{\n,\n+A}\hookrightarrow \CM_\nu^{\n}$ to pull back the fundamental class $|\n\rangle$ to the fundamental class in $\CM_{\rav\,\nu}^{\n,\n+A}$,  then intersecting with the fundamental class in $\CM_{\rav\,\nu}^{\n,\n+A}$ that represents $v_A$, and finally pushing forward to $\CM_\nu^{\n+A}$ via the isomorphism $\pi'$. In this case there is no equivariant delta function and we simply find
\be v_A|\n\rangle = |\n+A\rangle\qquad A<0 \,.\ee

This of course is the same action that we found more directly in Section \ref{sec:SQED-mono}. Since the correspondences \eqref{corresp-SQED} are so simple, the whole construction reduces to the sequence of maps described in \eqref{SQED-inj}; the map `$g$' in \eqref{SQED-inj} is $\pi'\circ \pi^{-1}$.

\subsection{Non-abelian theories}
\label{sec:nonab-mono}

The structure of correspondences and monopole operators in nonabelian theories is well illustrated by the example of SQCD.

\subsubsection{SQCD}
\label{sec:SQCD-mono}

Let us consider $U(K)$ gauge theory with $N$ fundamental hypermultiplets, as in Section \ref{sec:SQCD-moduli}. 
Recall that for $t_\R<0$, the nontrivial vortex moduli spaces $\CM_\nu^\n$ have $\n\geq 0$ with fixed points labelled by non-negative integers $k=(k_1,...,k_K)$ with $\sum_a k_a=\n$. After a flavor rotation, we may assume that the vacuum $\nu$ has $X^a{}_i = \delta^a{}_i$ with all other hypermultiplet fields vanishing. Then the fixed-point states correspond to
\be |\n,k\rangle \quad\leftrightarrow\quad X^a{}_i = \delta^a{}_i\, z^{k_a}\,,\quad Y=0\,, \ee
and the equivariant weight of the normal bundle to a fixed point is $\omega_{\n,k}$ as in \eqref{QCD-weights}.

The basic monopole operators $V_A$ of SQCD are labelled by cocharacters $A\in \text{Hom}(U(1),U(K))\simeq \Z^K$, and correspond to singular gauge transformations of the form
\be  g(z) = z^A = \text{diag}(z^{A_1},...,z^{A_K})\,. \label{gA-SQCD} \ee
Naively, this maps a fixed point $|\n,k\rangle$ to $|\n+\sum_a A_a,k+A\rangle$. However, this conclusion is clearly not gauge-invariant.
We must account for the fact that the singular transformation \eqref{gA-SQCD} may be composed with arbitrary regular gauge transformations $g\to g_0'gg_0$ when mapping one fixed point to another. The correspondences of Section \ref{sec:corresp} provide a precise way to do this.

We focus on the basic monopole operators $V_\pm$ of charge $A = (\pm 1,0,...,0)$. It is actually sufficient to understand the action of these operators: as discussed in \cite{BDG-Coulomb}, the entire Coulomb-branch algebra of SQCD is generated by $V_\pm$ and their dressed versions.

The operator $V_+$ (say) increases vortex number by one, so we should look at the correspondence space $\CM_{\rav\,\nu}^{\n,\n+1}$. Specifically, we are interested in the stratum $\CM_{\rav\,\nu}^{\n,\n+1;(1,0,...,0)} \subset \CM_{\rav\,\nu}^{\n,\n+1}$ consisting of triples $(X,g,X')$ such that $X'=gX$, modulo two copies of the gauge group $\CG\times \CG'$, with the extra condition that $g$ is in the $\CG\times \CG'$ orbit of $z^{(1,0,...,0)}=\text{diag}(z,0,...,0)$. Let us just call this stratum $\CM^{+1}$. It has maps 
\be \CM_\nu^\n \,\overset\pi\leftarrow\, \CM^{+1}\,\overset{\pi'}\rightarrow\, \CM_\nu^{\n+1}\,, \label{SQCD-strat}\ee
and $V_+$ corresponds to the equivariant fundamental class $1\!\!1_{\CM^{+1}}$.

The map $\pi$ in \eqref{SQCD-strat} is actually a surjection with regular compact fibers isomorphic to $\cp^{K-1}$.
(This implies that $\CM^{+1}$ is compact, which was expected because the cocharacter $(1,0,...,0)$ is minuscule.)
To justify this claim, let us choose a point $X\in \CM_\nu^\n$. Specifying $X$ fully breaks the gauge symmetry $\CG$. Then the fibers of $\pi$ at $X$ consists of gauge transformations of the form
\be   g_0'(z) z^{(1,0,...,0)} g_0(z)\,, \ee
where $g_0',g_0$ are regular, modulo the action of regular $\CG'$ transformations on the left. The action of $\CG'$ can be used to absorb $g_0'$ as well as \emph{most} of $g_0$. The gauge transformations $g_0$ that can be commuted to the left past $z^{(1,0,...,0)}$ (as regular gauge transformations) are of the form
\be g_0(z) = \begin{pmatrix} * & * & * & \cdots \\
  z(*) & * & * & \cdots \\
   & \ddots \\
  z(*) & * & * & \cdots \end{pmatrix}
\ee
where each `$*$' denotes a polynomial in $z$. The remaining $g_0$ that cannot be commuted to the left precisely parameterize a coset $\cp^{K-1}$.%
\footnote{This sort of analysis is very familiar in the study of the affine Grassmannian and its stratification by orbits of cocharacters $z^A$. For miniscule cocharacters, the orbits are ordinary Grassmannians.}

The map $\pi'$ in \eqref{SQCD-strat} is more complicated. Generically it is an injection, in the sense that its fibers above generic $X'\in \CM_\nu^{\n+1}$ are either empty or single points. However, above values of $X'$ that are fixed points of the $T_H\times U(1)_\epsilon$ action on $\CM_\nu^{\n+1}$, the fibers of $\pi'$ can be nontrivial. We will not  need to know about this to find the action of~$V_+$.

Now consider the various symmetries acting on the triples $(X,g,X')$. The flavor symmetry $T_H$ simultaneously acts on the columns of $X$ and $X'$ (viewed as $K\times N$ matrices). The symmetry $U(1)_\epsilon$ rotates $z$ as usual (and $X,X'$ with weight $\frac12$). In addition, constant gauge transformations $(g_0,g_0')\in \CG\times \CG'$ act as $(X,g,X')\mapsto (g_0X,g_0'gg_0^{-1},g_0'X')$.
Just as in the analysis of vortex moduli spaces, the gauge action is free. The $T_H\times U(1)_\epsilon$ action has isolated fixed points, provided that this action is compensating for by an appropriate gauge transformation.

Concretely, the fixed points of $T_H\times U(1)_\epsilon$ on the correspondence space $\CM^{+1}$ are of the form
\be X^a{}_i = z^{k_a}\delta^a{}_i\,,\qquad g = \text{diag}(1,...,\underset{b}z,...,1)\,,\qquad X'{}^a{}_i = z^{k_a+\delta_{ab}}\delta^a{}_i\,\ee
for all nonnegative vectors $k$ (such that $\n(k)=\n$) and all integers $1\leq b\leq K$. We should understand $k$ as labeling a fixed point on $\CM_\nu^\n$ and $b$ labeling a fixed point on the $\cp^{K-1}$ fiber of $\pi$. The relation between flavor parameters $m_\C,\epsilon$ and compensating gauge parameters $\varphi,\varphi'$ is
\be \begin{array}{c} \varphi_a+m_a+(k_a+\tfrac12)\epsilon=0\,,\qquad \varphi_a'+m_a+(k_a+\delta_{ab}+\tfrac12)\epsilon=0\,, \\[.1cm]
 \varphi_a = \varphi_a'+\delta_{ab}\epsilon\,. \end{array} \ee
Thus $\Tr\varphi-\Tr\varphi' = \epsilon$, reflecting the fact that vortex number is increased by one.

We need one more ingredient to describe the action of $V_+$ on equivariant cohomology. Recall that $V_+$ is realized as the equivariant fundamental class of $\CM^{+1}$. It is extremely useful to use the localization theorem to decompose this as a sum of fixed-point classes in the $\cp^{K-1}$ fibers
\be V_+ =  1\!\!1_{\CM^{+1}} = \sum_{b=1}^K v_b^+\,,\qquad v_b^+:=  \frac{1}{\prod_{a\neq b}(\varphi_a-\varphi_b)}\, 1\!\!1_b\,, \label{Vv-SQCD} \ee
where $1\!\!1_b$ denotes the fundamental class of $\CM_\nu^\n$ (thought of as the base of $\CM^{+1}$) times a point in the fiber. The $v_b^+$ were introduced in \cite{BDG-Coulomb} as ``abelianized'' monopole operators.

Now, the state $|\n,k\rangle = \frac{1}{\omega_{\n,k}} 1\!\!1_{\n,k}$ is a normalized fixed-point class in $H^*(\CM_\nu^\n)$. It pulls back via $\pi^{-1}$ to $\frac{1}{\omega_{\n,k}}$ times the fundamental class of the $\cp^{K-1}$ fiber of $\CM^{+1}$, sitting above the fixed point of $\CM_\nu^\n$. The product with $v_b^+$ then produces the normalized fundamental class of a fixed point in $\CM^{+1}$,
\be v_b^+\cdot \pi^* |\n,k\rangle\, =\, \frac{1}{\omega_{\n,k}}  \frac{1}{\prod_{a\neq b}(\varphi_a-\varphi_b)} 1\!\!1_{\n,k;b}\,.\ee
Since the fundamental class of a fixed point $1\!\!1_{\n,k;b}$ in $\CM^{+1}$ pushes forward via $\pi'$ to the fundamental class of the fixed point $1\!\!1_{\n+1,k+\bm\delta_b}$ in $\CM_\nu^{\n+1}$ (where $\bm \delta_b = (0,...,0,\underset{b}1,0,...,0)$), we finally find that
\be v_b^+|\n,k\rangle:= \pi'_*\big(v_b^+\cdot \pi^* |\n,k\rangle\big)\,=\, \frac{\omega_{\n+1,k+\bm\delta_b}}{\omega_{\n,k}}\frac{1}{\prod_{a\neq b}(\varphi_a-\varphi_b)} \,|\n+1,k+\bm\delta_b\rangle\,,\ee
and $V_+|\n,k\rangle = \sum_{b=1}^K v_b^+|\n,k\rangle$. Using the formula \eqref{QCD-weights} for the equivariant weights, this can easily be brought to the form
\be v_b^+|\n,k\rangle \,=\, \frac{P(\varphi_b'+\frac\epsilon2 )}{\prod_{a\neq b}(\varphi_a'-\varphi_b')} |\n+1,k+\bm\delta_b\rangle\,. \label{SQCD-Vplus} \ee

We can similarly find the action of the negatively charged operators $V_-$ by running backwards through the same correspondence $\CM^{+1}$. Using the same decomposition into abelianized monopole operators
\be V_- = \sum_{b=1}^K v_b^-\,, \ee
we find that
\be v_b^- \,|\n,k\rangle = \begin{cases} \ds \frac{1}{\prod_{a\neq b}(\varphi_a-\varphi_b)} |\n-1,k-\bm \delta_b\rangle &\text{if $\n-1$, $k-\bm\delta_b$ nonnegative} \\ 0 & \text{otherwise}\,. \end{cases} \label{SQCD-Vminus}\ee
The combined action of $v_a^+$ and $v_b^-$ on any vortex state is
\be v_b^+v_b^- = \frac{P(\varphi_b+\tfrac12\epsilon)}{\prod_{a\neq b}(\varphi_b-\varphi_a)(\varphi_a-\varphi_b-\epsilon)}\,,\qquad  
v_b^-v_b^+ = \frac{P(\varphi_b-\tfrac12\epsilon)}{\prod_{a\neq b}(\varphi_b-\varphi_a)(\varphi_a-\varphi_b+\epsilon)}\,. \label{va+-} \ee
This is the fundamental relation in the Coulomb-branch algebra that was derived more abstractly in \cite[Sec 5.3]{BDG-Coulomb} (with $\epsilon\to -\epsilon$). There the weights on the RHS were interpreted as one-loop corrections to the chiral ring, arising from hypermultiplets (numerator) and W-bosons (denominator).

The dressed monopole operators of SQCD can be very easily described in terms of the abelianized $v_b^\pm$. Namely, they all take the general form
\be V_{\pm,p} = \sum_{w\in W} p(w\cdot \varphi)v_{w\cdot b}^\pm \ee
for some polynomial $p$ in the fields $\varphi=(\varphi_1,...,\varphi_K)$, where the sum implements an averaging over the Weyl group. The dressed monopole operators can be understood as characteristic classes of various bundles on $\CM^{+1}$, and their action on vortex states derived accordingly.

We note that in \cite{BDG-Coulomb}, the Coulomb-branch algebra was also re-derived by relating the Coulomb branch of SQCD to a moduli space of singular monopoles \cite{HananyWitten} -- namely a moduli space of $K$ $PSU(2)$ monopoles with $N$ Dirac monopole singularities.
To see this connection it is convenient to introduce polynomial generating functions $Q(z) = \prod_{a=1}^k(z-\varphi_a)$ and
\be
U^+(z) = \sum_{a=1}^K u_a^+ \prod_{b \neq a}(z-\varphi_b )  \qquad U^-(z) = \sum_{a=1}^K u_a^- \prod_{b \neq a}(z-\varphi_b)  \, .
\label{poly-def}
\ee
where $u_a^+=v_a^+$ and $u^-_a=(-1)^K v^-_a$. The polynomials $U^\pm(z)$ are generating functions for dressed monopole operators. The relations~\eqref{va+-} can now be written in `quantum determinant' form
\be
Q(z-\tfrac{\ep}{2}) \tilde Q(z+\tfrac{\ep}{2}) - U^+(z-\tfrac{\ep}{2}) U^-(z+\tfrac{\ep}{2}) = P(z)
\label{qdet}
\ee
where $\tilde Q(z)$ is a generating function for dressed monopole operators with magnetic weight in the adjoint representation. In the limit $\epsilon \to 0$, we recover the coordinate ring of the moduli space of $k$ $PSU(2)$ monopoles with $N$ fundamental Dirac monopole singularities, written in terms of scattering data.

\subsubsection{Triangular quivers}

Let us now state the results of the corresponding computation in the case of a triangular quiver, with notation from Section~\ref{sec-triang-hilb}. Let us denote the monopole operators of fundamental and anti-fundamental magnetic charge at the $\al$-th node by $V_\al^\pm$. Then we find
\be
\varphi_{\al,a} = - m_{i_{\al,a}} - (k_{\al,a}+\tfrac{1}{2}) \ep \, ,
\ee
together with
\bea
V_\al^+ | \vec \n , \vec k \ra & = \sum_{a=1}^{K_\al} \frac{Q_{\al+1}(\varphi_{\al,a})}{\prod_{b \neq a}(\varphi_{\al,a}-\varphi_{\al,b})} \, | \vec \n+\delta_a , \vec k+\delta_{\al,a}\ra \\
V_\al^- | \vec \n , \vec k \ra & = \sum_{a=1}^{K_\al} \frac{Q_{\al-1}(\varphi_{\al,a})}{\prod_{b \neq a}(\varphi_{\al,b}-\varphi_{\al,a})} \, | \vec \n-\delta_a , \vec k-\delta_{\al,a}\ra \, .
\eea

These generators obey the following `quantum determinant' relation for each node $\al=1,\ldots,L-1$, independent of which state is acted upon,
\be
Q_\al(z-\tfrac{\ep}{2}) \tilde Q_\al(z+\tfrac{\ep}{2}) - U_\al^+(z-\tfrac{\ep}{2}) U_\al^-(z+\tfrac{\ep}{2}) = Q_{\al-1}(z)Q_{\al+1}(z) \, .
\label{qdet-quiver}
\ee
In the limit $\ep \to 0$ we recover the coordinate ring of the moduli space of $PSU(L+1)$ monopoles with $N$ singular monopoles in the fundamental representation. Indeed, the relations \eqref{qdet-quiver} are Pl\"ucker relations for the monopole scattering matrix~\cite{BDG-Coulomb}. This is the expected Coulomb branch chiral ring of the quiver.

\subsection{Recovering the Coulomb-branch algebra}
\label{sec:conv}

So far we have described an action on the equivariant cohomology $\CH=\oplus_\n H^*(\CM_\nu^\n)$ of vortex moduli spaces generated by the correspondences $\CM_{\rav\,\nu}^{\n,\n'}$. Our main claim is that this is an action of the quantized Coulomb-branch algebra $\C_\epsilon[\CM_C]$, which is a fundamental observable of the underlying 3d $\CN=4$ theory, independent of the particular boundary conditions that lead to vortices. We verified this above for SQED and SQCD.

We can make the claim a little more precise, by giving an intrinsic description of the algebra of correspondences.
Mathematically, this leads to a new ``definition'' of $\CM_\epsilon[\CM_C]$, complementary to the one proposed by Braverman-Finkelberg-Nakajima.

The basic idea is to construct an algebra intrinsically from the correspondence spaces $\CM_{\rav\,\nu}^{\n,\n'}$ (forgetting for the moment that they act on vortices), and to embed the Coulomb-branch algebra in it.
To this end, let us define a sum of equivariant cohomology groups
\be \CA_\nu\,:= \bigoplus_{\n,\n'\,\in\, \pi_1(G)} H^*(\CM_{\rav\,\nu}^{\n,\n'})\,, \ee
with the usual $T_H\times U(1)_\epsilon$ equivariance made implicit. This vector space has a standard ``convolution product'' that realizes the physical OPE of monopole operators. To see it, we introduce the double-correspondence space
\be \label{Mtrip} \CM_{\drav\,\nu}^{\n,\n',\n''} \,=\, \left\{\begin{array}{c} \text{$(\CE,\CE',\CE'';g,g')$ s.t. $\CE,\CE',\CE''\to\nu$ at $z=\infty$} \\
\text{and $\CE \overset{g}{\overset{\sim}{\to}} \CE'$,  $\CE' \overset{g'}{\overset{\sim}{\to}} \CE';$ on $\C^*$}\end{array}\right\}\Big/\CG\times \CG''\,, \ee
involving holomorphic data $\CE=(E,X,Y)$ on three copies of the $z$-plane $\C_z$, identified by potentially singular gauge transformations $g$ and $g'$. The space \eqref{Mtrip} has three maps to ordinary correspondence spaces, obtained by forgetting the data one one of the three copies of $\C$,
\be \raisebox{-.5in}{$\includegraphics[width=2.6in]{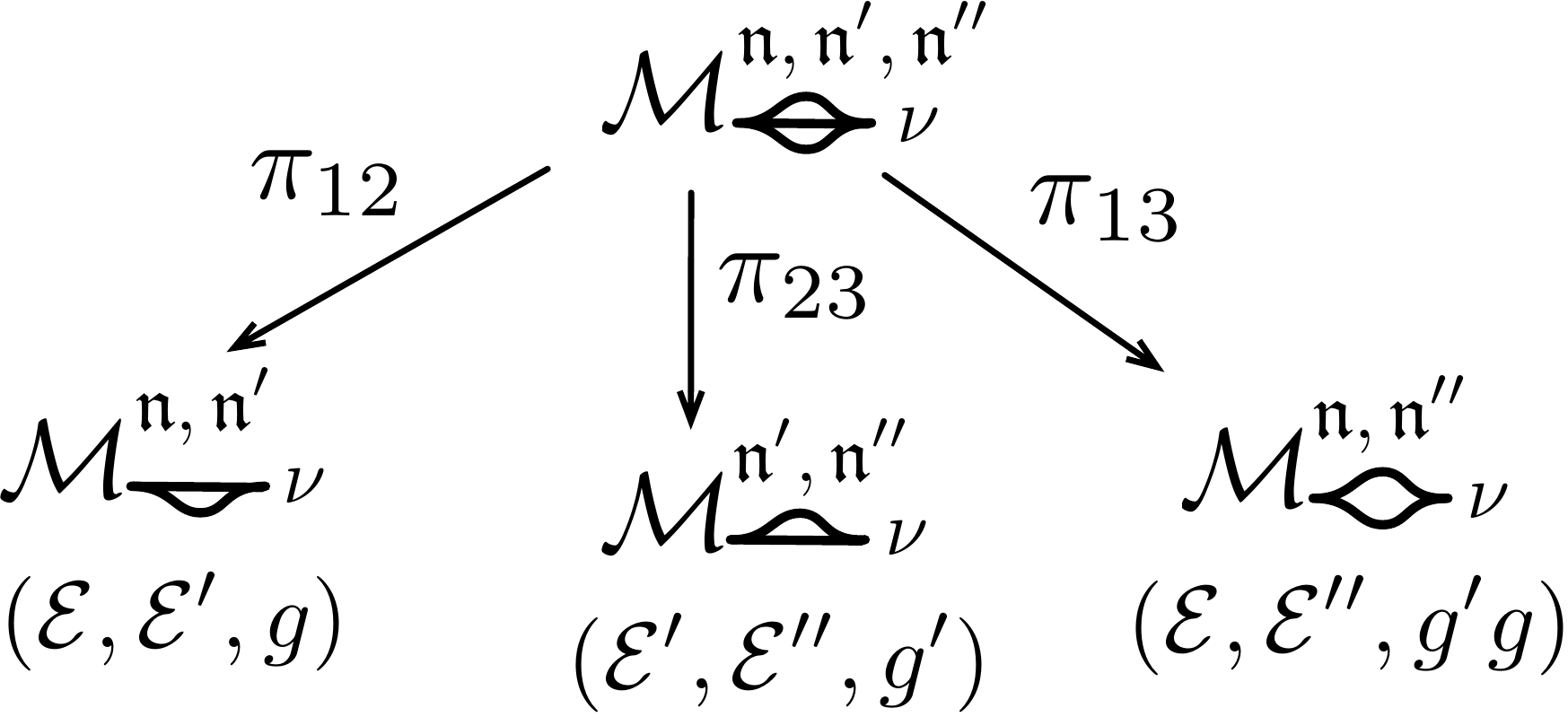}$} \label{conv} \ee
The convolution product is then defined by pushing and pulling, just as in \eqref{Maction},
\be \star\,: \;  \begin{array}{cccl}  H^*(\CM_{\rav\,\nu}^{\n,\n'})\times H^*(\CM_{\rav\,\nu}^{\n',\n''}) &\;\to&  H^*(\CM_{\rav\,\nu}^{\n,\n''}) \\
(\,\eta\quad\,,\; \quad\lambda\,) &\;\mapsto& \lambda\star\eta &= (\pi_{13})_*(\pi_{12}^*(\eta) \cdot \pi_{23}^*(\lambda)) \end{array}
\ee
The product can be extended to all of $\CA_\nu$ by defining it to be zero when vortex numbers are incompatible, \ie\ $\lambda\star \eta=0$ if $\eta\in H^*(\CM_{\rav\,\nu}^{\n,\n'})$ and $\lambda\in H^*(\CM_{\rav\,\nu}^{\n'',\n'''})$ with $\n'\neq \n''$.

The convolution product makes $\CA_\nu$ into an algebra. Moreover, by construction, the product is automatically compatible with the action of $\CA_\nu$ on vortices. In other words, $\lambda\cdot(\eta\cdot |\n,k\rangle) = (\lambda\star\eta)\cdot|\n,k\rangle$ for any vortex state $|\n,k\rangle$.

Our main claim can be rephrased as the statement that \emph{the Coulomb-branch algebra is embedded in $\CA_\nu$},
\be \boxed{\C_\epsilon[\CM_C] \;\overset{\iota}\hookrightarrow\; \CA_\nu} \,.\label{MC-nu}\ee
Physically, we are saying that all Coulomb-branch operators can be represented via their action on vortices, and that this representation is faithful. This is manifestly true in SQED, SQCD, and the various quiver theories that we discuss. A similar assertion appears in \cite[Section 6]{Nakajima-handsaw} in the case of triangular quivers (see Section \ref{sec-adhm} below).

\subsubsection{Orthogonal idempotents}
\label{sec:idemp}

The algebra $\CA_\nu$ above is actually much bigger than the Coulomb-branch algebra $\C_\epsilon[\CM_C]$. 
Indeed, a given monopole operator $V_A$ has an image in $H^*(\CM_{\rav\,\nu}^{\n,\n'})$ whenever $\n'-\n=\n(A)$, and thus has infinitely many images in $\CA_\nu$. Under the map in \eqref{MC-nu}, we must take the sum of all  images; but in $\CA_\nu$ it is also possible to consider them individually.

In order to speak about the individual images of a monopole operator, we introduce an infinite set of orthogonal projection operators or ``idempotents''
\be \{\,e_\n\,|\, \n\in \pi_1(G)\;\text{and $\CM_\nu^\n$ nonempty} \}\,,\ee
satisfying the orthogonality and completion relations
\be e_\n e_{\n'} = \delta_{\n,\n'}e_\n\,, \qquad \sum_{\n\,\in\,\pi_1(G)} e_\n = 1\,. \ee
In fact, these operators are already part of the algebra $\CA_\nu$. Namely, in every correspondence space $\CM_{\rav\,\nu}^{\n,\n}$ that leaves vortex-number unchanged there is a stratum $\CM_{\rav\,\nu}^{\n,\n;0}$ corresponding to the orbit of the trivial gauge transformation $g=1$ (so that $X=X'$ and $Y=Y'$). Then $e_\n$ is its fundamental class
\be e_\n \,:=\, 1\!\!1_{\ol{\CM_{\rav\,\nu}^{\n,\n;0}}} \,\in\, \CM_{\rav\,\nu}^{\n,\n}\,.\ee

Physically, the $e_\n$ are operators in the effective $\CN=4$ quantum mechanics obtained by placing a 3d $\CN=4$ theory in the $\Omega$-background with a vacuum $\nu$ at $z\to\infty$. Each $e_\n$ acts a projection to a subsector of the quantum mechanics whose states have fixed vortex number $\n$. Unlike ordinary Coulomb-branch operators, the $e_\n$ do \emph{not} admit a UV realization in the underlying 3d $\CN=4$ theory. They are additional operators that exist in the infrared.

Now, given some monopole operator $V_A$ that is represented as a sum of classes, say
\be V_A \,=\, \bigoplus_{\n,\n'\,\in\,\pi_1(G)} 1\!\!1_{\ol{\CM_{\rav\,\nu}^{\n,\n';A}}}\;\,, \ee
we can simply sandwich with the projection operators to obtain an individual image $e_{\n'} *V_A *e_{\n} = 1\!\!1_{\ol{\CM_{\rav\,\nu}^{\n,\n';A}}}\,\in\, H^*(\CM_{\rav\,\nu}^{\n,\n'})$. 

It is natural to conjecture that the convolution algebra $\CA_\nu$ is simply equivalent to the Coulomb-branch algebra together with the idempotents $e_\n$. In other words,
\be \boxed{ \CA_\nu \,\simeq\, \dot \C_\epsilon[\CM_C] }\,,\ee 
where
\be \dot\C_\epsilon[\CM_C] \,=\, \big( \C_\epsilon[\CM_C] \otimes \C\langle e_\n\rangle_{\n\in \pi_1(G)}\big)\big/\text{(relations)}\,, \ee
is obtained by adjoining the idempotents, subject to all the relations that exist when Coulomb-branch operators act on states of fixed vortex number.%
\footnote{Such an enhancement of an algebra with idempotents is especially familiar in the study of quantum groups and their categorification. It was introduced there by Lusztig \cite{Lusztig-qg}.} %
For example, in SQED with vacuum $\nu=\nu_1$, the additional relations set
\be  \begin{array}{c} \varphi\,e_\n = (-m_1-(\n+\tfrac12)\epsilon)\,e_\n \\
 e_{\n'}v_+e_{\n} = e_{\n} v_- e_{\n'} = 0\quad\text{if $\n'-\n\neq 1$} \\
 e_{\n} = 0\quad \text{if $\n<0$}\,. \end{array}  \ee

\subsection{Verma modules}
\label{sec:Verma}

We have proposed that the Hilbert space $\cH_\nu$ of a 3d $\cN=4$ gauge theory in an $\Omega$-background admits an action of the quantized Coulomb-branch algebra $\C_\epsilon[\CM_C]$. We would now like to argue that $\CH_\nu$ is a very special representation of $\C_\epsilon[\CM_C]$, namely a Verma module.

Let us first recall what it means to be a Verma module for the Coulomb-branch algebra.%
\footnote{This discussion is slightly heuristic. For more details see Secs. 5 and 7.2.3 of \cite{BDGH} or mathematical references, \eg\ \cite{BLPW-II, CPS-weight}.} %
The notion depends on a choice of real FI parameter $t_\R$, which we think of as generating an infinitesimal $\mathfrak u(1)_t$ isometry of the Coulomb branch as in \eqref{FIt}. This isometry makes $\C_\epsilon[\CM_C]$ into a graded algebra, such that the degree of any operator $\CO$ equals its weight (or charge) under $\mathfrak u(1)_t$. 
Concretely, all polynomials in the $\varphi$ fields have weight zero and each monopole operator $V_A$ (dressed or undressed) has weight $\langle t_\R,A\rangle$. Our assumption that $t_\R$ is generic means that the weight $\langle t_\R,A\rangle$ is nonzero whenever $A$ is nonzero.

We may decompose
\be \C_\epsilon[\CM_C] \,=\, \C_\epsilon[\CM_C]_< \oplus \C_\epsilon[\CM_C]_0 \oplus \C_\epsilon[\CM_C]_> \label{MC-decomp} \ee
into subspaces of operators with negative, zero, and positive weights, respectively. The space $\C_\epsilon[\CM_C]_0$ simply contains gauge-invariant polynomials in $\varphi$; whereas $\C_\epsilon[\CM_C]_<$ and $\C_\epsilon[\CM_C]_>$ contain monopole operators. A Verma module $M$ is characterized by the following properties:
\begin{enumerate}
\item $M$ is a weight module: it decomposes as a sum $M=\oplus_{\lambda} M_\lambda$ of finite-dimensional spaces $M_\lambda$ of fixed weight $\lambda$, such that for any $\CO\in \C_\epsilon[\CM_C]$  we have $\CO:M_\lambda\to M_{\lambda+\text{deg}(\CO)}$\,.
Physically, this means that $M$ preserves the Coulomb-branch flavor symmetry $\mathfrak u(1)_t$.
\item There is a maximal $\lambda_{\rm max}$ appearing in the sum $\oplus_{\lambda} M_\lambda$, and there exists a ``highest-weight vector'' $|0\rangle \in M_{\lambda_{\rm max}}$  that is annihilated by operators in $\C_\epsilon[\CM_C]_>$ and is an eigenvector for $\C_\epsilon[\CM_C]_0$.
\item The entirety of $M$ is freely generated from $|0\rangle$ by acting with $\C_\epsilon[\CM_C]_<$.
\end{enumerate}

The first property is already manifest for a Hilbert space of the form $\CH_\nu = \oplus_{\n\in \pi_1(G)}H^*(\CM_\nu^\n)$, since the decomposition by vortex number is equivalent to a decomposition into weight spaces. Explicitly, we may assign weight $\langle t_\R,\n\rangle$ to every state $|\n\rangle \in H^*(\CM_\nu^\n)$. Compatibility with the grading of the Coulomb-branch algebra is automatic, since a (potentially dressed) monopole operator $V_A$ sends $H^*(\CM_\nu^\n)\to H^*(\CM_\nu^{\n+\n(A)})$ and $\langle t_\R,\n+\n(A)\rangle = \langle t_\R,\n\rangle+\langle t_\R,A\rangle$.

For the second property we identify $|0\rangle$ as the unique zero-vortex state in $\CH_\nu$, \ie\ the fundamental class of the zero-vortex moduli space $\CM_\nu^0$. Recall that $\CM_\nu^0$ is simply a point, and describes a configuration in which all fields are fixed to their values in the vacuum $\nu$. Thus $|0\rangle$ is an eigenvector for all gauge-invariant polynomials $p(\varphi)$, which are simply set to their vacuum values.

Moreover, all vortex moduli spaces $\CM_\nu^\n$ with $\langle t_\R,\n\rangle>0$ are empty, identifying $|0\rangle$ as a unique \emph{highest}-weight vector. To see this, we observe that in the vacuum $\nu$ some combination of the $X$ and $Y$ hypermultiplet fields are necessarily nonzero, and moreover real moment-map equation $\mu_\R+t_\R=0$ (or, equivalently, the stability condition) requires that $t_\R$ can be written as a non-positive linear combination of weights of the nonvanishing $X$ and $Y$,
\be t_\R \,= \sum_{\mu\,\in\,\text{weights of $X,Y$ nonvanishing at $\nu$}}  \alpha_\mu \,\mu\,,\qquad \alpha_\mu\leq 0\,. \ee
Therefore, if $\langle t_\R,\n\rangle>0$ we must have $\langle \mu,\n\rangle<0$ for at least one $X$ or $Y$ that is nonvanishing in the vacuum. In a configuration of vortex number $\n$, this $X$ or $Y$ must be a) nonvanishing (in order to tend to $\nu$ as $z\to\infty$); b) regular at $z=0$; and c) a polynomial of negative degree. Since this is impossible, the moduli space $\CM_\nu^\n$ is empty.

SQED and SQCD provide simple examples of the highest-weight property. In both theories, we chose a negative FI parameter $t_\R<0$ and found in every vacuum various `$X$' fields had to be nonzero. Correspondingly, the nonempty vortex moduli spaces $\CM_\nu^\n$ all had $\n\geq 0$, which is to say $\langle t_\R,\n\rangle\leq 0$. The zero-vortex state $|0\rangle \in H^*(\CM_\nu^0)$ is the unique vector of maximal weight.

The intuition behind the third property is that any nontrivial vortex configuration can be created from $|0\rangle$ by acting with appropriate monopole operators. We can see this rather explicitly. Consider some nonzero vortex state $|\n,*\rangle$, represented as the (normalized) class of a fixed point $p\in \CM_\nu^\n$. Note that we necessarily have $\langle t_\R,\n\rangle<0$, due to the highest weight property.
The correspondence space $\CM_{\rav\,\nu}^{0,\n} \overset{\pi'}{\hookrightarrow} \CM_\nu^\n$ is a subset of the $\n$-vortex moduli space itself that includes all of the fixed points in $\CM_\nu^\n$. In particular, $\pi'{}^{-1}(p)$ is a fixed point of $\CM_{\rav\,\nu}^{0,\n}$, and its fundamental class $1\!\!1_{\pi'{}^{-1}(p)}$ corresponds to some monopole operator that precisely maps $|0\rangle$ to $|\n,*\rangle$. This monopole operator has negative weight $\langle t_\R,\n-0\rangle = \langle t_\R,\n\rangle$.

We remark that this (somewhat heuristic) argument only holds when complex masses $m_\C$ are generic. The complex masses enter the normalizations of vortex states as equivariant parameters; for special values of the masses, the relative normalizations of states may tend to zero, and the requisite monopole operators relating them  may not exist. For example, in SQED with the action \eqref{SQED-module}, we have
\be v_\n|0\rangle \,=\, \Big(\prod_{l=1}^\n P(-m_1-l\epsilon)\Big)|\n\rangle = \Big(\prod_{i=1}^N\prod_{l=1}^\n (m_i-m_1-l\epsilon)\Big)|\n\rangle\,.\ee
Generically, the prefactor $\prod_{l-1}^\n P(-m_1-l\epsilon)$ is nonzero and $|\n\rangle$ is created from $|0\rangle$ by acting with $v_\n=(v_+)^\n$ (times the inverse of this prefactor).%
\footnote{By generalizing this observation, one can actually show that when the complex masses are generic, every module satisfying (1) and (2) automatically decomposes as a direct sum of Verma modules.} %
However, if the masses are tuned so that some difference $m_i-m_1$ equals $l\epsilon$ for some $1\leq l\leq \n$, then the prefactor vanishes and there is no way to generate $|\n\rangle$ from $|0\rangle$. The case of specialized (or ``quantized'') masses is extremely interesting, and formed the context for much of \cite{BDGH}, but it is not directly relevant here.

\section{Boundary conditions and overlaps}
\label{sec:bound}

We now enrich the setup of the previous sections by adding boundary conditions $\CB$ that fill the $z$-plane at various times $t$, as shown in Figures \ref{fig:boundary}, \ref{fig:interval} in the introduction.
We are interested in boundary conditions that preserve a 2d $\CN=(2,2)$ supersymmetry algebra (as in Table~\ref{tab:SUSY} on page \pageref{tab:SUSY}) and a $U(1)_V$ vector R-symmetry. Such boundary conditions also preserve the two supercharges $Q$ and $Q'$ that we have been using to localize, and are compatible with the $\Omega$-background. Large families of boundary conditions of this type were studied in~\cite{BDGH}.

Roughly speaking, one expects a boundary condition $\CB$ at (say) $t=0$ to define a state $|\CB\rangle$ in the Hilbert space of our 3d $\CN=4$ theory on the cylinder, or equivalently the SUSY Hilbert space of the effective $\CN=4$ quantum mechanics. 
The main goal of this section is to analyze this state when $\CB$ is a ``Neumann-type'' boundary conditions, which preserve gauge symmetry on the boundary.
Using results of \cite{BDGH} (reviewed in Section \ref{sec:N}) we will find that $|\CB\rangle$ must satisfy certain relations of the form
\be V_A\,|\CB\rangle \sim p_A(\varphi)\,|\CB\rangle\,, \label{bdy-Whit} \ee
which identify it as a generalized Whittaker vector in the Verma module $\CH_\nu$. Physically, we would say that $|\CB\rangle$ is a coherent state, a generalized eigenstate of the monopole operators. In addition, using the description $\CH_\nu = \oplus_\n H^*(\CM_\nu^\n)$ as a sum of equivariant cohomology groups of vortex moduli spaces, we will explicitly identify $|\CB\rangle$ with an equivariant cohomology class. In simple cases, it will just be a weighted sum of fundamental classes of each $\CM_\nu^\n$. The fact that this class satisfies the equations \eqref{bdy-Whit} is rather nontrivial.

A 3d $\cN=4$ theory compactified on an interval with boundary conditions $\CB$ and $\CB'$ at either end leads to a 2d $\CN=(2,2)$ gauge theory. This setup is illustrated in Figure \ref{fig:interval} of the introduction. In section~\ref{sec:J}, we show that the partition function of this two-dimensional theory in $\Omega$-background, or vortex partition function, is an inner product of vectors in the Hilbert space of the three-dimensional theory,
\be \CZ_{\rm vortex} \,=\,\langle \CB'|\CB\rangle\,. \label{ZBB}\ee
The Whittaker-like equations \eqref{bdy-Whit} imply that $\CZ_{\rm vortex}$ must satisfy certain differential equations, often of hypergeometric type. In addition, we can use our construction to derive identities for expectation values of twisted chiral operators of the two-dimensional theory in $\Omega$-background. Taken together, these results constitute a `finite' version of the AGT correspondence.

\subsection{Boundaries and modules}
\label{sec:bdy-module}

We begin by describing more carefully the structure of boundary conditions.

\begin{figure}[htb]
\centering
\includegraphics[width=2.5in]{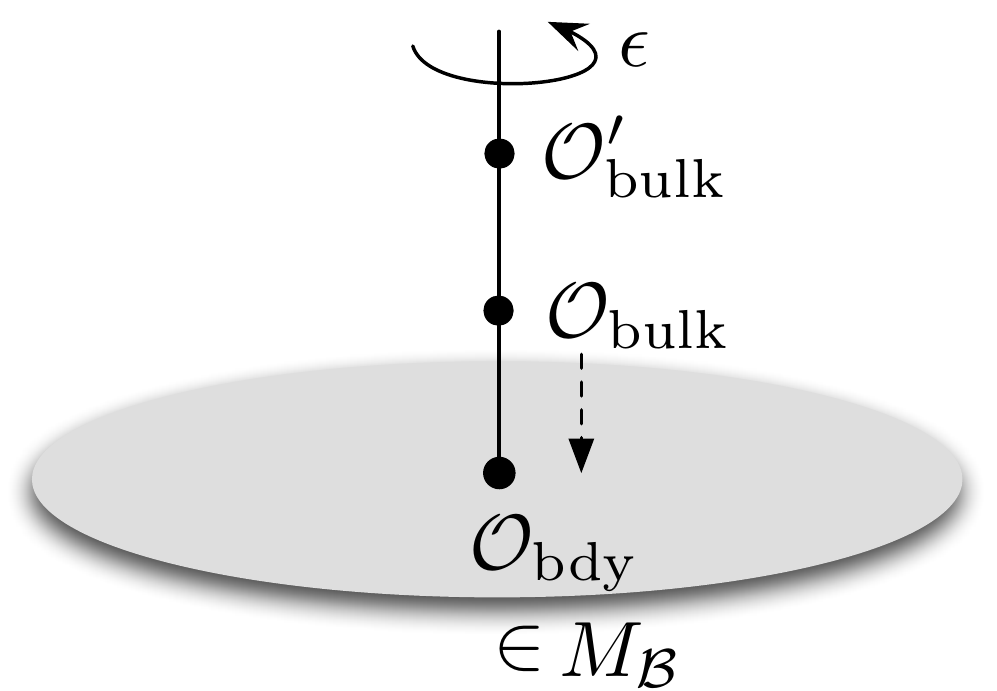}
\caption{The action of a bulk operator algebra on the vector space of boundary operators.}
\label{fig:module}
\end{figure}

The insertion of a $(2,2)$ boundary condition $\CB$ at $t=0$ in our setup has two main effects.
First, via the bulk-boundary OPE,
 the space $M_{\CB}$ of BPS local operators on the boundary (preserved by $Q$ and $Q'$) becomes a module for the algebra $\C_\epsilon[\CM_C]$ of local operators in the bulk (Figure \ref{fig:module}). This is an entirely local phenomenon, independent of the vacuum $\nu$ at $|z|\to\infty$ or any other features at large $z$. One of the main goals of \cite{BDGH} was to describe the module $M_\CB$ associated to a particular UV boundary condition.
 
Second, as we move away from $t=0$, any local operator $\CO_{\rm bdy}\in M_{\CB}$ on the boundary defines a state in the Hilbert space $\CH_\nu$ of the 3d theory; thus there is a map
\be M_\CB \,\overset{\ell}\to\, \CH_\nu\,. \label{MH-hom} \ee
As explained in previous sections, the Hilbert space $\CH_\nu$ is also a module for the Coulomb-branch algebra, and the map \eqref{MH-hom}
respects this action. In other words, it is a homomorphism of modules.

The precise map \eqref{MH-hom} depends on the details of the intersection between the boundary condition $\CB$ at $t=0$ and the vacuum boundary condition $\nu$ at spatial infinity. In particular, one could modify the map \eqref{MH-hom} by adding a line operator along the circle at $t=0$ and $z\to\infty$. We will not do so here.

Given a map \eqref{MH-hom}, the boundary condition $\CB$ defines distinguished state in the Hilbert space
\be |\CB\rangle := \ell(1) \,\in \CH_\nu\,, \label{ell1} \ee
which is the image of the identity operator `$1$' on the boundary. This is the state we seek to describe.

\subsection{Local operators on a Neumann b.c.}
\label{sec:N}

We want to consider boundary conditions $\CB$ involving Neumann boundary conditions for the vectormultiplets that preserve the gauge symmetry $G$ at the boundary. As discussed in \cite[Sec. 2]{BDGH}, the simplest boundary conditions of this type require an additional choice of $G$-invariant Lagrangian splitting of the hypermultiplet representation,
\be R\oplus \bar R \,\simeq\, L\oplus \bar L \ee
This splitting need not have anything to do with the reference splitting $R\oplus \bar R$. Let us write the hypermultiplet chiral fields as $(X_L,Y_L)\in L\oplus \bar L$. Then the boundary condition sets
\be Y_L\big|_\pd = 0\,,\ee
with Neumann boundary conditions for $X_L$, where $|_\pd$ denotes restriction to the boundary. For example, if $G=U(1)$, we have a binary choice of $X|_\pd=0$ or $Y|_\pd=0$ for each hypermultiplet.

The boundary condition also depends on a choice of boundary FI parameter and theta angle, which can be grouped into the twisted chiral combination
\be \xi = e^{t_{2d}+i\theta_{2d}}\,.\ee
Formally, $\xi\in \text{Hom}(G_\C,\C^*)$ is a character of $G_\C$. Given any cocharacter $A$, for example labeling a monopole operator, we denote by
\be \xi^A = e^{\langle t_{2d}+i\theta_{2d},A\rangle} \,\in\,\C^* \ee
under the natural pairing. We denote the Neumann boundary condition with Lagrangian splitting $L\oplus \bar L$ and boundary parameters $\xi$ as $\CN_{L,\xi}$.

The only twisted chiral operators that exist on a Neumann boundary condition are formed from the boundary values of gauge-invariant polynomials in the fields $\varphi$. These are completely unconstrained. Indeed, $\CN=(2,2)$ supersymmetry requires that if gauge symmetry is preserved at the boundary then $\varphi$ has a Neumann boundary condition $\pd_\perp \varphi\big|_\pd = 0$. Thus the space of local operators on any $\CN_{L,\xi}$ is
\be M_{L,\xi} = \{\text{gauge-invariant polys in $\varphi$}\}\,\simeq\, \C[\mathfrak t_\C/W]\,. \label{N-module} \ee

On the other hand, monopole operators are killed by a Neumann boundary condition. Classically, one expects their boundary values to be fixed by the boundary FI parameter and theta angle, $V_A\big|_\pd \sim \xi^A$. Quantum corrections modify this relation. To review how, we introduce abelianized monopole operators $v_A$ following \cite{BDG-Coulomb}, out of which nonabelian monopole operators are constructed. (The abelianized monopole operators were identified with fixed-point classes in correspondence spaces in \eqref{Vv-SQCD}. They also appear as fixed-point classes in the work of Braverman-Finkelberg-Nakajima \cite{BFN-II, BFN-III}.)
When brought to a boundary, an abelianized monopole operator satisfies
\be \label{vN-class}
 v_A\big|_\pd \,=\, \xi^A\; \frac{\ds \prod_{  \lambda\,\in\, L,\;\langle\lambda,A\rangle>0}  (\langle \lambda, \varphi+m_\C\rangle)^{\langle\lambda,A\rangle}  }
{ \ds\prod_{\alpha\,\in\,\text{roots},\;\langle \alpha,A\rangle>0}
 (\langle \alpha, \varphi\rangle)^{\langle\alpha,A\rangle}
}\,,
\ee
where the product in the numerator is over weights of $L$ (counted with multiplicity), and the product in the denominator is over roots of $G$. In the presence of $\Omega$-background, the relation is deformed to%
\footnote{This is related to Eqn (2.58) of \cite{BDGH} by reversing the sign $\epsilon\to -\epsilon$ of the $\Omega$-deformation.}
\be \label{vN-quant}
   v_A\big|_\pd \,=\, \xi^A\,\frac{\ds \prod_{  \lambda\,\in\, L,\;\langle\lambda,A\rangle>0}
       \prod_{l=0}^{\langle \lambda,A\rangle-1}(\langle \lambda, \varphi+m_\C\rangle + (l+\tfrac12)\epsilon)  }
{ \ds\prod_{\alpha\,\in\,\text{roots},\;\langle \alpha,A\rangle>0}
 \prod_{l=0}^{\langle \alpha,A\rangle-1}(\langle \alpha, \varphi\rangle + l\epsilon)
} \,=:\, \xi^A\, \frac{P_A^{\text{hyper}}(\varphi,m_\C)}{P_A^{\text{W}}(\varphi)}\,.
\ee

The various factors in \eqref{vN-class}, \eqref{vN-quant} were understood in \cite{BDGH} as quantum corrections arising from the hypermultiplets in $L$ and the W-bosons with positive charge under the $U(1)_A\subset G$ subgroup defined by $A$. These factors clearly resemble equivariant weights, and we will interpret them as such in Section \ref{sec:Whit}.

In terms of the module $M_{L,\xi}$ containing local operators on the boundary, the relation \eqref{vN-quant} specifies the action of bulk monopoles on the identity operator. Acting on more general polynomials $f(\varphi)\in M_{L,\xi}$, the bulk commutation relations $[v_A,\varphi] = \epsilon A$ imply that
\be v_A\cdot f(\varphi) = \xi^A\, \frac{P_A^{\text{hyper}}(\varphi,m_\C)}{P_A^{\text{W}}(\varphi)} \,f(\varphi+A\epsilon)\,. \label{M-diff} \ee
In nonabelian theories, the denominators $P_A^{\text{W}}$ are nontrivial, and abelianized monopole operators $v_A$ do not preserve the space of polynomials $f(\varphi)$. However, the actual nonabelian operators $V_A$, constructed as Weyl-invariant sums of the $v_A$, are expected to preserve the space of polynomials.

For example, in SQED with $N$ hypermultiplets, the $G$-invariant Lagrangian splittings are labelled by a sign vector $\varepsilon = (\varepsilon_1,\ldots,\varepsilon_N)$, such that
\be X_i \,\in\, L\quad \text{if} \quad \varepsilon_i=+ \,,\qquad Y_i\,\in\,L \quad \text{if} \quad \varepsilon_i=-\,. \label{L-SQED}\ee
Thus the basic Neumann boundary conditions can be labelled $\CN_{\varepsilon,\xi}$. The corresponding space of boundary operators simply consists of polynomials,
\be M_{\varepsilon,\xi} = \C[\varphi]\,, \ee
and the two basic monopole operators $v_\pm$ act on $f(\varphi)\in M_{\varepsilon,\xi}$ as
\be \label{Whit-SQED} \begin{array}{l} 
\ds v_+\cdot f(\varphi) = \xi\,\hspace{-.2cm}\prod_{\text{$i$ s.t. $\varepsilon_i=+$}} \hspace{-.3cm}(\varphi+m_i+\tfrac\epsilon 2)\; f(\varphi+\epsilon)\,, \\[.3cm]
\ds v_-\cdot f(\varphi) = \xi^{-1}\,\hspace{-.3cm}\prod_{\text{$i$ s.t. $\varepsilon_i=-$}} \hspace{-.3cm}(-\varphi-m_i+\tfrac\epsilon 2)\; f(\varphi-\epsilon)\,. \end{array} \ee
It is easy to check that the algebra relations \eqref{SQED-module} are obeyed, up to a sign that can be absorbed in the definition of $v_-$.

\subsubsection{Whittaker modules}
\label{sec:Whit-module}

The module $M_{L,\xi}$ defined above is a generalization of what is known as a Whittaker module in the representation theory of complex semi-simple Lie algebras \cite{Kostant-Whittaker}.

If $\mathfrak g$ is a complex semi-simple Lie algebra, let $\mathfrak g = \mathfrak n^- \oplus \mathfrak t \oplus \mathfrak n^+$ be its decomposition into positive and negative nilpotent subalgebras and a Cartan.
Then a Whittaker module $M$ is characterized by two properties: 
\begin{itemize}
\item $M$ contains an eigenvector $w$ of $\mathfrak n^+$ with nonzero eigenvalues $\xi$.%
\footnote{Note that the commutator subalgebra $[\mathfrak n^+,\mathfrak n^+]\subset \mathfrak n^+$ necessarily annihilates $w$. The requirement that ``$\xi$ is nonzero'' actually means that $\mathfrak n^+/[\mathfrak n^+,\mathfrak n^+]$ acts with generic, nonzero eigenvalues.}
\item $M$ is freely generated from $w$ by acting with $\mathfrak n^-\oplus \mathfrak t$.
\end{itemize}
This is very different from a highest-weight Verma module, which would be generated from a vector $v$ such that $\mathfrak n^+\cdot v = 0$.

The space $M_{L,\xi}$ of local operators on a Neumann boundary condition described above is somewhat similar to a Whittaker module.
Recall from Section \ref{sec:Verma} and \eqref{MC-decomp} that in the presence of real FI parameters, the Coulomb-branch algebra decomposes into positive, zero, and negatively graded subalgebras $\C[\CM_C]_<\oplus \C[\CM_C]_0\oplus\C[\CM_C]_>$. If the FI parameters are generic, then all monopole operators $V_A$ belong to $\C[\CM_C]_<$ or $\C[\CM_C]_>$ while polynomials in $\varphi$ belong to $\C[\CM_C]_0$. As a module for the Coulomb-branch algebra, $M_{L,\xi}$ 
\begin{itemize}
\item contains a unique identity operator $1$ that satisfies $V_A\cdot 1\in \C[\CM_C]_0\cdot 1$ for any $V_A\in \C[\CM_C]_>$; and
\item is freely generated from the identity $1$ by acting with $\C[\CM_C]_<\oplus \C[\CM_C]_0$.
\end{itemize}
In this sense $M_{L,\xi}$ is a generalization of a standard Whittaker module.

The generalized Whittaker modules we encounter here also have a nice geometric characterization \cite[Sec 2.5.1]{BDGH}. Namely, if we send $\epsilon\to 0$, the equations \eqref{vN-class} obeyed at the boundary define a holomorphic Lagrangian section of the Coulomb-branch integrable system $\CM_C \to \mathfrak t_\C/W$. This section is called the support $\text{Supp}(M_{L,\xi})$ of the module.

\subsection{Whittaker states}
\label{sec:Whit}

Next we combine the Neumann boundary condition $\CN_{L,\xi}$ with the vacuum boundary condition $\nu$ at $|z|\to\infty$ for all $t$. No extra data is needed in this case to specify what happens on the circle at infinity where the two boundaries intersect --- we simply require fields there to obey both the $\CN_{L,\xi}$ conditions and to sit in the vacuum $\nu$. Thus, we expect to find a canonical map of modules $\ell:M_{L,\xi}\to\CH_\nu$ as in \eqref{MH-hom}.

We are interested in finding the image of the identity operator \eqref{ell1} under this map, \ie\ the state $|\CN_{L,\xi}\rangle \in \CH_\nu$ created by the boundary condition. Specifying this state actually fixes the entire map, because $M_{L,\xi}$ is generated from the identity (by acting with polynomials in $\varphi$'s) and the map $\ell$ commutes with the action of $\C_\epsilon[\CM_C]$.

Since the identity operator $1\in M_{L,\xi}$ satisfies the Whittaker-like relations \eqref{vN-quant}, the state $|\CN_{L,\xi}\rangle \in \CH_\nu$ must satisfy the same relations --- now with $v_A$ and $\varphi$ interpreted as elements of the convolution algebra (singular gauge transformations) as in \eqref{MC-nu} acting on vortices. Explicitly,
\be \label{Whit-condition} v_{A}\,| \CN_{L,\xi}\rangle \,=\, \xi^A \frac{P_{A}^{\text{hyper}}(\varphi,m_\C)}{P_{A}^{\text{W}}(\varphi)}\,| \CN_{L,\xi}\rangle\,. \ee
This indirectly characterizes $|\CN_{L,\xi}\rangle$.
However, there is also a direct definition of $|\CN_{L,\xi}\rangle$ coming (classically) from looking at solutions to the BPS equations in the presence of a Neumann boundary condition, and (quantum mechanically) from evaluating the path integral in the presence of a Neumann boundary condition.

Let's begin with the BPS equations. Since $\CD_tX=\CD_t Y=0$ we see that if the hypermultiplets $Y_L\in \bar L$ are set to zero on the Neumann boundary at $t=0$, they will continue to be zero for all $t$. Therefore, evolving in time from the Neumann boundary, we will only be able to reach vortex configurations ``supported on $L$.'' Algebraically, for any given vortex number $\n$, we find a restricted moduli space
\begin{align} \label{L-vortex}
 \CM_{\nu,L}^\n &\,=\, \{\text{$(E,X_L)_{\cp^1}$ s.t. $X_L\overset{z\to\infty}{\longrightarrow}G_\C\cdot\nu$}\}/\CG_\C \\
 &\,\subseteq\, \CM_\nu^\n\,, \notag 
\end{align}
where $E$ is an algebraic $G_\C$-bundle on $\cp^1$, trivialized at infinity, and $X_L$ is a section of an associated $L$-bundle. (The moment-map condition $\mu_\C=0$ is automatically obeyed because $Y_L=0$ and $L\oplus\bar L$ is a Lagrangian splitting.) Put differently, \eqref{L-vortex} describes based maps from $\cp^1$ to the Higgs-branch stack $[\mu_\C^{-1}(0)/G_\C]$ supported on the Lagrangian $[L/G_\C]$. Notice that the space $\CM_{\nu,L}^\n$ will be empty unless the orbit $G_\C\cdot\nu$ of the chosen vacuum is contained in $L$.

Quantum mechanically, the localized path integral should produce a corresponding state
\be  |\CN_{L,\xi}\rangle \,:=\, \xi^{\varphi/\epsilon}\, \sum_{\n\,\in\, \pi_1(G)}  1\!\!1_{\CM_{\nu,L}^\n} \,, \label{NLxi}  \ee
where $1\!\!1_{\CM_{\nu,L}^\n} \in H^*(\CM_\nu^\n)$ denotes the Poincar\'e dual of the fundamental class of the subvariety $\CM_{\nu,L}^\n\subset \CM_\nu^\n$. (As usual, we work in $T_F\times U(1)_\epsilon$ equivariant cohomology, but suppress these groups.)

The prefactor $\xi^{\varphi/\epsilon}$ does require a little explanation. This is a contribution to the path integral coming from the twisted superpotential on the boundary, which in the $\Omega$-background takes the form
\be \frac1\epsilon \CW\,=\, \frac1\epsilon\langle t_{2d}+i\theta_{2d}, \varphi\rangle\,. \label{twistedW}\ee
This exponentiates to $\xi^{\varphi/\epsilon}$.
Note that the contraction $\langle t_{2d}+i\theta_{2d}, \varphi\rangle$ is naturally gauge-invariant.
For example, if $G = U(K)$, \eqref{twistedW} is $\tfrac1\epsilon (t_{2d}+i\theta_{2d})\Tr\varphi$.
Acting within a sector of fixed vortex number $\n$, we simply have $\varphi \sim -\n\epsilon+\text{const.}$, where the constant depends on the weights of the flavor symmetry acting on the tangent space to the vacuum $\nu$. Therefore, we could also write
\be |\CN_{L,\xi}\rangle \,=\, \xi^{\text{const}(m_\C,\epsilon)}\, \sum_{\n\,\in\, \pi_1(G)} \xi^{-\n}\,  1\!\!1_{\CM_{\nu,L}^\n}\,. \label{NLxi-n}  \ee
Now the weight $\xi^{-\n}$ is a familiar contribution coming from a topological term $-(t_{2d}+i\theta_{2d})\int_\C F$ in the localized action on the boundary, \cf\ \cite{Shadchin-2d, DGH}.

Mathematically, \eqref{NLxi} may be taken as a definition of the state created by the Neumann boundary condition. It is then a nontrivial conjecture that this state satisfies the Whittaker-like conditions \eqref{Whit-condition}.

In addition to the Whittaker-like conditions, the state $|\CN_{L,\xi}\rangle$ also satisfies some very simple differential equations coming from varying the boundary parameter $\xi$. To be explicit, let us choose a basis for the character lattice such that $\xi=(\xi_1,...,\xi_r)$, and expand $\xi^{\varphi/\epsilon} = \prod_\alpha (\xi_\alpha)^{\varphi_\alpha/\epsilon}$. Then it is obvious from \eqref{NLxi} that
\be \label{var-xi} \epsilon\,\xi_\alpha \frac{\pd}{\pd\xi_\alpha} |\CN_{L,\xi}\rangle = \varphi_\alpha\,|\CN_{L,\xi}\rangle\,.\ee
For example, in the case of a $U(K)$ gauge theory the equation would read $\epsilon\,\xi\frac{d}{d\xi}|\CN_{L,\xi}\rangle = (\Tr\varphi)|\CN_{L,\xi}\rangle$. This differential equation is completely independent of the vacuum $\nu$ or even the Lagrangian splitting $L$. 
It reflects a fundamental property of the module $M_{L,\xi}$ of local operators on the Neumann boundary condition, discussed in further detail in \cite[Sec. 2.5.4]{BDGH}.

\subsubsection{Example: SQED}
\label{sec:N-SQED}

The Lagrangian splittings involved in a Neumann boundary condition are labelled by a sign vector $\varepsilon$ as in \eqref{L-SQED}, and a given Lagrangian $L$ contains the vacuum $\nu_1$ if and only if $\varepsilon_1=+$, that is if $X_1\in L$. Thus
\be \text{$\CM_{\nu,L}^\n$ nonempty}\;\Leftrightarrow\; X_1\in L.\ee

In the extreme case $\varepsilon=(+,...,+)$, the space $\CM_{\nu,L}^\n$ is the entire vortex moduli space $\CM_\nu^\n = \C^{\n N}$ and therefore
\be |\CN_{(+,...,+),\xi}\rangle \,=\, \xi^{\varphi/\epsilon} \sum_{\n\geq 0}  1\!\!1_{\CM_{\nu}^\n} \,=\, \big(\xi^{-\frac{m_1}{\epsilon}-\frac{1}{2}}\big) \sum_{\n\geq 0} \xi^{-\n}\,|\n\rangle\,. \label{SQED+Whit} \ee
Given the action of monopole operators $v_\pm$ in \eqref{SQED-module}, this state clearly satisfies the Whittaker conditions
\be v_-|\CN_{(+,...,+),\xi}\rangle \,=\, \xi^{-1} |\CN_{(+,...,+),\xi}\rangle\,,\quad v_-|\CN_{(+,...,+),\xi}\rangle \,=\, \xi\, P(\varphi+\tfrac\epsilon2) |\CN_{(+,...,+),\xi}\rangle\,.\ee
It also satisfies the differential equation
\be \epsilon\,\xi\frac{d}{d\xi} |\CN_{(+,...,+),\xi}\rangle = \varphi\,|\CN_{(+,...,+),\xi}\rangle\,. \label{var-xi-SQED1} \ee

More generally, if $\varepsilon$ is some sign vector with $\varepsilon_1=+$, then $\CM_{\nu,L}^\n$ is a linear subspace of the vortex moduli space, $\C^{\n N_+}$ where $N_+$ is the number of `$+$' in $\varepsilon$. The corresponding vector,
\be \label{SQEDgenWhit}
 |\CN_{\varepsilon,\xi}\rangle \,=\, \big(\xi^{-\frac{m_1}{\epsilon}-\frac{1}{2}}\big) \sum_{\n\geq 0} \xi^{-\n} \Big(\prod_{\varepsilon_i=-}\prod_{l=0}^{\n-1}(\varphi+m_i+(l+\tfrac12)\epsilon)\Big)|\n\rangle\, ,
\ee
contains extra equivariant weights for the Euler class of the normal bundle to $\CM_{\nu,L}^\n$. This state satisfies the generalized Whittaker conditions \eqref{Whit-SQED}, namely
\be v_\pm|\CN_{\varepsilon,\xi}\rangle = \xi^\pm \prod_{\varepsilon_i=\pm} (\pm(\varphi+m_i)+\tfrac\epsilon2) |\CN_{\varepsilon,\xi}\rangle\,. \ee
Every single $|\CN_{\varepsilon,\xi}\rangle$ obeys \eqref{var-xi-SQED1} as well.

Finally, if $\varepsilon_1=-$, we simply have $|\CN_{\varepsilon,\xi}\rangle=0$. In this case there is no nontrivial solution to the Whittaker-like conditions in $\CH_\nu$.
For example, if $\varepsilon=(-,...,-)$ we would be looking for a state of the form $|\CN_{(-,...,-),\xi}\rangle = \sum_{\n\geq 0}\alpha_\n |\n\rangle$ that obeys $v_+ |\CN_{(-,...,-),\xi}\rangle = \xi |\CN_{(-,...,-),\xi}\rangle$. The image of $v_+$ does not contain $|0\rangle$, so $\alpha_0=0$. By induction, this forces all the remaining $\alpha_\n=0$.

\subsection{Overlaps and vortex partition functions}
\label{sec:J}

Finally, we construct ``sandwiches'' of Neumann boundary conditions. Suppose we place our theory on an interval $[0,t']$, with one Neumann boundary condition $\CN_{L,\xi}$ at $t=0$ and a second $\CN_{L',\xi'}$ at $t=t'$ (Figure \ref{fig:interval} of the introduction). Combined with an $\Omega$-background and a fixed vacuum $\nu$ at $|z|\to\infty$, the system effectively becomes zero-dimensional and should have a well-defined partition function $\CZ$. There are two ways to describe it:
\begin{itemize}
\item[1)] Reducing first to quantum mechanics (say, in the limit of large $t'$), we find that each boundary condition defines states $|\CN_{L,\xi}\rangle$ and $\langle \CN_{L',\xi'}|$ in the Hilbert space $\CH_\nu$ and its dual. The partition function is the inner product of these states
\be \CZ = \langle \CN_{L',\xi'}|\CN_{L,\xi}\rangle_{\CH_\nu}\,. \label{overlap}\ee
\item[2)] Alternatively, we may first collapse the interval $[0,t']$ to zero size, obtaining a 2d $\CN=(2,2)$ theory $\CT_{L,L'}$.
It has a well-studied standard partition function in the $\Omega$-background, sometimes called its vortex partition function
\be \CZ= \CZ_{\rm vortex}[\CT_{L,L'},\nu]\,. \label{Zvortex} \ee
\end{itemize}
The equivalence of these two perspectives follows from the fact that in the BPS sector of the 3d theory that contributes to the partition function (\ie\ in the cohomology of $Q$ and $Q'$) $t$-dependence is trivialized, so the actual length of the interval $[0,t']$ is irrelevant.

Let us spell out some of the details of these constructions. First consider the vortex partition function. The 2d theory $\CT_{L,L'}$ is a gauged linear sigma model, with gauge group $G$ and chiral matter in the representation $L\cap L'$. 
It has a complexified FI parameter equal to the difference of the boundary FI parameters $2\pi i\tau=(t_{2d}+i\theta_{2d})-(t_{2d}'+i\theta_{2d}')$ on the two boundaries at $t=0$ and $t=t'$. In the limit $t'\to 0$, the dependence on the 3d FI parameter $t_\R$ disappears. At low energies, the theory flows to a sigma-model to the Higgs branch
\be \CY_{L,L'} \,\simeq\, (L\cap L')^{\rm stab}/G_\C\,,\ee
where the stability condition depends on the FI parameter $\tau$. Assuming that the difference $t_{2d}-t_{2d}'$ is aligned with the 3d FI parameter $t_\R$, and that $L$ and $L'$ are both compatible with the 3d vacuum $\nu$, then $\CY_{L,L'}$ can be identified with a complex submanifold of the 3d Higgs branch that contains $\nu$,
\be \nu \,\in\, \CY_{L,L'} \,\subset\, \CM_H\,.\ee

The 2d vortex partition function in this case is also known as the equivariant J-function of $\CY_{L,L'}$, appearing in Gromov-Witten theory, \cf\ \cite{GiventalKim, GiventalLee, CoatesGivental}. It takes the form
\begin{align} \label{Jfun} \CZ_{\rm vortex}[\CT_{L,L'},\nu]
 &\,=\, {\text{(const)}}\cdot \sum_{\n\,\in\, \pi_1(G)} q^{-\n} \int_{\CM_{\nu;L,L'}^\n} 1\!\!1_{\CM_{\nu;L,L'}^\n}\,,  \end{align}
where $1\!\!1_{\CM_{\nu;L,L'}^\n}$ denotes the fundamental class of the vortex moduli space
\begin{align} \CM_{\nu;L,L'}^\n &\,=\, \{\text{$\n$-vortex moduli space of $T_{L,L'}$ with vacuum $\nu$ at infinity}\} \\
&\,\simeq\, \{\text{maps to the stack $[(L\cap L')/G_\C]$ of degree $\n$, tending to $\nu$ at infinity}\}\,. \notag
\end{align}
Here $q=e^{2\pi i\tau} = \xi/\xi'$ is the exponentiated 2d FI parameter, and there may be an additional constant prefactor analogous to that in \eqref{NLxi-n} above.

Let us compare this to the inner product of Whittaker states \eqref{overlap}. The state $|\CN_{L,\xi}\rangle$ was expressed as a weighted sum \eqref{NLxi} of fundamental classes of moduli spaces $\CM_{\nu,L}^\n \subset \CM_\nu^\n$, containing vortices supported on the image of $L$ in the 3d Higgs branch. Similarly, we the dual state is
\be \langle \CN_{L',\xi'}|\,= \sum_{\n'\in \pi_1(G)} 1\!\!1_{\CM_{\nu,L'}^{\n'}}\, (\xi')^{\varphi/\epsilon}
 \,=\, \text{(const)}\sum_{\n'\in \pi_1(G)} (\xi')^{\n'} 1\!\!1_{\CM_{\nu,L'}^{\n'}}\,.  \label{dual-Whit} \ee
It belongs to a dual Hilbert space that still takes the form $\CH_\nu^*\simeq\oplus_\n H^*(\CM_\nu^\n)$, but on which monopole operators act ``from the bottom,'' in an opposite representation. For example, a correspondence such as \eqref{Vmaps} of Section \ref{sec:corresp} acts by pulling back via $\pi'$ and pushing forward via $\pi$. In analogy with \eqref{Whit-condition}, the dual state $\langle \CN_{L',\xi'}|$ should satisfy the Whittaker-like conditions
\be \langle \CN_{L',\xi'}|\, v_{A} = \langle \CN_{L',\xi'}|\,(\xi')^A\, \frac{P_{-A}^{\text{hyper}}(\varphi,m_\C)}{P_{-A}^{\text{W}}(\varphi)}\,. \label{dWhit-condition} \ee

The inner product of two equivariant cohomology classes (as in Section \ref{sec:fixed}) is given by taking the cup product and integrating over the entire 3d vortex moduli space $\CM_\nu$. However, since
\be \CM_{\nu,L}^\n \cap \CM_{\nu,L'}^{\n'} \,=\, \begin{cases}  \CM_{\nu;L,L'}^\n  & \n=\n' \\ \oslash & \text{otherwise}\end{cases}\,,\ee
the inner products are just
\be \langle 1\!\!1_{\CM_{\nu,L'}^{\n'}}|1\!\!1_{\CM_{\nu,L}^{\n}}\rangle \,=\, \delta_{\n,\n'} \int_{\CM_\nu^\n} 1\!\!1_{\CM_{\nu;L,L'}^\n} \,=\,  \delta_{\n,\n'} \int_{\CM_{\nu;L,L'}^\n} 1\!\!1_{\CM_{\nu;L,L'}^\n}\,, \ee
and we rather explicitly obtain an equivalence $\langle \CN_{L',\xi'}|\CN_{L,\xi}\rangle = \CZ_{\text{vortex}}[\CT_{L,L'};\nu]$, with the expected identification $q=\xi/\xi'$. 

For example, following the discussion of Section \ref{sec:N-SQED} for SQED, we could choose both Lagrangian splittings so that $L=L'$ and $\varepsilon=\varepsilon' = (+,...,+)$. Then the 2d theory $\CT_{L,L'}$ is a $U(1)$ gauge theory with chiral multiplets $(X_1,...,X_N)$ of charge $+1$, and $\CY_{L,L'}= \cp^{N-1}$ is just the base of the 3d Higgs branch $\CM_H=T^*\cp^{N-1}$. In this case the inner product of Whittaker states \eqref{SQED+Whit} is
\begin{align} \Big(\frac\xi{\xi'}\Big)^{-\frac{m_1}{\epsilon}-\frac12} \sum_{\n,\n'\geq 0} (\xi')^{\n'}\xi^{-\n}\langle \n'|\n\rangle
 &\,=\, q^{-\frac{m_1}{\epsilon}-\frac12} \sum_{\n\geq 0} \frac{q^{-\n}}{\prod_{l=0}^{\n-1} P(-m_1+(l-\n)\epsilon)} \notag \\
 &\hspace{-1in}\,=\, q^{-\frac{m_1}{\epsilon}-\frac12} \sum_{\n\geq 0} \frac{q^{-\n}}{(-\epsilon)^{\n N} \n!\big(\frac{m_1-m_2}{\epsilon}+1\big)_\n\cdots \big(\frac{m_1-m_N}{\epsilon}+1\big)_\n}\,, \label{SQED-sand+} \end{align}
where
\be (x)_n := x(x+1)\cdots(x+n-1) \label{Poch} \ee
is the Pochhammer symbol. This is the equivariant J-function of $\cp^{N-1}$.

More generally, consider $L=L'$ but $\varepsilon=\varepsilon'=(+,*,...,*)$ with both $N_+$ plus signs and  $N_-$ plus minus signs. Then the 2d theory $\CT_{L,L'}$ is a $U(1)$ gauge theory with $N_+$ chiral multiplets $\{X_i\}_{\varepsilon_i=+}$ of charge $+1$ and $N_-$ chiral multiplets $\{Y_i\}_{\varepsilon_i=-}$ of charge $-1$. Its 2d Higgs branch $\CY_{L,L'}\simeq \CO(-1)^{\oplus N_-}\to \cp^{N_+-1}$ is the conormal bundle of a Schubert cell $\cp^{N_+-1}$ in the base of the 3d Higgs branch. The Whittaker-like state $|\CN_{\varepsilon,\xi}\rangle$ is given by \eqref{SQEDgenWhit}, while the dual state is
\be \langle \CN_{\varepsilon,\xi'}| \,=\,   (\xi')^{\frac{m_1}{\epsilon}+\frac{1}{2}} \sum_{\n\geq 0} \langle\n|\, (\xi')^{\n} \Big(\prod_{\varepsilon_i=-}\prod_{l=0}^{\n-1}(\varphi+m_i+(l+\tfrac12)\epsilon)\Big)\,. \ee
The inner product then gives
\begin{align} \label{SQED-sand} \langle \CN_{\varepsilon,\xi'}|\CN_{\varepsilon,\xi}\rangle &\,=\, q^{-\frac{m_1}{\epsilon}-\frac12} \sum_{\n\geq 0}
 \frac{\prod_{\varepsilon_i=-} \big(\frac{m_1-m_i}{\epsilon}+1\big)_\n }
  {\prod_{i>1,\,\varepsilon_i=+} \big(\frac{m_1-m_i}{\epsilon}+1\big)_\n} \frac{q^{-\n}}{(-\epsilon)^{\n(N_+-N_-)}\n!}\,.
\end{align}
This happens to be a generalized hypergeometric function. If we arrange the signs so that $\varepsilon_i=+$ for $i\leq N_+$ and $\varepsilon_i=-$ for $i>N_+$, we can write it as
\be  \langle \CN_{\varepsilon,\xi'}|\CN_{\varepsilon,\xi}\rangle \,=\, q^{-\frac{m_1}{\epsilon}-\frac12} {}_{\;N_+\!}F{}_{N_-}\!\!\left[\begin{array}{c} \frac{m_1-m_{N_++1}}{\epsilon}+1,...\,,\frac{m_1-m_{N}}{\epsilon}+1 \\
\frac{m_1-m_2}{\epsilon}+1,...\,,\frac{m_1-m_{N_+}}{\epsilon}+1\end{array};\, \frac{q}{(-\epsilon)^{N_+-N_-}}\right]\,.
\ee
The expression \eqref{SQED-sand} is also known to be the vortex partition function of the 2d theory $\CT_{L,L'}$~\cite{AGGTV, DGH}.

One could also consider boundary conditions with $L\neq L'$. We only give the simplest, most dramatic example. If $\varepsilon=(+,...,+)$ but $\varepsilon'=(+,-,...,-)$, the 2d theory $\CT_{L,L'}$ is a $U(1)$ theory with a single chiral $X_1$ of charge one and has a trivial 2d Higgs branch. Our two Whittaker-like states are $|\CN_{\varepsilon,\xi}\rangle = \xi^{-\frac{m_1}{\epsilon}-\frac12}\sum_{\n\geq0} \xi^{-\n}|\n\rangle$ and $\langle \CN_{\varepsilon',\xi'}| = (\xi')^{\frac{m_1}{\epsilon}+\frac12}\sum_{\n\geq 0}\langle\n|\, (\xi')^\n \prod_{i>1} \prod_{l=0}^{\n-1} (\varphi+m_i+(l+\tfrac12)\epsilon)$, and due to many cancellations the inner product just gives
\be \langle \CN_{\varepsilon',\xi'}|\CN_{\varepsilon,\xi}\rangle \,=\, q^{-\frac{m_1}{\epsilon}-\frac12} \sum_{\n=0}^\infty \frac{q^{-\n}}{(-\epsilon)^\n\,\n!} \,=\, q^{-\frac{m_1}{\epsilon}-\frac12} e^{ -1/(\epsilon\, q)}\,, \ee
which is the vortex partition function of the simple 2d theory.

\subsubsection{Differential equations}
\label{sec:diff}

In the preceding SQED example, we found that overlaps of vortex states take the form of generalized hypergeometric functions. These functions famously satisfy a differential equation in the parameter $q$. In fact, as discussed in the introduction, more general 2d vortex partition functions (or equivariant J-functions) are all expected to satisfy differential equations in $q$. The differential equations can be explicitly derived from the central relation $\CZ_{\rm vortex}(q;...) = \langle \CN_{L',\xi'}|\CN_{L,\xi}\rangle$ and the defining properties of Whittaker states.

Schematically, the idea is to first observe that $q=\xi/\xi'$, so that $q\pd/\pd q=\xi \pd/\pd \xi$, and to use the relations \eqref{var-xi} to write
\be f\Big( \epsilon\, q\frac{\pd}{\pd q}\Big)\,\CZ_{\rm vortex} \,=\, \langle \CN_{L',\xi'}|  f\Big( \epsilon\, \xi\frac{\pd}{\pd \xi}\Big)|\CN_{L,\xi}\rangle\,=\, \langle \CN_{L',\xi'}| f(\varphi)|\CN_{L,\xi}\rangle \label{dqphi} \ee
for any polynomial $f$ in the logarithmic derivatives $q\pd/\pd q$. Then we recall that monopole operators in the Coulomb-branch algebra satisfy relations of the form $V_{-A}V_{A} = f_A(\varphi)$. (Here we really are being schematic, as for general nonabelian theories such a relation might involve a sum over monopole operators with various dressing factors on the LHS. Also, we are suppressing the dependence on complex masses $m_\C$ and $\epsilon$.)
On the other hand, due to the Whittaker-like conditions \eqref{Whit-condition}, \eqref{dWhit-condition}, the monopole operators act on Whittaker states to give
\be V_A\,|\CN_{L,\xi}\rangle \,=\, \xi^A p_A(\varphi)\,|\CN_{L,\xi}\rangle\,,\qquad \langle \CN_{L',\xi'}|\,V_{-A} \,=\, \langle \CN_{L',\xi'}\,|(\xi')^{-A} p_A'(\varphi)\,. \ee 
for some $p_A$ and $p_A'$. Putting all this together, we find that
\be \begin{array}{rl} \ds f_A\Big( \epsilon\, q\frac{\pd}{\pd q}\Big)\CZ_{\rm vortex} &\,=\,\langle \CN_{L',\xi'}| V_{-A}V_A|\CN_{L,\xi}\rangle \\
 &\,=\,\langle \CN_{L',\xi'}| q^A p_A'(\varphi)p_A(\varphi) |\CN_{L,\xi}\rangle \\[.2cm]
 &\,=\, \ds q^A\,  p_A'\Big( \epsilon\, q\frac{\pd}{\pd q}\Big)p_A\Big( \epsilon\, q\frac{\pd}{\pd q}\Big)\CZ_{\rm vortex}\,,\end{array}
\ee
or 
\be \label{diff-eq}  \Big[  f_A\Big( \epsilon\, q\frac{\pd}{\pd q}\Big) - q^A\,  p_A'\Big( \epsilon\, q\frac{\pd}{\pd q}\Big)p_A\Big( \epsilon\, q\frac{\pd}{\pd q}\Big)\Big] \CZ_{\rm vortex}[\CT_{L,L'},\nu;q] = 0\,. \ee
These are the equations we seek. In principle, there is such an equation for every cocharacter $A$, but only finitely many equations are independent.

To illustrate the procedure explicitly, consider SQED with boundary conditions $\varepsilon=\varepsilon'=(+,...,+)$. Recall that $v_+v_- = P(\varphi+\tfrac\epsilon2)$. Thus
\be \label{SQED-diffeq}
 \begin{array}{rl} \ds P\Big(\epsilon\, q\frac{\pd}{\pd q}+\frac\epsilon2\Big) \CZ_{\rm vortex}(q) &\,=\, \langle \CN_{\varepsilon,\xi'}| P(\varphi+\tfrac\epsilon2) |\CN_{\varepsilon,\xi}\rangle \\
& \,=\, \langle \CN_{\varepsilon,\xi'}|v_+v_- |\CN_{\varepsilon,\xi}\rangle \\
& \,=\, \langle \CN_{\varepsilon,\xi'}| \xi' \xi^{-1} |\CN_{\varepsilon,\xi}\rangle \\
& \,=\, q^{-1}\,\CZ_{\rm vortex}(q)\,,
\end{array}
\ee
whence $\big[\prod_{i=1}^N \big( \epsilon\,q\,\pd/\pd q + m_i + \tfrac\epsilon2\big) -q^{-1}\big]\CZ_{\rm vortex}(q)=0$, which is indeed the hypergeometric equation satisfied by \eqref{SQED-sand+}. Notice that the derivation of this equation did not actually depend on the choice of vacuum $\nu$; the $N$ different choices of vacuum produce the $N$ linearly independent solutions to the hypergeometric equation.

More generally, if $\varepsilon=\varepsilon'=(+,*,...,*)$, analogous manipulations lead to
\begin{align} \prod_{i=1}^N \Big( \epsilon\,q\frac{\pd}{\pd q}\ + m_i + \frac\epsilon2\Big)\, \CZ_{\rm vortex} &\,=\, q^{-1} \prod_{\varepsilon_i=-} \Big( \epsilon\,q\frac{\pd}{\pd q}\ + m_i - \frac\epsilon2\Big)^2\,\CZ_{\rm vortex} \\
&\hspace{-1in}\,=\, \prod_{\varepsilon_i=-} \Big( \epsilon\,q\frac{\pd}{\pd q}\ + m_i + \frac\epsilon2\Big)\, q^{-1}\, \prod_{\varepsilon_i=-} \Big( \epsilon\,q\frac{\pd}{\pd q}\ + m_i - \frac\epsilon2\Big)\, \CZ_{\rm vortex}
\,. \notag \end{align}
For $i\neq 1$, the operators $\epsilon\,q\,\pd/\pd q+m_i+\tfrac\epsilon2$ are invertible as long as mass parameters are generic, so this equation reduces to
\be  \bigg[\prod_{\varepsilon_i=+}  \Big( \epsilon\,q\frac{\pd}{\pd q}\ + m_i + \frac\epsilon2\Big) - q^{-1}\prod_{\varepsilon_i=-} \Big( \epsilon\,q\frac{\pd}{\pd q}\ + m_i - \frac\epsilon2\Big)\bigg]\,\CZ_{\rm vortex}(q) = 0\,,\ee
which is the hypergeometric differential equation governing \eqref{SQED-sand}. 

\subsubsection{Quantization of 2d twisted-chiral rings}
\label{sec:ring}

In addition to taking overlaps of basic Neumann boundary conditions to produce 2d vortex partition functions, we may consider insertions of any bulk Coulomb branch operator $\CO\in \C_\epsilon[\CM_C]$
\be \langle \CN_{L',\xi'}|\CO|\CN_{L,\xi}\rangle \,=\, \langle \CO_{2d}\rangle\,. \ee
This computes the expectation value of a particular twisted chiral operator $\CO_{2d}$ in the $\Omega$-deformed 2d gauge theory $\CT_{L,L'}$. In fact all operators in the 2d twisted-chiral ring can be created this way, by `sandwiching' a 3d Coulomb-branch operator between Neumann boundary conditions. Moreover, the differential equations \eqref{diff-eq} that we derived above can be reinterpreted as relations in a quantized version of the 2d twisted-chiral ring.

We outline a bit of this structure here. We emphasize, however, that very few of the actual results are new. The differential equations of Section \ref{sec:diff} and many of their interpretations were discussed in the introduction.
Expectation values of twisted-chiral-ring operators in the $\Omega$-background have recently been computed explicitly by \cite{ClossetCremonesiPark} using localization methods (see also the related \cite{NS-curved}). We are simply offering a new perspective on these relations, coming from the overlaps of boundary conditions.

Let us first recall that the 2d theory $\CT_{L,L'}$ has a Higgs branch $\CY_{L,L'}$ that (for suitable values of the FI parameter) may be viewed as a complex submanifold of the 3d Higgs branch $\CM_H$. In the absence of the $\Omega$-deformation, the 2d theory has a twisted-chiral ring $\CR_{2d}$ generated by gauge-invariant polynomials in the complex scalar fields $\varphi$, viewed as fields in the 2d gauge multiplet. (They descend from the 3d complex scalar $\varphi$.) Schematically,
\be \CR_{2d} \,=\, \C[\varphi]^G\big/\text{(relations)}\,. \label{2d-relations} \ee
The relations depend on complex masses $m_\C$ (twisted masses in the 2d theory) and on the exponentiated, complex FI parameters $q$. This ring can be identified as the equivariant quantum cohomology ring of the 2d Higgs branch,
\be \CR_{2d} \,\simeq\, QH^*_{T_H}(\CY_{L,L'})\,. \label{QH}\ee

When an $\Omega$-background is turned on, one can still consider expectation values of operators $\CO_{2d}\in \CR_{2d}$, but the ring structure is destroyed. Intuitively, this is because operators are forced to live at the origin and there is no longer a notion of OPE. Nevertheless, there is still a way to make sense of the relations in \eqref{2d-relations}. Recall from \eqref{dqphi} that inserting an operator $\varphi$ in the $\Omega$-deformed partition function has the same effect as acting on the bare partition function with a differential operator
\be \hat \varphi := \epsilon\, q\frac{\pd}{\pd q} \qquad\longleftrightarrow\qquad \text{insertion of $\varphi$}\,.\ee
We can also ``act'' on the partition function with $q$ itself, simply as multiplication. Together, $\hat\varphi$ and $q$ generate a quantum algebra $\C[\hat\varphi,q]^G$, with relations
\be [\hat\varphi,q] = \epsilon\,q\,. \ee
(We retain the superscript ${}^G$ to emphasize that we are only considering gauge-invariant polynomials in $\hat\varphi$.) In the limit $\epsilon\to 0$, we can simply interpret $q$ as a number and $\hat\varphi\to \varphi$ as the usual twisted-chiral ring generator. Thus $\C[\hat\varphi,q]^G\to \C[\varphi]^G$ becomes the usual algebra of 2d twisted-chiral operators, before relations are imposed.

In the presence of $\Omega$-background, the analogues of twisted-chiral ring relations are precisely the differential equations that we found in Section \ref{sec:diff}. These equations, schematically of the form $p(\hat \varphi,q)\cdot \CZ_{\rm vortex}(q)=0$, generate a \emph{left ideal} $\CI_{2d}$ in the algebra $\C[\hat\varphi,q]^G$ --- this ideal is just the set of all differential operators that annihilate the vortex partition function. Thus the analogue of the twisted-chiral ring in the $\Omega$-background is the left $\C[\hat\varphi,q]^G$-module generated by $\CZ_{\rm vortex}$, namely
\be \widehat \CR_{2d} \,=\, \widehat{QH^*_{T_H}}(\CT_{L,L'}) \,:=\, \C[\hat\varphi,q]^G\big/\CI_{2d}\,.\ee
In the limit $\epsilon\to 0$, the ideal $\CI_{2d}$ just becomes the usual (commutative) ideal of relations in the twisted-chiral ring, and $\widehat \CR_{2d}\to \CR_{2d}$.

Our interpretation of vortex partition functions as overlaps of boundary states provides an interesting construction of the differential equations in $\CI_{2d}$, coming from relations in the 3d Coulomb-branch algebra together with the Whittaker-like conditions obeyed by the boundary states.

For example, in SQED with $N$ hypermultiplets and boundary conditions $\varepsilon=\varepsilon'=(+,...,+)$, we saw above that the 2d theory $\CT_{L,L'}$ is a $U(1)$ theory with $N$ chiral multiplets of charge one, whose Higgs branch is $\CY_{L,L'}=\cp^{N-1}$. The equivariant quantum cohomology ring of $\cp^{N-1}$ is $(N-1)$ dimensional,
\be \CR_{2d} \,=\, \C[\varphi]\big/\big((\varphi+m_1)(\varphi+m_2)...(\varphi+m_N)-q^{-1}\big)\,. \ee
The differential equation derived in \eqref{SQED-diffeq} ``quantizes'' this ring, promoting it to a right module for the the algebra $\C[\hat\varphi,q]$,
\be \widehat \CR_{2d} \,=\, \C[\hat\varphi,q]\big/\big((\hat\varphi+m_1+\tfrac\epsilon2)(\hat\varphi+m_2+\tfrac\epsilon2)...(\hat\varphi+m_N+\tfrac\epsilon2)-q^{-1}\big)\,. \ee
In \eqref{SQED-diffeq} we explicitly derived this module starting from the 3d Coulomb-branch relations $v_+v_-=(\varphi+m_1+\tfrac\epsilon2)...(\varphi+m_N+\tfrac\epsilon2)$ together with the Whittaker conditions on boundary states.

\subsection{Example: SQCD}
\label{sec:bound-sqcd}

Consider SQCD with $G=U(K)$ and $N$ fundamental hypermultiplets $(X,Y)$, as in Sections \ref{sec:SQCD-moduli} and \ref{sec:SQCD-mono}. We take $t_\R<0$ and choose the usual vacuum $X^a{}_i=\delta^a{}_i$ at infinity. Choosing the Lagrangian $L$ to contain all the $X$'s, the boundary state corresponding to $\CN_{L,\xi}$ obeys the conditions
\bea
v_a^- | \CN_{L,\xi} \ra & =  \xi\frac{1}{\prod_{b \neq a}(\varphi_b - \varphi_a )} | \CN_{L,\xi} \ra \\
v_a^+ | \CN_{L,\xi} \ra & =  \xi^{-1}\frac{ \prod_{j=1}^N(\varphi_a+m_j+\tfrac{\ep}{2}) }{ \prod_{b \neq a}(\varphi_a - \varphi_b ) }| \CN_{L,\xi} \ra\,, 
\eea
and can be written explicitly in terms of fixed-point classes as
\be
| \CN_{L,\xi} \ra =  \sum_{\n\geq 0}\sum_{k} \xi^{\Tr\varphi/\epsilon} |\n,k\rangle = \xi^{-\sum_{1\leq i\leq K} m_i/\epsilon - K/2} \sum_{\n\geq 0}\sum_{k} \xi^{-\n} |\n,k\rangle \, .
\ee
Again, this reflects the fact that Neumann boundary conditions are compatible with all vortex configurations.

The overlap of Neumann boundary conditions with the same Lagrangian $L$ is
\be
\la \CN_{L,{\xi'}} |\CN_{L,\xi}\ra = q^{-\sum_{1\leq i\leq K} m_i/\epsilon - K/2}  \sum_{\n\geq 0}\sum_k \frac{q^{-\n}}{ \omega_{\n,k}} 
\label{vpf-sqcd}
\ee
where $\omega_{\n,k}$ is the usual equivariant tangent-space weight \eqref{QCD-weights}.
This is the vortex partition function of 2d $U(K)$ gauge theory with $N$ chiral multiplets in the fundamental representation. It (roughly) counts holomorphic maps to the two-dimensional Higgs branch $\CY_{L,L} \simeq \text{Gr}(K,N)$.

It is illuminating to write the constraints on $|\CN_{L,\xi}\rangle$ in terms of the generating functions for dressed monopole operators. In terms of the polynomial generating functions $U^\pm(z)$ the condition is
\bea
U^-(z)& = - \xi^{-1} \\
U^+(z)& = \xi \left[ P(z+\tfrac{\ep}{2}) \; \mathrm{mod} \; Q(z) \right] 
\eea
which is compatible with the quantum determinant relation~\eqref{qdet}. In the limit $\ep\to0$, this defines a nice holomorphic lagrangian in the moduli space of $K$ $PSU(2)$ monopoles with $N$ Dirac singularities.

\section{Vortex quantum mechanics}
\label{sec-adhm}

In sections~\ref{sec:setup} and~\ref{sec:Hilbert}, we argued that a 3d $\cN=4$ gauge theory in an $\Omega$-background in the $x^{1,2}$-plane localizes to an infinite-dimensional supersymmetric quantum mechanics on the $x^3$-axis. The Hilbert space of supersymmetric vacua of this theory decomposed as a direct sum
\be
\cH_\nu = \bigoplus_{\n \in \pi_1(G)} \cH_\nu^\n\; ,
\ee
where each summand is given by the equivariant cohomology of a moduli space of vortices $\cM_\nu^\n$ labelled by a vortex number $\n \in \pi_1(G)$. In this section, we first describe each summand in isolation as an gauged supersymmetric quantum mechanics $Q(\nu,\n)$ with a \emph{finite}-dimensional target, whose Higgs branch is the moduli space of vortices~$\cM_\nu^\n$. This quantum mechanics is known from the brane construction of $\cM_\nu^\n$\,: for three-dimensional triangular quiver gauge theories, they are `handsaw' quiver varieties.
Monopole operators are then realized as a family of interfaces between these vortex quantum mechanics, which we construct in detail.

One way to think about $Q(\nu,\n)$ is as an effective description of the deep-infrared limit of the original 3d $\CN=4$ theory, in an $\Omega$-background, with boundary condition $\nu$\,:
\be \text{3d $\CN=4$}\quad \overset{\text{deep-IR}}\leadsto \quad \bigoplus_\n Q(\nu,\n)\,.\ee
The vortex numbers $\n$ label superselection sectors of the deep-infrared theory. We then correct the deep-infrared description by adding back in the monopole operators, which must take the form of interfaces between different sectors $Q(\nu,\n)$ and $Q(\nu,\n')$. This is similar in spirit to classic constructions of Cecotti-Vafa \cite{CV-tt*, CV-class} and more recently \cite{GMW-infrared}, which analyzed massive 2d $\CN=(2,2)$ theories by first approximating them as a direct sum of vacua, then correcting the approximation with solitons (interfaces) among the vacua.

In this section, we proceed straight to examples, first SQCD, and then triangular quiver gauge theories.

\subsection{SQCD}

\subsubsection{Brane construction}
\begin{figure}[htp]
\centering
\includegraphics[height=4cm]{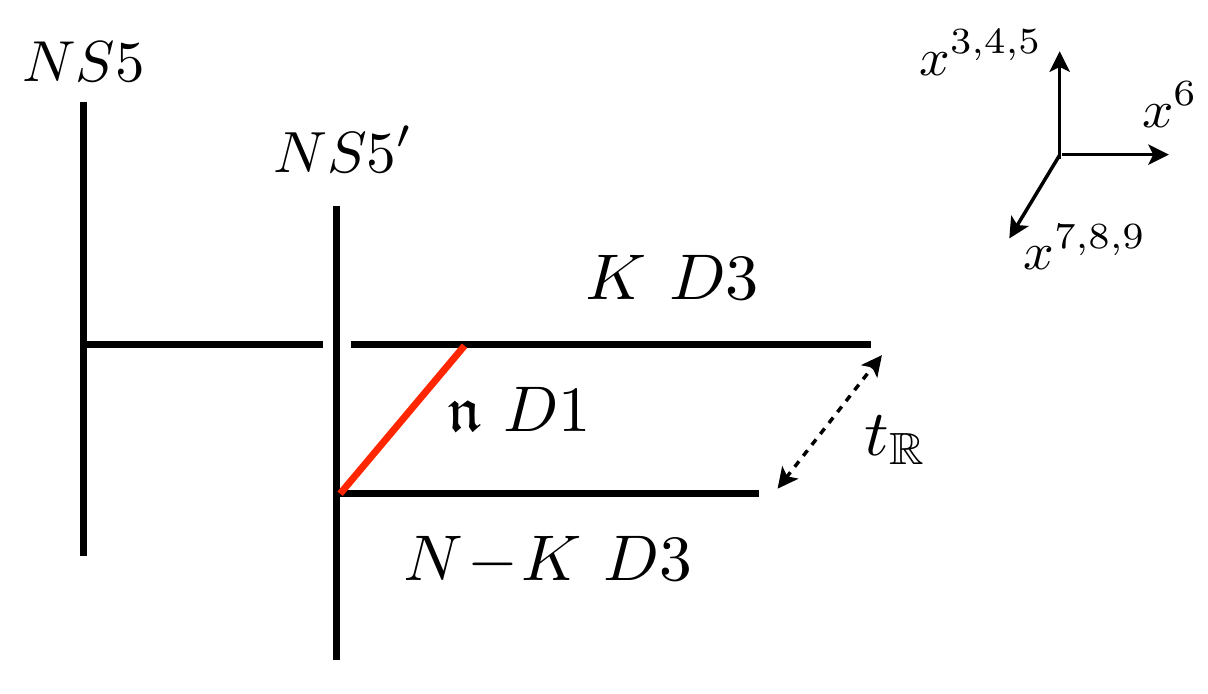}
\caption{The brane construction of vortices in 3d SQCD. The vortices are the D1-branes (red line).}
\label{fig:vortexSQCD}
\end{figure}

We first consider $U(K)$ gauge theory with $N$ fundamental hypermultiplets. The brane construction of the moduli space $\cM_\nu^{\n}$ of vortices is known from the work of Hanany and Tong~\cite{HananyTong-branes}. The brane set-up consists of D3-branes with worldvolume 0126 and two NS5-branes with worldvolume 012345 separated in the $x^6$ direction. In this construction, the $U(K)$ gauge theory arises from a stack of $K$ D3-branes suspended between two NS5-branes and the fundamental hypermultiplets are provided by $N$ semi-infinite D3-branes ending on the right-hand NS5-brane, as drawn in Figure~\ref{fig:vortexSQCD}. Turning on a real FI parameter $t_{\mathbb{R}}$ corresponds to translating the right-hand NS5-brane (NS5') along $x^7$, while generic complex masses $m_1,\ldots,m_N$ correspond to separating the D3-branes in the $x^4,x^5$-directions. There are $N \choose k $ distinct configurations for the D3 branes, which are in 1-1 correspondence with the isolated massive vacua $\nu_I$ described in Section \ref{sec:SQCD-moduli}.

Now we can consider $\n$ D1-branes connecting the $K$ D3-branes and the NS5'-brane. They become the vortices with magnetic flux $\n$ in the 3d gauge theory.  The low energy dynamics of the D1-branes can be described by an $\mathcal{N}=(2,2)$ supersymmetric gauged quantum mechanics%
\footnote{Throughout this section, we use the notation of two-dimensional supersymmetry to describe different types of one-dimensional gauged quantum mechanics and zero-dimensional gauged matrix models. We simply mean that the latter theories can be obtained by dimensional reduction of a two-dimensional gauge theory of the appropriate type.} %
with gauge group $U(\mathfrak{n})$ with $K$ fundamental chiral multiplets $(q,\psi_{\dot a})$, $N-K$ anti-fundamental chiral multiplets $(\tilde{q},\tilde\psi_{\dot a})$, and an adjoint chiral multiplet $(B,\chi_{\dot a})$. The fields of the vectormultiplet are a gauge field $A_t$, gauginos $\lambda_{\dot a}$, $\bar\lambda_{\dot a}$, three scalars $\phi^I$ and an auxiliary field $D$. Turning off the complex mass parameters, the vortex quantum mechanics would have R-symmetry $SU(2)_C\times  U(1)_H$, and $\dot a$ and $I$ are doublet and triplet indices of $SU(2)_C$, respectively. The flavor symmetry is $[U(K)\times U(N-K)]/U(1) \times U(1)_\epsilon$, acting as
\be \label{QCD-handsaw-flavor} \begin{array}{c|ccc} & q,\psi & \tilde q,\tilde\psi & B,\chi \\\hline
  U(K) & \mathbf{K} & \mathbf{1} &  \mathbf{1} \\
  U(N-K) &  \mathbf{1} & \overline{\mathbf{N-K}} &  \mathbf{1} \\
  U(1)_\epsilon & \tfrac12 & \tfrac12 & 1 \end{array}
\ee
The $U(1)_\epsilon$ here is the symmetry \eqref{def-U1e} associated to the $\Omega$-background.

The Lagrangian for the vector multiplet and a chiral multiplet is given by
\be\begin{array}{rl}
   \mathcal{L}_{\rm vec} &=  \Tr\big(\frac{1}{2}D_t\phi^I D_t\phi^I + i \bar\lambda D_t\lambda - \frac{1}{2}D^2 + \frac{1}{4}[\phi^I,\phi^J]^2  + \bar\lambda \sigma^I [\phi^I,\lambda]\big)  \ , \\[.2cm]
   \mathcal{L}_{\rm chi} &=  |D_t q|^2 + i \bar\psi D_t \psi - q^\dagger \phi^I\phi^Iq  -i  q^\dagger Dq - \bar\psi \sigma^I \phi^I \psi +  \sqrt{2}i \bar\psi \bar\lambda q-   \sqrt{2}iq^\dagger \lambda\psi  \,,
\end{array}\ee
where $\sigma^I$ are $SU(2)_C$ Pauli matrices.
The supersymmetry transformation of the vector multiplet is given by
\begin{align}
    \delta A_t &= i\varepsilon \bar\lambda - i\bar\varepsilon \lambda \ , \quad \delta\phi^I = i\varepsilon \sigma^I\bar\lambda + i \bar\varepsilon\sigma^I\lambda \ , \cr
    \delta \lambda_{\dot a} &= \sigma^I_{\dot a \dot b}\varepsilon^{\dot b} \left(-D_t\phi^I + \frac{1}{2}\epsilon_{IJK}[\phi^J,\phi^K]\right) + i\varepsilon_{\dot a} D \ , \cr
    \delta \bar\lambda_{\dot a} &= \sigma^I_{\dot a\dot b}\bar\varepsilon^{\dot b} \left(D_t\phi^I + \frac{1}{2}\epsilon_{IJK}[\phi^J,\phi^K] \right) -i\bar\varepsilon_{\dot a} D \ , \cr
    \delta D &= \bar\varepsilon D_t\lambda +\varepsilon D_t \bar\lambda+i[\phi^I,\varepsilon\sigma^I\bar\lambda]  + i[\phi^I,\bar\varepsilon\sigma^I\lambda] \ ,
\end{align}
and each chiral multiplet transforms under SUSY as
\begin{align}
    \delta q &= \sqrt{2} \varepsilon \psi \ , \quad \delta q^\dagger = - \sqrt{2}\bar\varepsilon\bar\psi \ , \cr
    \delta \psi_{\dot a} & = -i\sqrt{2} \bar\varepsilon_{\dot a} D_t q - \sqrt{2} \sigma^I_{\dot a \dot b}\bar\varepsilon^{\dot b} \phi_I q \ , \cr
    \delta \bar\psi_{\dot a} &= i\sqrt{2}\varepsilon_{\dot a} D_t q^\dagger - \sqrt{2}\sigma^I_{\dot a \dot b}\varepsilon^{\dot b} q^\dagger \phi^I \ .
\end{align}
We introduce the notation $\phi \equiv \phi^1+i\phi^2$ for the complex scalar of charge $+1$ under $U(1)_C \subset SU(2)_C$.

Complex masses and the $\Omega$-background parameter have a common origin in the vortex quantum mechanics as twisted masses for flavor symmetries. Introducing them modifies the supersymmetry transformations schematically by $\phi \to \phi + m_{\mathbb{C}}+\epsilon$.
For simplicity, we will choose the vacuum $\nu$ labelled by $\{1,\ldots,K\} \subset \{1,\ldots,N\}$; then parameters $\{m_1,\ldots,m_K\}$ (of the 3d $\CN=4$ theory) are twisted masses for the fundamental chirals, while $\{m_{K+1},\ldots,m_{N}\}$ are twisted masses for the anti-fundamental chirals. As expected already from the 3d theory, twisted masses break the R-symmetry of the quantum mechanics from $SU(2)_C$ to $U(1)_C$. The vortex quantum mechanics also has a 1d FI parameter $\zeta$, which is identified with the inverse of the 3d gauge coupling as $\zeta\sim 1/ g^2$.

\subsubsection{Hilbert space}

Keeping the 1d FI parameter finite and setting the twisted masses to zero, the supersymmetric quantum mechanics has a Higgs branch of vacua parametrized by the scalar fields in the chiral multiplets subject to the D-term constraint,
\begin{equation}
	[B,B^\dagger] + qq^\dagger -\tilde{q}^\dagger\tilde{q} = \zeta \ ,
\end{equation}
with $\phi^I=0$, modulo $U(\mathfrak{n})$ gauge transformations. This defines a K\"ahler manifold of complex dimension $\mathfrak{n}N$, which coincides with the moduli space of vortices $\cM_\nu^\n$.

Turning back on the twisted masses $m_1,\ldots, m_N$ and $\ep$, the Higgs branch of the supersymmetric quantum mechanics is lifted to the fixed-point set of $T_H \times U(1)_\ep$. The fixed points are found by solving the D-term constraint and
\begin{equation} \label{SQM-QCD-vac}
	 (\phi +m_i + \tfrac{\epsilon}{2} )q^i = 0 \ , \quad -\tilde{q}^j (\phi + m_j - \tfrac{\epsilon}{2} )= 0 \ , \quad [\phi, B] + \epsilon B = 0  \ , \quad [\phi,\phi^\dagger]=0 \ ,
\end{equation}
for all $i=1,\cdots,K$ and $j =K+1,\cdots,N$.
The fixed points are labelled by non-negative integers $k=(k_1,\cdots, k_K)$ such that $\sum_a k_a=\mathfrak{n}$. The explicit solution to \eqref{SQM-QCD-vac} corresponding to each such $k$ is
\be \label{phi-fp}
 \begin{array}{c} \phi=\phi_1\oplus ...\oplus \phi_K\;\;\text{with}\;\; \phi_a = -\text{diag}\big(m_a+\tfrac12\epsilon,m_a+\tfrac32\epsilon,...,m_a+(k_a-\tfrac12)\epsilon\big)\,,\\[.1cm]
 q^i = \begin{cases}(0,...,0,\overbrace{\sqrt{k_a\zeta}}^i,0,...,0) & i=1+\sum_{a=1}^bk_a \,\text{for some $b$} \\
 0 & \text{oth.} \end{cases}, \qquad \tilde q=0\,, \\[.1cm]
 B= B_1\oplus...\oplus B_K\;\;\text{with}\;\; B_a = \begin{pmatrix} 0 &0  & \cdots & 0&0 \\
    \sqrt{(k_a-1)\zeta} & 0 & \cdots & 0&0 \\
    0& \sqrt{(k_a-2)\zeta} & \cdots & 0&0 \\
    &&\ddots &  \\
    0&0&\cdots & \sqrt{\zeta} &0 \end{pmatrix}\,.
\end{array}
\ee

The Hilbert space of the vortex quantum mechanics is the equivariant cohomology of $\cM_\nu^\n$ with respect to the action of $T_H \times U(1)_\ep$. Each equivariant fixed point contributes a state $| \n , k\ra$, normalized as usual such that
\be
\langle \n,k | \n',k' \rangle = \delta_{\n,\n'}\delta_{k,k'} / \omega_{\n,k} \, ,
\label{QCD-norm-ADHM}
\ee
where $\omega_{\n,k}$ is the equivariant weight of the fixed point~\eqref{QCD-weights}. We will derive this result in the following section by computing the partition function of the vortex quantum mechanics on an interval with the insertion of an `identity' interface. Taking a direct sum over the Hilbert spaces of the supersymmetric vortex quantum mechanics with $\n \geq 0$, we recover the full Hilbert space of the effective $\cN=4$ quantum mechanics described in Section~\ref{sec:nonab-moduli}.

\subsubsection{Interfaces}

We now discuss monopole operators in the vortex quantum mechanics. In our setting, a monopole operator is represented as an interface interpolating between a pair of vortex quantum mechanics with different gauge groups.
We will focus on the monopole operators $v_{\pm}$ that change the vortex number by one unit. The other monopole operators $v_{n}$ with $|n|>1$ can be constructed by concatenation.  

In particular, we consider a $U(\n)$ vortex quantum mechanics on the half-line $t<0$ and a $U(\n')$ vortex quantum mechanics on $t>0$ with Neumann-type boundary conditions at $t=0$. Without loss of generality we assume that $\n'\ge \n$. We then couple the theories at $t=0$ by adding boundary matrix degrees of freedom and appropriate superpotential couplings.
In the case $\n' = \n$ this will construct an `identity' interface allowing the computation of the norms $\la \n,k| \n,k\ra$. In the case $\n' = \n + 1$, this will allow us to compute correlation functions $\la \n +1,k' | v_+ | \n,k \ra$ and $\la \n,k| v_- | \n +1,k'\ra$ of monopole operators between pairs of vortex states. 
 
Our proposal for the interfaces is guided by the mathematical construction of Hecke correspondences for the handsaw quiver varieties in \cite{Nakajima-handsaw}.  The first step is to impose a Neumann-type boundary condition at $t=0$ for the two theories on $t<0$ and $t>0$. This boundary condition is given by
\be A_t = \phi^3 = 0 \ , \quad \partial_t \phi = \partial_t B = \partial_t q = \partial_t \tilde{q} = 0\,. \ee
Supersymmetry then requires that
\be\bar\varepsilon\lambda = \bar\varepsilon\sigma^3\lambda = 0 \ , \quad \varepsilon \sigma^I \chi = \varepsilon \sigma^I \psi = \varepsilon\sigma^I\tilde\psi = 0 \qquad (I=1,2)\,. \ee
One can easily check that the 1d action with this boundary condition is invariant under two supersymmetries parametrized by $\varepsilon^+$ and $\bar{\varepsilon}^+$. They correspond to the bulk supersymmetries $Q,Q'$ from \eqref{QQ'} mutually preserved by boundary conditions and vortices. 
Each chiral multiplet in the supersymmetric quantum mechanics leaves a $\mathcal{N}=(0,2)$ chiral multiplet at the boundary.

The second step is to add matrix-model degrees of freedom at $t=0$ preserving the $\mathcal{N}=(0,2)$ supersymmetry with appropriate superpotential couplings. In particular, we will introduce a bi-fundamental chiral multiplet and three Fermi multiplets at $t=0$, whose lowest components transform in the representations
\be\begin{array}{rl}
  \upsilon \ &: \ {\rm chiral \ multiplet \ in \ } (\bf{\bar{\n}'},\bf{\n},\bf{1},\bf{1}) \ , \cr
  \gamma \ & : \ {\rm Fermi \ multiplet \ in \ } (\bf{\bar{\n}'},\bf{\n},\bf{1},\bf{1}) \ , \cr
  \eta \ & : \ {\rm Fermi \ multiplet \ in \ } (\bf{1},\bf{\n},\bf{\overline{K}},\bf{1}) \ , \cr
  \tilde\eta \ & : \ {\rm Fermi \ multiplet \ in \ } (\bf{\bar{\n}'},\bf{1},\bf{1},\bf{N\!-\!K}) \ ,
\end{array}\ee
under the $U(\n')\times U(\n)\times U(K)\times U(N-K)$ symmetries of the supersymmetric quantum mechanics at $t<0$ and $t>0$.
The interactions at the interface are specified by $\cN=(0,2)$ superpotentials for the Fermi multiplets
\begin{equation}
  E_\gamma = \upsilon B' - B\upsilon \ , \quad  E_\eta = \upsilon q' - q \ , \quad  E_{\tilde{\eta}} =  \tilde{q}' - \tilde{q}\upsilon \ ,
\end{equation}
together with $J_\gamma =J_\eta = J_{\tilde\eta}=0 $. Here the primed and unprimed fields correspond to the quantum mechanics on $t>0$ and $t<0$ respectively.
Note that the flavor symmetries on either side of the interface are naturally identified by the superpotentials.

Keeping the 1d FI parameters $\zeta$ finite and setting the twisted masses to zero, the system has a Higgs branch parametrized by the chiral fields on either side and the bifundamental field $\upsilon$ at the interface, subject to the relations
\begin{align}
  &[B,B^\dagger] +qq^\dagger - \tilde{q}^\dagger \tilde{q} - \zeta = 0 \ ,\cr
  &[B',(B')^\dagger] +q'(q')^\dagger - (\tilde{q}')^\dagger \tilde{q}' - \zeta = 0 \ , \cr
  &\upsilon B' = B\upsilon \ , \quad \upsilon q' = q  \ , \quad \tilde{q}' = \tilde{q}\upsilon \ ,
  \label{qm-int-rel}
\end{align}
together with $\phi=\phi'=0$, and modulo the action of $U(\n)$ and $U(\n')$ gauge transformations. The first two lines are the standard D-term contributions from the vortex quantum mechanics, whereas the third line sets to zero the `E-type' $\cN=(0,2)$ superpotentials at the interface. 

This defines a finite dimensional K\"ahler quotient $\mathcal{Z}_\nu^{\n',\n}$ with natural projections onto both $\cM_\nu^{\n'}$ and $\cM_\nu^{\n}$. The complex dimension is
\begin{eqnarray}
\n N + \n' N+ \n \n' - \n \n' - \n K - \n'(N-K)   = \n N+(\n'-\n)K \ ,
\end{eqnarray}
where the first two summands on the left correspond to the dimension of the moduli space of $\n$ and $\n'$ vortices respectively, while the third summand comes from the bi-fundamental scalar $\upsilon$ and the last three from the superpotential constraints at the interface. We consider two cases:
\begin{itemize}
\item $\n'=\n$: the complex dimension is $\n N$ and $\mathcal{Z}_\nu^{\n,\n}$ is the diagonal in $\cM_\nu^\n \times \cM_\nu^\n$. This is the expected result for an identity interface. In particular, the bifundamental field $\upsilon$ is simply a complex gauge transformation.
\item $\n'=\n+1$: the complex dimension is $N(\n+1)+K$. This is consistent with 
\be
\mathcal{Z}_\nu^{\n+1,\n} = \cM^{\n+1,\n}_{\rav,\nu} \times \C \; ,
\label{Cfactor}
\ee
where $\cM^{\n+1,\n}_{\rav,\nu}$ is the correspondence of Section~\ref{sec:corresp} and the factor $\mathbb{C}$ parameterizes the position of the monopole operator in the $z$-plane. In the correspondence $\cM^{\n+1,\n}_{\rav,\nu}$ the position was fixed to $z=0$. We will therefore need to remove the contributions from the factor of $\C$ in order find precise agreement.
\end{itemize}

We now compute the partition function of the quantum mechanical system with a supersymmetric vacuum $|\n,k\ra$ at some $t_1<0$, a supersymmetric vacuum $|\n',k'\ra$ at some $t_2 >0$ and an interface at $t=0$. The saddle point of the path integral corresponds to the equivariant fixed point of the correspondence $\mathcal{Z}_{\nu}^{\n+1,\n}$ labelled by the pair of partitions $k$, $k'$. Note that the equivariant fixed points of $\mathcal{Z}_{\nu}^{\n+1,\n}$ are labelled by a pair of equivariant fixed points for $\cM_\nu^{\n'}$ and $\cM_\nu^{\n}$ with the value of the bifundamental chiral $\upsilon$ determined by the final line of~\eqref{qm-int-rel}.

The 1-loop contributions from fluctuations around the saddle point contains contributions from three sources: a) the quantum mechanics on $t>0$; b) the quantum mechanics on $t<0$; and c) the matrix model degrees of freedom at $t=0$. 
First, the contributions coming from a) and b) are given by 
\begin{equation}
Z^{\rm 1-loop}_{t>0} = Z^{\rm 1-loop}_{\n',k'} \ , \quad Z^{\rm 1-loop}_{t<0} = Z^{\rm 1-loop}_{\n,k}
\end{equation}
where
\begin{equation}
Z^{\rm 1-loop}_{\n,k} =
\frac{\prod^\mathfrak{n}_{I\neq J}(\phi_I - \phi_J)}{\prod_{I,J=1}^\mathfrak{n} (\phi_I-\phi_J+\epsilon)} \prod_{I=1}^\mathfrak{n} \frac{1}{\prod_{i=1}^K (\phi_I + m_i+\frac{\epsilon}{2}) \prod_{i=K+1}^N(-m_i-\phi_I+\frac{\epsilon}{2})} \ .
\end{equation}
The additional contribution from the matrix degrees of freedom at $t=0$ is given by
\begin{equation}
	Z^{\rm 1-loop}_{t=0} = \prod_{I=1}^{\n}\prod_{J=1}^{\n'}\frac{(\phi_I-\phi'_J+\epsilon)}{(\phi_I-\phi'_J)}\cdot\prod_{I=1}^{\n}\prod_{i=1}^K(\phi_I+m_i+\frac{\epsilon}{2}) \cdot\prod_{J=1}^{\n'}\prod_{j=K+1}^N(-m_j - \phi'_J+\frac{\epsilon}{2}) \ ,
\end{equation}
Then the localized partition function with the interface can be expressed as the residue of a product of the 1-loop contributions evaluated at the supersymmetric vacua~\eqref{phi-fp} corresponding to $k$ and $k'$ for $\phi_I$ and $\phi_I'$ respectively.
\begin{equation}
	Z_{(\n',k')\times (\n,k)} = {\rm Res}_{\phi,\phi'}\, Z^{\rm 1-loop}_{t>0} \cdot Z^{\rm 1-loop}_{t=0} \cdot Z^{\rm 1-loop}_{t<0} \ .
\end{equation}
Note that the contribution from the matrix degrees of freedom cancels all the poles of $\phi_I$ corresponding to the supersymmetric vacuum labelled by $k$, and instead introduce new poles at $\phi_I-\phi_J' = 0$. Therefore, the the partition $k$ should be a subset of the partition $k'$, i.e. $k_a\le k_a'$ for all $a$. Otherwise the partition function becomes trivial, which implies that the full system with the interface has no corresponding supersymmetric vacuum. 

Plugging the saddle point values of $\phi_I$ and $\phi_I'$ into the 1-loop contributions (and removing poles and zeros), we find that $Z^{\rm 1-loop}_{t>0} =1/\omega_{\n,k}$ and $Z^{\rm 1-loop}_{t<0} =1/\omega_{\n',k'}$ where 
\begin{equation}\label{QCD-weights-ADHM}
\omega_{\n,k} = \prod_{i=1}^K\prod_{s=1}^{k_i} \left[ \, \prod_{j=1}^K(m_j-m_i+(1+k_j-s)\epsilon)\prod_{j=K+1}^N(m_i-m_j+s\epsilon) \, \right] \, .
\end{equation}
is an equivalent expression for the the equivariant weight~\eqref{QCD-weights}. An similar computation leads to the contribution
\begin{align}\label{eq:1-loop-interface}
  Z^{\rm 1-loop}_{t=0} = \prod_{i,j=1}^K\prod_{s=1}^{k_i}\left(m_j-m_i+(1+k_j'-s)\epsilon\right) \cdot\prod_{i=1}^K\prod_{j=K+1}^{N}\prod_{s=1}^{k_i'}\left(m_i-m_j+s\epsilon\right) \ .
\end{align}
Multiplying these contributions gives the desired partition function. When $\n'=\n+1$ we further multiply by a factor of $\epsilon$ to remove the contribution from $\mathbb{C}$ in~\eqref{Cfactor} corresponding to the position of the monopole.

Let us first consider $\n' = \n$. In this case, we are inserting an identity interface at $t=0$ and therefore our computation result should reproduce the overlap $\la \n,k'| \n,k\ra$. Indeed, the contribution from the matrix degrees of freedom is given by $Z^{\rm 1-loop}_{t=0} = \delta_{k,k'}\omega_{\n,k}$ and therefore we reproduce the normalization~\eqref{QCD-norm-ADHM}.

In the case $\n' = \n+1$, we expect the partition function to reproduce the correlation function of monopole operators $\la \n+1,k'| v^+_a | \n,k\ra$. This partition function vanishes unless $k' = k+\bm\delta_a$ for some $1\leq a \leq K$ as discussed above. Putting all contributions together, we find
\begin{align}\label{eq:correlation-Vplus}
  & \la \n+1,k+\bm\delta_a | v_a^+ | \n,k\ra \\
  &= \prod_{i\neq a}^K \frac{1}{(m_i\!-\!m_a\!+\!(k_i\!-\!k_a)\epsilon)} \prod_{i=1}^K  \prod_{s=1}^{k_i}\frac{1}{\prod_{j=1}^K(m_j\!-\!m_i\!+\!(1\!+\!k_j\!-\!s)\epsilon)\prod_{j=K+1}^{N}(m_i\!-\!m_j\!+\!s\epsilon)} \ .\nonumber
\end{align}

This result can also be interpreted as the correlation function with the monopole operator $v_a^-$. So we have the relation
\begin{equation}
	\la \n,k | v_a^- | \n+1,k+\bm\delta_a\ra  =  \la \n+1,k+\bm\delta_a | v_a^+ | \n,k\ra \ .
\end{equation}

Using these correlation functions, we can extract explicit forms of the actions of monopole operators on vortex states. The monopole operators  act by
\begin{equation}
	v_a^+ | \n,k \rangle = C^+_a | \n+1,k+\bm\delta_a \rangle \ ,
	\qquad v_a^- | \n,k \rangle = C^-_a | \n-1,k-\bm\delta_a \rangle \ ,
\end{equation}
with some coefficients $C^\pm_a$. One can easily compute the coefficients by sandwiching the vortex states $\langle \n+1,k+\bm\delta_a |$ and  $\langle \n,k |$ on these relations. Some simple algebra leads to
\begin{align}
	&C_a^+ = \frac{ \langle \n+1,k+\bm\delta_a | v_a^+ | \n,k \rangle}{ \langle \n+1,k+\bm\delta_a| \n+1,k+\bm\delta_a \rangle}
  	=\frac{\prod_{i=1}^{N}(m_a-m_i+(k_a+1)\epsilon)}{\prod_{i\neq a}^K(m_a-m_i+(k_a-k_i+1)\epsilon)} \ .
\end{align}
Therefore the monopole operator $v_a^+$ acts as
\begin{equation}
	v_a^+ | \n,k \rangle = \frac{\prod_{i=1}^{N}(m_a-m_i+(k_a+1) \epsilon)}{\prod_{i\neq a}^K(m_a-m_i+(k_a -k_i+1)\epsilon)} | \n+1,k+\bm\delta_a\rangle \ ,
\end{equation}
A similar computation leads to the action of the monopole operator $v_a^-$ as
\begin{align}
  & v_a^-|\n,k\rangle = \prod_{i\neq a}^K\frac{1}{(m_i-m_a+(k_i-k_a+1)\epsilon)} | \n-1,k-\bm\delta_a \rangle \ .
\end{align}
We assert that $v_a^-|\n,k\ra=0$ if $k_a$ is zero because our system with the interface has no such vacuum and thus the correlation function of $v^-_a$ becomes zero.

If we define an operator such as
\begin{equation}
	\varphi_a | \n,k \rangle = (-m_a - k_a\epsilon - \tfrac{\epsilon}{2}) |\n,k\rangle \ ,
\end{equation}
then the monopole operators can be simplified as
\begin{equation}
	v_a^+| \n,k \rangle = \frac{P(\varphi_a+\frac{\epsilon}{2})}{\prod_{b\neq a}^K(\varphi_b-\varphi_a)}| \n+1,k+\bm\delta_a \rangle \ , \quad v_a^- | \n,k \rangle = \frac{1}{\prod_{b\neq a}^K(\varphi_a-\varphi_b)} | \n-1,k-\bm\delta_a\rangle  \ .
\end{equation}
These actions of the monopole operators perfectly agree with the actions in (\ref{SQCD-Vplus}) and (\ref{SQCD-Vminus}) computed using the correspondence $\CM_{\rav,\nu}^{\n,\n+1}$.

\subsection{Triangular quiver}

\begin{figure}[htp]
\centering
\includegraphics[height=4cm]{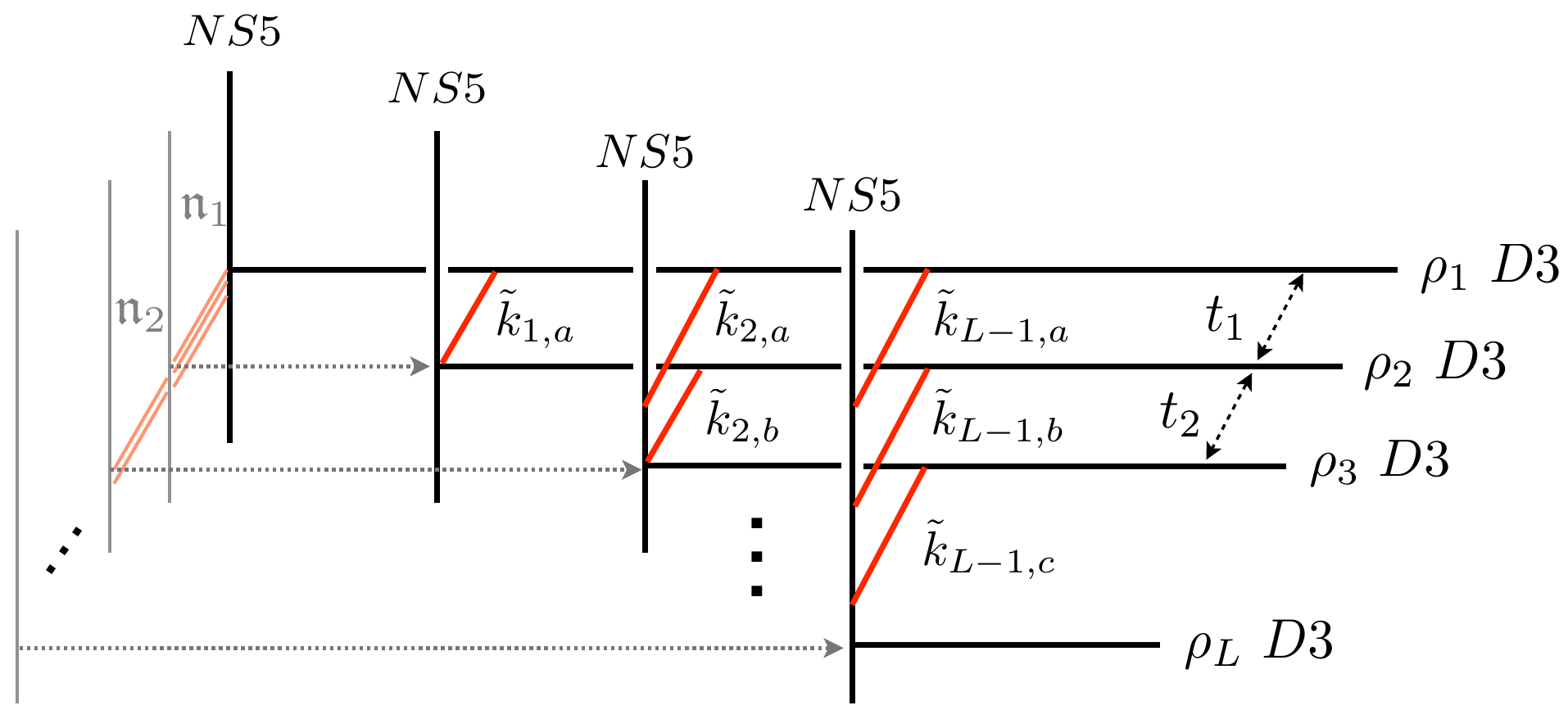}
\caption{The brane construction of vortices in the triangular quiver theory for the partition $\rho=[\rho_1,\rho_2,\cdots,\rho_L]$. The vortices are the $\tilde{k}_{\alpha,a}$ D1-branes (red lines) with $k_{\alpha,a} = \sum_{\upsilon=\alpha}^{L-1} 
\tilde{k}_{\alpha,a}$.}
\label{fig:vortextriangular}
\end{figure}

We now turn to the 3d triangular quiver gauge theory discussed in Section~\ref{sec-triang-hilb} and the vortex quantum mechanics in this theory.
We have a simple brane construction for the triangular quiver theory. When the theory is fully Higgssed with the real FI parameters $t_\alpha\ (1\le \alpha < L)$, it can be engineered by the brane system with $L$ NS5-branes and $N$ D3-branes with $\sum_{\alpha=1}^L\rho_\alpha=N$ in Figure~\ref{fig:vortextriangular}. 

The 3d field theory has $N_\rho = N!/(\rho_1!\cdots \rho_L!)$ supersymmetric massive vacua when generic hypermultiplet masses are turned on. In the brane system, a field theory vacuum labelled by nested subsets $\mathcal{I}_\alpha=\{i_{\alpha,1},\cdots,i_{\alpha,K_\alpha}\}$ define in \eqref{nested-subsets} is mapped to a configuration with a choice of $\rho_\alpha$ D3-branes among the total $N$ D3-branes attached at the $\alpha$-th NS5-brane. Given a supersymmetric vacuum, vortex particles are provided by  D1-branes suspended between one of the NS5-branes and a D3-brane.

The dynamics of the vortices can be described by the gauge theory living on the D1-branes. At low energy with finite FI parameters, the theory on the D1-branes reduces to a $\mathcal{N}=(2,2)$ supersymmetric quantum mechanics given by so-called `{\it handsaw quiver}' gauge theory. The quiver presentation can be easily read off from the brane configuration. This handsaw quiver theory has been studied extensively in the mathematical literature, for example in \cite{2010arXiv1009.0676F,Nakajima-handsaw}. The moduli space of vortices in the 3d triangular quiver theory agrees with the Higgs branch of this quantum mechanics.

\begin{figure}[htp]
\centering
\includegraphics[height=3.5cm]{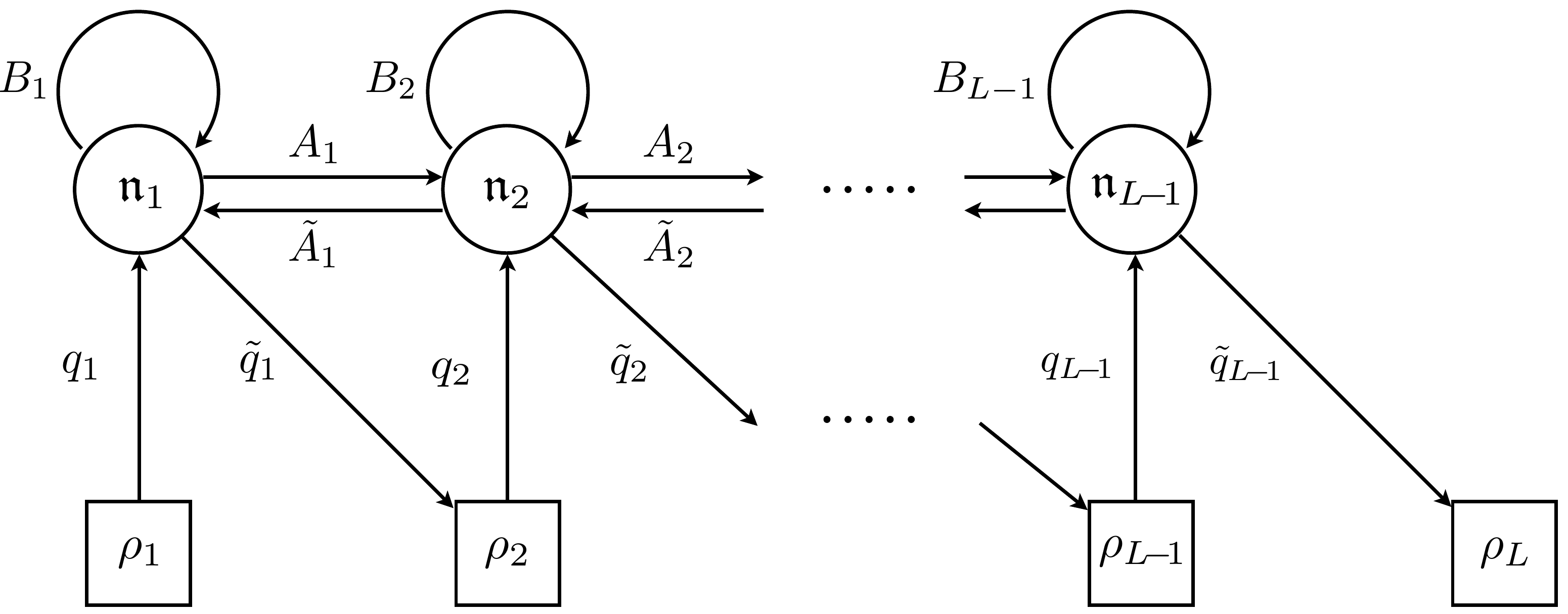}
\caption{Hand-saw quiver theory.}
\label{fig:handsaw}
\end{figure}

The handsaw quiver theory, as illustrated in Figure \ref{fig:handsaw},
has gauge group $G_{QM}=\prod_{\alpha=1}^{L-1}U(\n_\alpha)$ and flavor group $\prod_{\alpha=1}^{L}U(\rho_\alpha)$. Each gauge node has chiral multiplets with scalar components $q_\alpha$ in a bi-fundamental representation $(\n_\alpha, \bar{\bm \rho}_\alpha)$ and $\tilde{q}_\alpha$ in $({\bm \rho}_{\alpha+1}, \bar\n_\alpha)$ under the gauge and flavor groups; together with $B_\alpha$ in the adjoint representation of the gauge group $U(\n_\alpha)$. 
Two adjacent nodes are connected by bi-fundamental chiral multiplets with scalar components $A_\alpha$ and $\tilde{A}_\alpha$. The theory has additional superpotential couplings
\begin{equation}
	\mathcal{W}_\alpha = {\rm Tr} \ \tilde{A}_\alpha \left(A_\alpha B_\alpha - B_{\alpha+1}A_\alpha + q_{\alpha+1}\tilde{q}_\alpha\right) \ , \quad (\alpha=1,\cdots,L\!-\!2) \ .
	\label{triangular-interface-W}
\end{equation}
The flavor charges for the bi-fundamental fields $A_\alpha$ and $\tilde{A}_\alpha$ are fixed by these superpotentials.

The moduli space of vortices in the 3d gauge theory coincides with a Higgs branch $\mathcal{M}_H$ of vacua in the quantum mechanics parametrized by the scalar fields $q_\alpha,\tilde{q}_\alpha,B_\alpha$, and $A_\alpha$, with $\phi_\alpha=\tilde{A}_\alpha=0$. These scalars are subject to the D-term and the F-term constraints:
\begin{equation}
	\mathcal{M}_H = \{ \mu_\alpha = \zeta_\alpha\, ,\, A_\alpha B_\alpha - B_{\alpha+1}A_\alpha + q_{\alpha+1}\tilde{q}_\alpha=0\}/G_{QM} \ 
\end{equation}
where $\zeta_\alpha$ are the FI parameters for the gauge group $G_{QM}$. 
The complex dimension of the Higgs branch is
\begin{equation}
	{\rm dim}_{\mathbb{C}} \mathcal{M}_H = \sum_{\alpha=1}^{L-1} \n_\alpha(\rho_\alpha+\rho_{\alpha+1}) \ .
\end{equation}
This agrees with the moduli space of vortices in the 3d gauge theory.

Let us now turn on generic twisted masses $m_1,\cdots, m_N$ for the flavor symmetry and $\epsilon$ for $U(1)_\epsilon$. 
The Higgs branch of vacua will be lifted to a set of isolated fixed points under the action $T_H\times U(1)_\epsilon$, which solve the deformed BPS equations
\be\begin{array}{l}
	[\phi_\alpha, B_\alpha] + \epsilon B_\alpha = 0 \;, \quad \phi_{\alpha+1}A_i - A_i\phi_\alpha = 0 \;, \\
	 (\phi_\alpha+ m_{i_{\alpha,a}} + \tfrac{\epsilon}{2})q^{a}_\alpha = 0
	\;, \quad -\tilde{q}_\alpha^{b}(\phi_\alpha+m_{i_{\alpha+1,b}} - \tfrac{\epsilon}{2}) = 0 \;, \\
	 {}[\phi_\alpha,\phi_\alpha^\dagger]=0 \;,
\end{array}\ee
for $K_{\alpha-1}<a\le K_\alpha$ and $K_\alpha < b \le K_{\alpha+1}$, as well as D-term and F-term constraints.
Generalizing \eqref{phi-fp}, one finds that solutions are labelled by multi-vectors $\vec{k}=\{k_{\alpha,a}\}$ with $\sum_{a=1}^{K_\alpha}k_{\alpha,a}=\n_\alpha$ and $k_{\alpha,a}\ge k_{\alpha+1,a}$.
For each such solution, the complex scalars $\phi_\alpha$ are diagonalized, with blocks
\be \label{phi-fp-triangular}
 \phi_\alpha = \phi_\alpha^{(1)}\oplus...\oplus\phi_\alpha^{(K_\alpha)}\,,\qquad \phi_\alpha^{(a)} = -\text{diag}\big(m_{i_{\alpha,a}} +\tfrac12\epsilon, m_{i_{\alpha,a}} +\tfrac32\epsilon,..., m_{i_{\alpha,a}}+(k_{\alpha,a}-\tfrac12)\epsilon\big)\,. \ee

The equivariant weight of the tangent space to the Higgs branch $\mathcal{M}_H$ can be easily computed again using the equivariant index theorem.
We find that the inverse of the residue of
\begin{eqnarray}
	&& \prod_{\alpha=1}^{L-1} \frac{\prod_{I\neq J}^{\n_\alpha}(\phi_{\alpha,I}\!-\!\phi_{\alpha,J})}
	{\prod_{I, J=1}^{\n_\alpha}(\phi_{\alpha,I}\!-\!\phi_{\alpha,J}\!+\!\epsilon)} \times 
	\prod_{\alpha=1}^{L-2}\prod_{I=1}^{\n_\alpha}\prod_{J=1}^{\n_{\alpha+1}}\frac{(\phi_{\alpha+1,J}-\phi_{\alpha,I}+\epsilon)}{(\phi_{\alpha+1,J}-\phi_{\alpha,I})}
	  \nonumber \\
	&&  \times \prod_{\alpha=1}^{L-1}\prod_{I=1}^{\n_\alpha}\frac{1}{\prod_{a=K_{\alpha\!-\!1}+1}^{K_\alpha}(\phi_{\alpha,I}\!+\!m_{i_{\alpha,a}}\!+\!\frac{\epsilon}{2})\prod_{a=K_\alpha+1}^{K_{\alpha\!+\!1}}(-m_{i_{\alpha+1,a}}\!-\!\phi_{\alpha,I}\!+\!\frac{\epsilon}{2})} \ ,
\end{eqnarray}
at \eqref{phi-fp-triangular} gives rise to the equivariant weight 
\begin{align}
	\omega_{\vec{n},\vec{k}} &=\prod_{\alpha=1}^{L-1} \prod_{a=1}^{K_\alpha}\prod_{s=1}^{k_{\alpha,a}} \frac{\prod_{b=1}^{K_\alpha}(m_{i_{\alpha,b}}\!-\!m_{i_{\alpha,a}}\!+\!(k_{\alpha,b}\!-\!s\!+\!1)\epsilon)\prod_{b=K_\alpha+1}^{K_{\alpha\!+\!1}}(m_{i_{\alpha,a}}\!-\!m_{i_{\alpha+1,b}}\!+\!s\epsilon)} 
	{\prod_{b=1}^{K_{\alpha\!-\!1}}(m_{i_{\alpha\!-\!1,b}}\!-\!m_{i_{\alpha,a}}\!+\!(k_{\alpha\!-\!1,b}-\!s\!+\!1)\epsilon)} \nonumber \\
	&=\prod_{\alpha=1}^{L-1} \prod_{a<b}^{K_\alpha}\frac{m_{i_{\alpha,b}}-m_{i_{\alpha,a}}}{\varphi_{\alpha,a}-\varphi_{\alpha,b}} \prod_{a=1}^{K_\alpha}\prod_{l=0}^{k_{\alpha,a}-1}
	\frac
	{\tilde{Q}_{\alpha+1}(\varphi_{\alpha,a}+(l+\tfrac{1}{2})\epsilon)}
	{Q_{\alpha-1}(\varphi_{\alpha,a}+(l+1)\epsilon)} \ ,
\end{align}
where we define $\varphi_{\alpha,a} = -m_{i_{\alpha,a}}-(k_{\alpha,a}+\tfrac{1}{2})$.
Thus the quantum mechanics result precisely reproduces the previous result in (\ref{triangular-eqweight}).

\subsubsection{Interface}\label{sec:interface-triangular}

We now construct 1d quantum mechanical systems for monopole operators in the triangular quiver theories. The 1d systems can realized by particular interfaces in the vortex quantum mechanics. By analogy with the construction of interfaces in the SQCD case, we will first consider two 1d handsaw quiver theories with vortex numbers $\vec\n$ and $\vec\n'$ living on the half-lines $t<0$ and $t>0$ respectively, and glue these two theories by adding extra boundary degrees of freedom and turning on boundary interactions at $t=0$.

For the theory at $t<0$, we first give Neumann-type boundary conditions to the vectormultiplets and the chiral multiplets with scalars $B_\alpha,A_\alpha, q_\alpha,\tilde{q}_\alpha$. For the chiral multiplets of $\tilde{A}_\alpha$, we will choose the Dirichlet-type boundary condition that sets
\begin{equation}
	\tilde{A}_\alpha = (\tilde\chi_\alpha)_+ = 0\qquad \text{at $t=0$} \ ,
\end{equation}
and leaves $\mathcal{N}=(0,2)$ Fermi multiplets for $\Lambda_\alpha\equiv(\tilde\chi_\alpha)_-$ with boundary interactions
\begin{equation}
	J_{\Lambda_\alpha} = A_\alpha B_\alpha - B_{\alpha+1}A_\alpha + q_{\alpha+1}\tilde{q}_\alpha \ , \quad E_{\Lambda_\alpha} = 0
\end{equation}
at $t=0$. We consider similar boundary conditions for the 1d theory on $t>0$.
Two vortex theories with these boundary conditions will be connected by new degrees of freedom at $t=0$.

We propose that
the new boundary degrees of freedom consists of the $\mathcal{N}=(0,2)$ chiral and Fermi multiplets whose lowest components transform in the representations
\begin{align}
  \upsilon_\alpha \ &: \ {\rm chiral \ multiplet \ in \ } ({\bf \bar{\n}'}_\alpha,{\bf\n}_\alpha,{\bf 1}) \ , \quad
  \tilde\upsilon_\alpha \ : \ {\rm chiral \ multiplet \ in \ } ({\bf\n'}_\alpha,{\bf\bar\n}_{\alpha+1},{\bf1}) \ , \cr
  \gamma_\alpha \ & : \ {\rm Fermi \ multiplet \ in \ } ({\bf\bar{\n}}'_\alpha,{\bf\n}_\alpha,{\bf1}) \ , \quad
  \tilde\gamma_\alpha \  : \ {\rm Fermi \ multiplet \ in \ } ({\bf\bar{\n}'}_\alpha,{\bf\n}_{\alpha+1},{\bf1}) \ , \cr
  \eta_\alpha \ & : \ {\rm Fermi \ multiplet \ in \ } ({\bf1},{\bf\n}_\alpha,{\bf\bar{n}}_\alpha) \ , \quad
  \tilde\eta_\alpha \  : \ {\rm Fermi \ multiplet \ in \ } ({\bf\bar{\n}'}_\alpha,{\bf1},{\bf n}_{\alpha+1}) \ , \ 
\end{align}
under the $U(\n'_\alpha) \times U(\n_\alpha)\times U(K_\alpha)$ symmetry groups. These multiplets couple to the boundary conditions of the 1d bulk fields through the zero-dimensional superpotentials for the Fermi multiplets given by
\begin{eqnarray}\label{eq:boundary-superpotential2}
	&&E_{\gamma_\alpha} = \upsilon_\alpha B_\alpha' - B_\alpha\upsilon_\alpha \ , \quad E_{\tilde\gamma_\alpha} = \upsilon_{\alpha+1}A'_\alpha-A_\alpha\upsilon_\alpha \ , \quad E_{\eta_\alpha} = \upsilon_\alpha q'_\alpha-q_\alpha \ , \quad E_{\tilde{\eta}_\alpha} = \tilde{q}'_\alpha - \tilde{q}_\alpha\upsilon_\alpha \ , \nonumber \\
	&& J_{\gamma_\alpha} = \tilde{\upsilon}_\alpha A_\alpha - A'_{\alpha-1}\tilde{\upsilon}_{\alpha-1} \ , \quad J_{\tilde\gamma_\alpha} = -\tilde{\upsilon}_\alpha B_{\alpha+1} + B'_{\alpha}\tilde\upsilon_\alpha \ , \quad J_{\eta_\alpha} = \tilde{q}'_{\alpha-1}\tilde\upsilon_{\alpha-1} \ , \quad J_{\tilde{\eta}_\alpha} = \tilde\upsilon_\alpha q_{\alpha+1} \ . \qquad \quad
\end{eqnarray}
Here, the primed fields are the bulk fields on $t>0$.
A priori, these superpotentials break all supersymmetries since they do not obey the SUSY constraint $\sum_a E_a\cdot J_a = 0$. The non-zero terms in the constraint can be compensated by the superpotentials for the fermi multiplets $\Lambda_\alpha,\Lambda_\alpha'$ coming from the Dirichlet b.c. of the 1d bulk chiral multiplets, if we modify them as
\begin{eqnarray}
	&&J_{\Lambda_\alpha} = A_\alpha B_\alpha - B_{\alpha+1}A_\alpha + q_{\alpha+1}\tilde{q}_\alpha \ , \quad E_{\Lambda_\alpha} = \upsilon_\alpha \tilde{\upsilon}_\alpha \ , \nonumber \\
	&&J_{\Lambda'_\alpha} = A_\alpha' B_\alpha' - B_{\alpha+1}'A'_\alpha + q_{\alpha+1}'\tilde{q}'_\alpha \ , \quad E_{\Lambda'_\alpha} = -\tilde\upsilon_\alpha \upsilon_{\alpha+1} \ .
\end{eqnarray}
Then the full system with the interface  and the boundary conditions at $t=0$ preserves $\mathcal{N}=(0,2)$ supersymmetry.

We would like to remark that the extra degrees of freedom and the $J$-type superpotentials in~(\ref{eq:boundary-superpotential2}) are introduced by following the construction of the handsaw quiver varieties and Hecke correspondence in~\cite{Nakajima-handsaw}. However, the construction in~\cite{Nakajima-handsaw} does not tell us the $E$-type superpotentials in~(\ref{eq:boundary-superpotential2}). Without the $E$-type superpotentials, there would be extra $U(1)$ flavor symmetries acting on $\tilde\upsilon_\alpha$, which appear to be absent in the Hecke correspondence in~\cite{Nakajima-handsaw}. In order to remove these extra flavor symmetries, we turn on the $E$-type superpotentials as in~(\ref{eq:boundary-superpotential2}), which seems to be a unique choice for this purpose.

The Higgs branch of this system is parametrized by the 1d scalar fields $B_\alpha,A_\alpha,q_\alpha,\tilde{q}_\alpha$ and $B_\alpha,A_\alpha,q_\alpha,\tilde{q}_\alpha$ and the boundary scalar $\upsilon_\alpha$ satisfying the D-term constraints and the superpotential constraints given by
\begin{align}
	&A_\alpha B_\alpha - B_{\alpha+1}A_\alpha + q_{\alpha+1}\tilde{q}_\alpha =  A_\alpha' B_\alpha' - B_{\alpha+1}'A'_\alpha + q_{\alpha+1}'\tilde{q}'_\alpha = 0 \ , \cr
	&\upsilon_\alpha B'_\alpha - B_\alpha \upsilon_\alpha = \upsilon_{\alpha+1}A'_\alpha - A_\alpha \upsilon_\alpha = 0 \ , \quad \upsilon_\alpha q'_\alpha = q_\alpha \ , \quad \tilde{q}'_\alpha=\tilde{q}_\alpha\upsilon_\alpha \ ,
\end{align}
together with $\tilde\upsilon_\alpha = 0$. We claim that the Higgs branch of the handsaw quiver theory with the interface coincides with the moduli space of the vortices in the 3d triangular quiver theory interacting with the monopole operator.

The localized partition function of our system with the interface can be written as
\begin{equation}
	Z_{(\vec{\n}',\vec{k}')\times (\vec{\n},\vec{k})} = Z_{t>0}^{\rm 1-loop} \cdot Z^{\rm 1-loop}_{t=0} \cdot Z_{t<0}^{\rm 1-loop} \ ,
\end{equation}
where $ Z_{t>0}^{\rm 1-loop}$ and $ Z_{t<0}^{\rm 1-loop}$ are the contributions from the 1d bulk fields with the boundary conditions given by the inverse of equivariant weights at the fixed points $|\vec\n',k'\ra$ and $|\vec\n,k\ra$, respectively, 
\begin{equation}
	Z_{t>0}^{\rm 1-loop} = 1/  \omega_{\vec{n}',\vec{k}'} \ , \quad Z_{t<0}^{\rm 1-loop} = 1/\omega_{\vec{n},\vec{k}} \ .
\end{equation}
The matrix degrees of freedom at $t=0$ contributes to a factor of
\begin{eqnarray}
	Z^{\rm 1-loop}_{t=0} &=& 
	 \prod_{\alpha=1}^{L-1}\prod_{I}^{\n_\alpha}\prod_{J}^{\n'_\alpha}\frac{(\phi_{\alpha,I}-\phi_{\alpha,J}'+\epsilon)}
	{(\phi_{\alpha,I}-\phi_{\alpha,J}')} \cdot 
	\prod_{\alpha=1}^{L-2}\prod_{I=1}^{\n'_\alpha}\prod_{J=1}^{\n_{\alpha+1}}\frac{(\phi_{\alpha+1,J}-\phi'_{\alpha,I})}{(\phi_{\alpha+1,J}-\phi'_{\alpha,I}\!+\!\epsilon)}
	  \nonumber \\
	&& \times \prod_{\alpha=1}^{L-1}\left[\prod_{I=1}^{\n_\alpha}\prod_{a=K_{\alpha\!-\!1}+1}^{K_\alpha}(\phi_{\alpha,I}\!+\!m_{i_{\alpha,a}}\!+\!\frac{\epsilon}{2}) \cdot \prod_{I=1}^{\n'_\alpha}\prod_{a=K_\alpha+1}^{K_{\alpha\!+\!1}}(-m_{i_{\alpha,a}}\!-\!\phi'_{\alpha,I}\!+\!\frac{\epsilon}{2})\right]\ ,
\end{eqnarray}
where $\phi_{\alpha,J}$ and $\phi_{\alpha,J}'$ takes the values at the fixed points $|\vec\n',k'\ra$ and $|\vec\n,k\ra$ respectively. We simplify further and obtain
\begin{equation}
	Z^{\rm 1-loop}_{t=0}=  \prod_{\alpha=1}^{L-1} \prod_{a=1}^{K_\alpha}
	\frac
	{\prod_{s=1}^{k_{\alpha,a}} \prod_{b=1}^{K_{\alpha}}(m_{i_{\alpha,b}}\!-\!m_{i_{\alpha,a}}\!+\!(k'_{\alpha,b}\!-\!s\!+\!1)\epsilon)\prod_{s=1}^{k'_{\alpha,a}} \prod_{b=K_\alpha+1}^{K_{\alpha\!+\!1}}(m_{i_{\alpha\!+\!1,b}}\!-\!m_{i_{\alpha,a}}\!-\!s\epsilon)}
	{\prod_{s=1}^{k_{\alpha,a}} \prod_{b=1}^{K_{\alpha\!-\!1}}(m_{i_{\alpha\!-\!1,b}}\!-\!m_{i_{\alpha,a}}\!+\!(k'_{\alpha\!-\!1,b}\!-\!s\!+\!1)\epsilon)} \ .
\end{equation}
As we discussed in the previous section, the partition function vanishes unless $\vec{k}\subset \vec{k}'$.

In the case $\vec\n' = \vec\n$, our interface defines an identity interface. So the partition function should give the norm of the vortex state $|\vec{\n},\vec{k}\ra$. Indeed,
\be
	\la\vec{\n},\vec{k}'|\vec{\n},\vec{k}\ra = Z_{(\vec{\n},\vec{k}')\times(\vec{\n},\vec{k})} =  \delta_{\vec{k}',\vec{k}} /\omega_{\vec\n,\vec{k}} \ ,
\ee
which reproduces the correct normalization for our vortex states.

On the other hand, when $\vec{\n}' = \vec{\n} +\bm\delta_\alpha$ and $\vec{k}' = \vec{k} +\bm\delta_{\alpha,a}$, the partition function will compute the correlation functions of monopole operators $v^\pm_{\alpha,a}$ such as
\begin{equation}
	\la \vec{\n}+\bm\delta_\alpha,\vec{k}+\bm\delta_{\alpha,a} | v^+_{\alpha,a}| \vec{\n},\vec{k}\ra = \la \vec{\n},\vec{k} | v^-_{\alpha,a}| \vec{\n}+\bm\delta_\alpha,\vec{k}+\bm\delta_{\alpha,a}\ra =  Z_{(\vec{\n}+\bm\delta_{\alpha},\vec{k}+\bm\delta_{\alpha,a})\times(\vec\n,\vec{k})} \ .
\end{equation}

Using the correlation functions, one can easily compute the action of the monopole operators on the vortex states. We find that the monopole operators act by
\begin{align}
	v_{\alpha,a}^+|\vec\n , \vec{k} \rangle &= \frac{\langle \vec\n + \vec\delta_{\alpha},\vec{k}+\vec{\delta}_{\alpha,a} \,|v_{\alpha,a}^+|\vec\n,\vec{k}\rangle}{\langle \vec\n + \vec\delta_{\alpha},\vec{k}+\vec{\delta}_{\alpha,a} |\vec\n + \vec\delta_{\alpha},\vec{k}+\vec{\delta}_{\alpha,a}\rangle} |\vec\n + \vec\delta_{\alpha},\vec{k}+\vec{\delta}_{\alpha,a}\rangle \cr
	&=\frac{Q_{\alpha+1}(\varphi_{\alpha,a})}{\prod_{b\neq a}(\varphi_{\alpha,a}-\varphi_{\alpha,b})}|\vec\n + \vec\delta_{\alpha},\vec{k}+\vec{\delta}_{\alpha,a}\rangle \ , \cr
	v_{\alpha,a}^-|\vec\n, \vec{k} \rangle &= \frac{\langle \vec\n ,\vec{k} |v_{\alpha,a}^-|\vec\n - \vec\delta_{\alpha}, \vec{k} - \vec\delta_{\alpha,a}\rangle}{\langle \vec\n - \vec\delta_{\alpha}, \vec{k} - \vec\delta_{\alpha,a} |\vec\n - \vec\delta_{\alpha}, \vec{k} - \vec\delta_{\alpha,a}\rangle} |\vec\n - \vec\delta_{\alpha}, \vec{k} - \vec\delta_{\alpha,a}\rangle \cr
	&=\frac{Q_{\alpha-1}(\varphi_{\alpha,a})}{\prod_{b\neq a}(\varphi_{\alpha,a}-\varphi_{\alpha,b})}|\vec\n - \vec\delta_{\alpha}, \vec{k} - \vec\delta_{\alpha,a}\rangle \ ,
\end{align}
and $v^-_{\alpha,a}|\vec\n,\vec{k}\ra=0$ if $k_{\alpha,a}=0$. Therefore the monopole operators satisfy the following relations
\begin{equation}
	v^+_{\alpha,a}v^-_{\alpha,a} = \frac{Q_{\alpha+1}(\varphi_{\alpha,a})Q_{\alpha-1}(\varphi_{\alpha,a}\!+\!\epsilon)}{\prod_{b\neq a}(\varphi_{\alpha,a}\!-\!\varphi_{\alpha,b})(\varphi_{\alpha,a}\!-\!\varphi_{\alpha,b}\!-\!\epsilon)} \ , \quad v^-_{\alpha,a}v^+_{\alpha,a} = \frac{Q_{\alpha+1}(\varphi_{\alpha,a}\!-\!\epsilon)Q_{\alpha-1}(\varphi_{\alpha,a})}{ \prod_{b\neq a}(\varphi_{\alpha,a}\!-\!\varphi_{\alpha,b})(\varphi_{\alpha,a}\!-\!\varphi_{\alpha,b}\!+\!\epsilon)} \ .
\end{equation}

\subsection{Equivalence to vortex moduli space}
\label{sec:handsaw-equiv}

In \cite[Section 3]{Nakajima-handsaw} Nakajima explicitly described the equivalence of the moduli space of handsaw quivers and the moduli space of vortices in a triangular quiver gauge theory. We briefly sketch how his argument works for $\n$ vortices in $U(K)$ gauge theory with $N \geq K$ hypermultiplets.

Recall that for SQCD the Higgs branch is the cotangent bundle to the Grassmannian of $K$-planes in $\C^N$. We  choose the vacuum $\nu$ corresponding to the standard inclusion $\C^K \hookrightarrow \C^N$, \ie\ the subset $\{1,...,K\}\subset \{1,...,N\}$.
As explained earlier, the supersymmetric quantum mechanics describing vortices consists of a $U(\n)$ vectormultiplet, together with an adjoint chiral multiplet, $K$ chiral multiplets in the fundamental representation of $U(\n)$ and $(N-K)$ chiral multiplets in the anti-fundamental representation of $U(\n)$. This is a handsaw quiver with a single gauge node.

The bosonic components of the chiral multiplets can be represented as complex matrices
\begin{itemize}
\item A $\n \times \n$ matrix $B$
\item A $\n \times K$ matrix $q$
\item A $(N-K) \times \n$ matrix $\tilde q$\, .
\end{itemize}
and an element $g \in GL(\n)$ of the complexified gauge group acts by
\be
B \to g B g^{-1} \qquad q \to g q \qquad \tilde q \to \tilde q g^{-1}.
\ee
In this case the $F$-term equation is vacuous and the real moment map equation coming from the $D$-term equations can be replaced with the following stability condition: a tuple $(B, q, \tilde{q})$ is stable if there is no proper $B$-stable subspace of $\C^{\n}$ containing the image of $q$. Then the moduli space is equivalent to
\be
\{ (B, q, \tilde{q}) \text{ stable} \} / GL(\n).
\ee

Recall that in Section \ref{sec:stack} a point in the vortex moduli space $\CM_\nu^\n$ is defined to be a map from $\cp^1$ to the Higgs stack $[\CM_H]$ of degree $\n$ sending $\infty$ to $\nu$. More concretely, this is just a rank $K$ vector bundle $E$ with%
\footnote{A direct comparison with Section \ref{sec:Hilbert} requires replacing $\n\to -\n$. This is simply a matter of convention.} %
$c_1(E) = -\n$ trivialized at $\infty$ together with an inclusion of sheaves $X: E \hookrightarrow \C^N \otimes_{\C} \CO_{\cp^1}$ such that $X|_{{\infty}}$ is the standard inclusion $\C^K \hookrightarrow \C^N$. Recall that an inclusion of locally free sheaves may only fail to be injective on a finite number of fibers. To recover the moduli matrix write $X$ in terms of the inhomogeneous coordinate $z$ on $\cp^1$. The gauge is fixed by the condition on $X|_{\infty}$.

From the matrices $(B, q, \tilde q)$ we can define a diagram
\be
\begin{array}{ccc}
 \C^\n \otimes_{C} \CO_{\cp^1}(1) \\
 \overset{\alpha}{}\uparrow \\
  (\C^K \oplus \C^\n) \otimes_{\C} \CO_{\cp^1} & \overset{\beta}{\longrightarrow} & (\C^K \oplus \C^{N-K}) \otimes_{\C} \CO_{\cp^1}
\end{array}
\ee
where $\alpha = \begin{bmatrix} q &\; z\!-\!B \end{bmatrix}$ and $\beta = 1 \oplus \tilde{q}$. By the stability condition we have $\text{im }\alpha|_p = \C^\n$ for all $p \in \cp^1$. Thus $E = \ker \alpha$ is a rank $K$ vector bundle with $c_1(E) = -\n$. 

The map $\beta$ induces a map $X: E \to \C^N \otimes_{\C} \CO_{\cp^1}$. To see that $X$ has the right behavior at infinity notice that $\beta|_{\infty} = 1 \oplus \tilde{q}$ and that $\alpha|_{\infty} = \begin{bmatrix} 0 & 1 \end{bmatrix}$ so $E|_{\infty} = \C^K \oplus 0$. The map $X$ is an injection of sheaves because
\be
\text{ker } X|_p \cong \text{ker } \begin{bmatrix} z(p) - B \\ \tilde{q} \end{bmatrix}
\ee
so $X|_p$ can only fail to be injective when $z(p)$ is one of the finitely many eigenvalues of $B$.

Now let us consider two supersymmetric quantum mechanics with $\n' = \n+1$ and an interface between them such that
\be
\upsilon B' = B \upsilon \qquad \upsilon q' = q \qquad \tilde{q}' = \tilde{q} \upsilon.
\ee
for some $\upsilon: \C^{\n'} \to \C^{\n}$. As a consequence of the stability condition corresponding to the $D$-terms the map $\upsilon$ is surjective. The first two equations tell us that the diagram
\be
\begin{array}{ccc}
  \C^{\n'} \otimes_{\C} \CO_{\cp^1}(1) & \overset{\upsilon}{\longrightarrow} &  \C^{\n} \otimes_{\C} \CO_{\cp^1}(1)\\
  \overset{\alpha'}{}\uparrow && \overset{\alpha}{}\uparrow\\
   (\C^K \oplus \C^{\n'}) \otimes_{\C} \CO_{\cp^1} & \overset{1\oplus \upsilon}{\longrightarrow} & (\C^K \oplus \C^{\n}) \otimes_{\C} \CO_{\cp^1}
\end{array}
\ee
commutes. Thus we have an induced map $i: E' \hookrightarrow E$. The third equations tells us that $X' = X \circ i$. The quotient $E/E'$ is supported at a single point which is the position of the vortex created by the interface.

\section{Case study: abelian quiver}
\label{sec:ab-quiv}

\begin{figure}[htp]
\centering
\includegraphics[height=2.5cm]{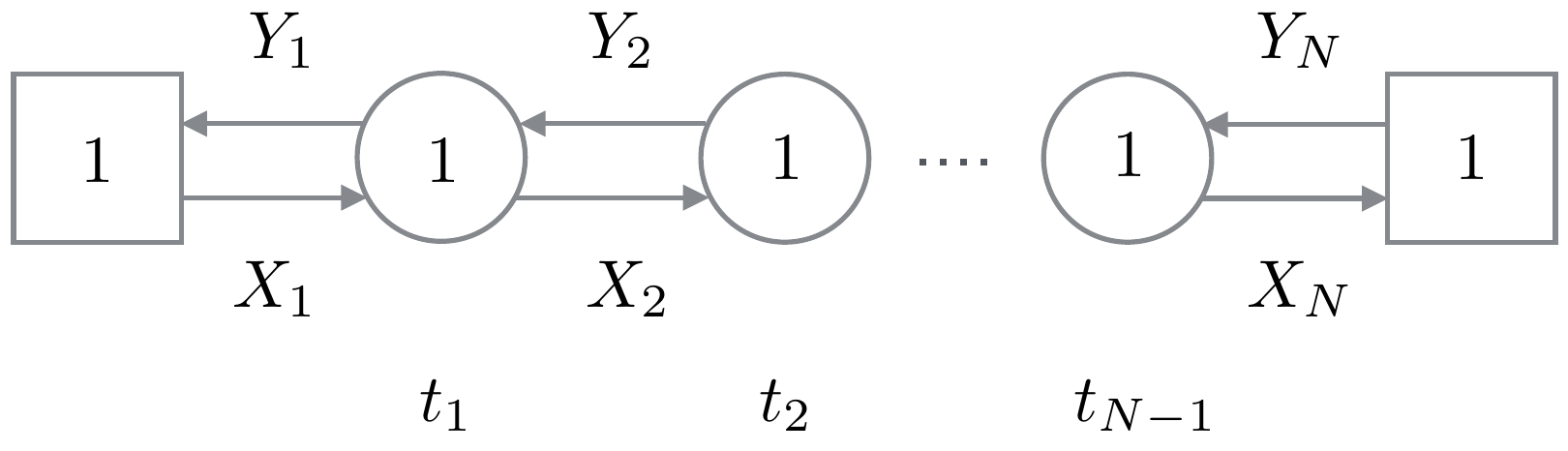}
\label{fig-ab-quiv}
\caption{Quiver for the mirror of SQED with $N$ hypermultiplets.}
\end{figure}

In this section, we consider the simplest example of our constructions that has \emph{not} already appeared in the mathematics literature: the abelian quiver gauge theory shown in Figure~\ref{fig-ab-quiv}. This is the 3d mirror of SQED.

Refering to Figure~\ref{fig-ab-quiv}, we see that the gauge group is $U(1)^{N-1}$. We denote the ``bifundamental'' hypermultiplet fields by $(X_j,Y_j)$ with $j=1,\ldots,N$. The moment map constraints for the gauge symmetry are
\bea
z_j - z_{j+1} & = 0 \qquad && z_j := X_j Y_j \\
 Z_j -Z_{j+1} & = -t_j \qquad && Z_j := |X_j|^2-|Y_j|^2
\eea
with $j = 1,\ldots, N-1$ and $t_j = t_{\mathbb{R},j}$ is real FI parameter at the $j$-th node of the quiver. We will assume that $t_j <0$ for all nodes of the quiver. The Higgs branch is a resolution of the singularity $\C^2/\mathbb{Z}_N$.

\begin{figure}[htp]
\centering
\includegraphics[height=2.5cm]{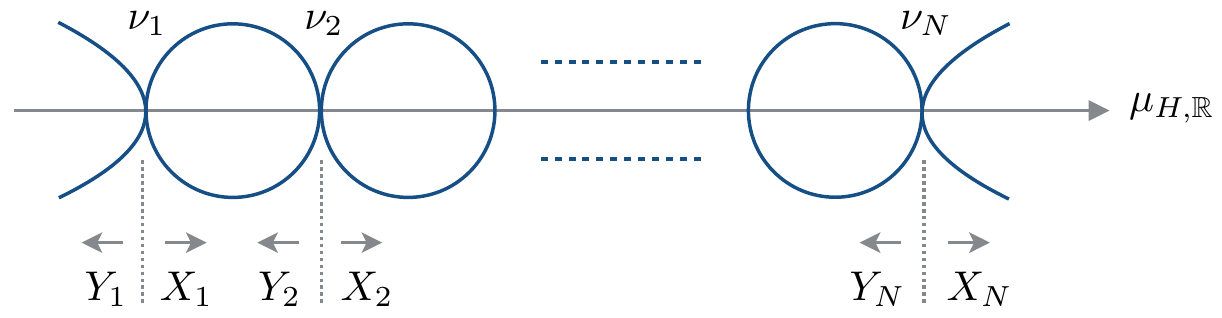}
\caption{The Higgs branch of abelian quiver on the slice $\mu_{H,\C}=0$.}
\label{fig-ab-quiv-hb}
\end{figure}

The moment maps for the flavor symmetry $G_H = U(1)$ are given by $\mu_{H,\C} = z_1+z_N$ and $\mu_{H,\R} = Z_1+Z_N$. Solving for $z_j$ and $Z_j$ in terms of the flavor moment maps and FI parameters, we find
\be
z_j = \frac{1}{2}\mu_{H,\C} \qquad
Z_j =\frac{1}{2}\left( \mu_{H,\R} + \sum_{n=1}^{j-1} t_n - \sum_{n=j}^{N-1} t_n \right) \, .
\ee
The Higgs branch can be described as a circle fibration over $\mathbb{R}^3$ with the base parametrized by the moment maps $\mu_{H,\C}$ and $\mu_{H,\R}$ and fibers rotated by $G_H = U(1)$. The circle fibers degenerate when $z_i = 0$ and $Z_i = 0$; these are fixed points of the flavor symmetry, \ie\ locations of the $N$ massive vacua $\{ \nu_i\}$ in the presence of a generic mass deformation. The slice $\mu_{\C,H}$ is illustrated in Figure~\ref{fig-ab-quiv-hb}.  With our convention $t_j <0$, the positions of these vacua on the slice $\mu_{H,\C} = 0$ are ordered such that $\mu_{H,\R}(\nu_i) < \mu_{H,\R}(\nu_{i+1})$. In each vacuum,
\be
\nu_i \quad : \quad \begin{cases} Y_{j} = 0 & j = 0,\ldots, i  \\
X_j = 0 & j = i ,\ldots, N \end{cases} \qquad .
\ee 
Furthermore, it is straightforward to check that, on the slice $\mu_{H,\C}=0$,
\begin{itemize}
\item $X_j = 0$ \quad for \quad  $\mu_{H,\R} \leq \mu_{H,\R}(\nu_j)$
\item $Y_j = 0$ \quad for \quad  $\mu_{H,\R} \geq \mu_{H,\R}(\nu_j)$ \, .
\end{itemize}

Generalized vortex solutions are characterized by a degree vector $\n = (\n_1,\ldots,\n_{N-1})$. With a supersymmetric vacuum $\nu_i$ at infinity, there are in general two nontrivial types of solutions, or ``chambers'' in the moduli space. We must always have $\n_1<\ldots<\n_{i-1}$ and $\n_i> \ldots >\n_{N-1}$. The two chambers are distinguished by the relative size of $\n_{i-1}$ and $\n_{i}$. In the case $\n_{i-1} < \n_i$ the nonvanishing holomorphic fields are
\bea
& X_j(z) \quad \text{monic of degree} \quad \n_{j} - \n_{j-1} \quad \mathrm{for} \quad  j = 1,\ldots,i-1 \\
& X_i(z) \quad \text{of degree} \quad \n_{i} - \n_{i-1} \\
& Y_j(z) \quad \text{monic of degree} \quad \n_{j-1} - \n_{j} \quad \mathrm{for} \quad  j = i+1,\ldots,N \\
\eea
where we define for convenience $\n_0 =\n_{N}=0$. The case with $\n_{i-1}>\n_i$ is similar except that $Y_i(z)$ (rather than $X_i(z)$) is now turned on and of degree $\n_{i-1}-\n_i$. Here we concentrate on the first chamber. In what follows we introduce the notation $\n_{i,j}:=\n_i-\n_j$.

For each fixed vortex number $\n$, there is a unique equivariant fixed point in the moduli space $\CM_\nu^\n$, leading to a unique state $|\n\rangle \in \CH_\nu^\n = H^*(\CM_\nu^\n)$. (This is a general property of abelian theories.) For $\n = (\n_1,\ldots,\n_{N-1})$, the fixed point is given by
\bea
X_j(z) & = z^{\n_{j,j+1}} \quad j = 1,\ldots,i-1 \\
X_i (z) & = 0 \\
Y_j(z) & =  z^{\n_{j-1,j}} \quad  j = i+1,\ldots,N-1
\eea
with 
\be
\varphi_j = \begin{cases} 
m-\n_j\ep-\tfrac{j \ep}{2} & \quad \mathrm{if}\quad  j=0,\ldots,i-1 \\
-m-\n_j \ep - \tfrac{(N-j)\ep}{2} & \quad \mathrm{if} \quad j = i,\ldots, N 
\end{cases}
\ee
where $m$ is the mass parameter corresponding to the $G_H =U(1)$ flavor symmetry. Note that it is convenient to introduce the notation $\varphi_0 = m$ and $\varphi_N = - m$. The corresponding equivariant weight is
\bea
w_\n & = \prod_{j=1}^{i-1} \prod_{\ell=0}^{\n_{j,j-1}-1} (\ell-\n_{j,j-1})\ep  \\
\times & \prod_{\ell = 0}^{\n_{i,i-1}-1} \left[ -2m+(\ell-\n_{i,i-1})\ep -\frac{1}{2}(N-2i+1) \ep \right] \\
\times & \prod_{j=i+1}^{N} \prod_{\ell=0}^{\n_{j-1,j}-1} (\ell+\n_{j,j-1})\ep \, .
\label{linear-weights}
\eea

As in Section~\ref{sec:SQED-mono}, we can model monopole operators as singular gauge transformations. For example, the monopole operator $u_j^+$ at the $j$-th node acts by
\begin{alignat}{2}
X_j(z) & \to z X_j(z)  \qquad &X_{j+1}(z) \to z^{-1} X_{j+1}(z) \\
Y_j(z) & \to z^{-1} Y_j(z)  \qquad &Y_{j+1}(z) \to z Y_{j+1}(z)
\end{alignat}
leaving the other polynomials unchanged.

For the monopole operator $u^+_j$ we find an action on equivariant cohomology
\be
u^+_j| \, \n \, \ra = | \, \n +\delta_j \, \ra  \begin{cases} (-\n_{j,j-1}-1)\ep &  j=1,\ldots,i-1 \\
(-\n_{i,i+1}-1)\ep  (-2m-(\n_{i,i-1}+1)\ep -\frac{1}{2}(N-2i+1) \ep)  & j = i \\
(-\n_{j,j+1}-1) \ep & j=i+1,\ldots, N
\end{cases}
\label{abelian-Vplus}
\ee
and for the monopole operator $u^-_j$
\be
u^-_j| \, \n \, \ra = | \, \n -\delta_j \, \ra  \begin{cases} (-\n_{j+1,j}-1)\ep &  j=1,\ldots,i-2 \\
-2m+(-\n_{i,i-1}-1)\ep -\frac{1}{2}(N-2i+1) \ep  & j = i-1 \\
1 & j = i \\
 (-\n_{j-1,j}-1) \ep & j=i+1,\ldots, N-1\, .
\end{cases}
\label{abelian-Vminus}
\ee 

We can now check the commutator algebra generated by these operators. The commutators of $u^\pm_j$ with $\varphi_k$ are straightforward: they depend only on the weights of $\varphi_j$ acting on a state $| \, \n\, \ra$ being linear in $-\n_j\ep$. From this we find
\be
[ \, u^\pm_j \, , \, \varphi_k \, ] = \pm \ep \, \delta_{jk} \, u^\pm_j
\ee
In order to write the remaining commutation relations, it is convenient to introduce the notation
\be
h_j = 2 \varphi_j-\varphi_{j-1}-\varphi_{j+1} + c_j \ep
\ee
where $c_j$ are some constants. We will need to choose non-vanishing constants $c_{i-1}=-c_i=\frac{1}{2}$. With this notation we find that the algebra has relations
\be
[ \, u^\pm_j \, , \, h_k \, ] = \pm \ep \, A_{jk} \, u^\pm_k \qquad [ u^+_j , u^+_k ] = \begin{cases} - \ep \delta_{jk} h_j & j=1,\ldots, i-1 \\ + \ep \delta_{jk} h_j & j = i,\ldots, N-1
\end{cases}
\ee
where $A_{jk}$ is the Cartan matrix of $\mathfrak{sl}_N$. The entire Coulomb-branch algebra is a central quotient of $U(\mathfrak{sl}_N)$, with (in particular) the Casimir elements fixed to be certain polynomials in the mass $m$. The very same algebra was computed by more abstract methods in \cite[Section 6.6.2]{BDG-Coulomb}; it is a quantization of the Coulomb branch $\CM_C$, which is a minimal nilpotent orbit in~$\mathfrak{sl}_N$.

\subsection{Vortex quantum mechanics}
\begin{figure}[htp]
\centering
\includegraphics[height=4cm]{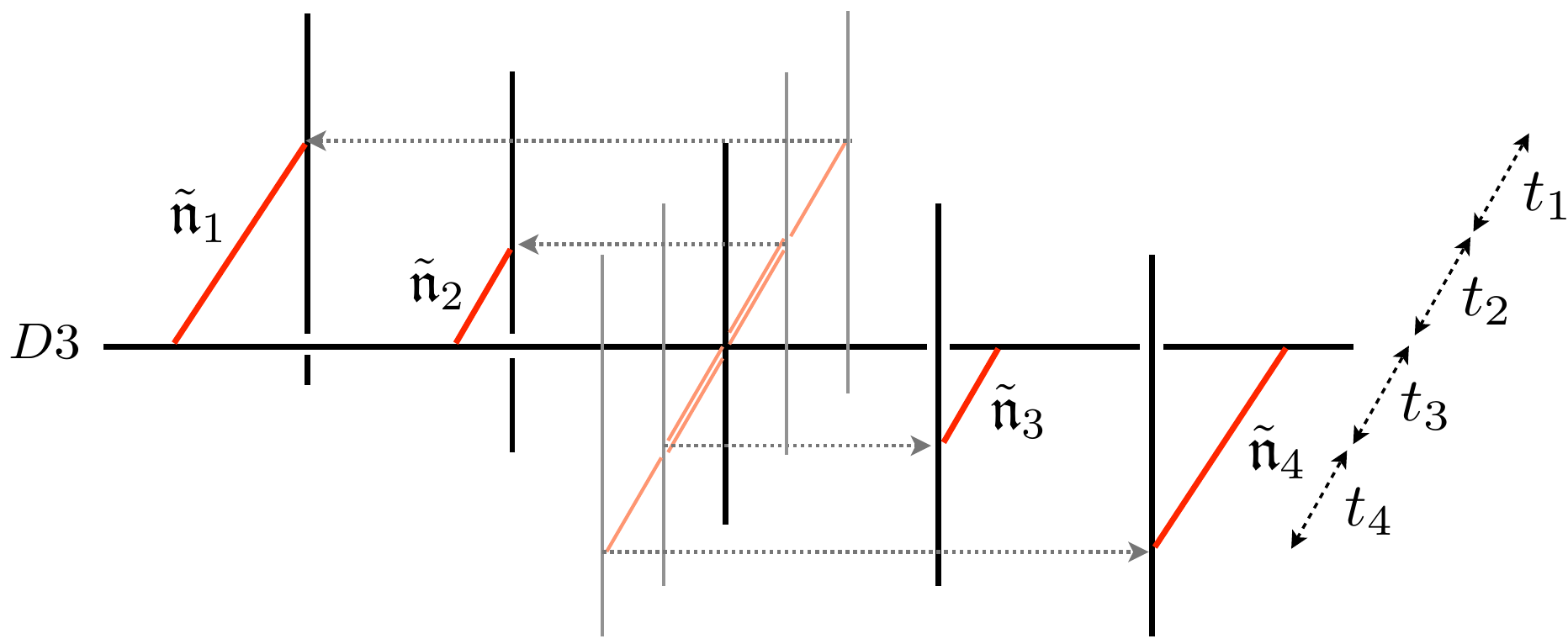}
\caption{The brane construction of vortices in the $A_N$ linear quiver theory at $i$-th vacuum. The vortices are the $\tilde{\n}_{j}$ D1-branes (red lines). This figure illustrates the vortices in the $A_4$ quiver theory at the 3-rd vacuum.}
\label{fig:vortexlinear}
\end{figure}

We now reproduce the same result using interfaces for vortex quiver quantum mechanics, as in Section~\ref{sec-adhm}. We first note that the theory admits a simple brane construction with $N+1$ NS5-branes and an infinite D3-brane which goes across the NS5-branes. The system with finite real FI parameters $t_i$ can be illustrated by the brane configuration in Figure \ref{fig:vortexlinear}.

The theory has $N$ massive vacua $\{\nu_i\}$. The $i$-th vacuum corresponds to the configuration where the D3-brane touches the $i$-th NS5-brane. Vortex particles in the gauge theory are the D1-branes suspended between the D3-brane and one of the NS5-branes. The number of D1-branes $\tilde{\n}_j$ is related to the vortex number $\n_j$ of the $j$-th gauge node as $\n_j = \sum_{k=1}^j\tilde{\n}_k$ for $j<i$ and $\n_j=\sum_{k=j}^{N}\tilde{\n}_k$ for $j\ge i$.

\begin{figure}[htp]
\centering
\includegraphics[height=4cm]{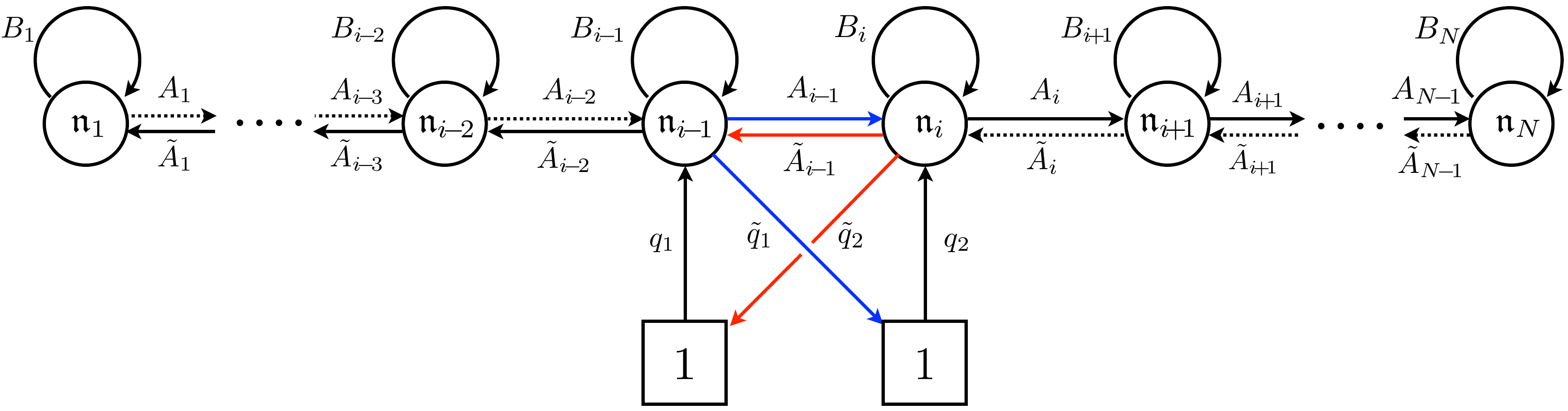}
\caption{The 1d vortex quantum mechanics in the $A_N$ linear quiver theory at $i$-th vacuum. The solid arrows and the red arrows represent the chiral multiplets whose the scalar fields parametrize the Higgs branch at the first chamber when $\n_{i\!-\!1}<\n_i$, whereas the solid arrows and the blue arrows represent those at the second chamber when $\n_{i\!-\!1}>\n_i$.}
\label{fig:linearQM}
\end{figure}

The vortex quantum mechanics describing the moduli space of the vortices can be read off from the brane configuration. The low energy theory living on the D1-branes is the 1d $\mathcal{N}=(2,2)$ linear quiver gauge theory with gauge group $\otimes_{i=1}^N U(\n_i)$, with $\n_1<\cdots<\n_{i-1}$ and $\n_i>\cdots>\n_N$, given in Figure~\ref{fig:linearQM}. Each quiver node has an adjoint chiral multiplet $B_j$ and two adjacent nodes, say the $j$-th and $j+1$-th nodes, are connected by two bi-fundamental chiral multiplets $A_j$ and $\tilde{A}_j$. In addition, at  the $i$-th vacuum, the $i\!-\!1$-th node couples to a fundamental chiral multiplet $q_1$ and an anti-fundamental chiral multiplet $\tilde{q}_1$, and $i$-th node has  a fundamental $q_2$ and an anti-fundamental $\tilde{q}_2$ chiral multiplets.
This theory has superpotentials of the form
\begin{align}
	\mathcal{W}_j &= {\rm Tr}\, \tilde{A}_j\left( A_jB_j - B_{j+1} A_j \right) \quad {\rm for} \quad j\neq i-1 \ ,\cr
	\mathcal{W}_{i-1} &= {\rm Tr}\, \tilde{A}_{i-1}\left( A_{i-1}B_{i-1} - B_k A_{i-1} + q_2\tilde{q}_1 \right) + A_{i-1}q_1\tilde{q}_2 \ .
\end{align}
We propose that the Higgs branch of this 1d quantum mechanics agrees with the moduli space of vortices in the 3d $A_N$ abelian quiver theory.

The 1d gauge theory has $U(1)_1\times U(1)_2 \times U(1)_\epsilon$ flavor symmetry. $q_1$ and $\tilde{q}_1$ have $U(1)_1$ charge $\pm1$, respectively and $q_2$ and $\tilde{q}_2$ have $U(1)_2$ charge $\pm1$, respectively. For $U(1)_\epsilon$ symmetry, $B_i,q_1,q_2$ have $+1,+\tfrac{1}{2},+\tfrac{1}{2}$ charges of the $U(1)_\epsilon$ symmetry and $\tilde{q}_1,\tilde{q}_2$ are singlets. We can identify the off-diagonal rotation of $U(1)_1\times U(1)_2$ with $G_H=U(1)$ flavor symmetry of the 3d abelian quiver theory.

Recall that the vortex moduli space has in general two chambers characterized by the relative size of $\n_{i-1}$ and $\n_i$ at the $i$-th vacuum, i.e. $\n_{i-1}<\n_i$ or $\n_{i-1}>\n_i$. Accordingly, we have two distinguished moduli spaces of the Higgs branch in the vortex quantum mechanics depending on the relative size of the ranks of two gauge groups at the $i\!-\!1$-th and $i$-th nodes.

For now we assume the FI parameters, which are proportional to the 3d gauge couplings, to be positive for all gauge nodes. It restricts us to the Higgs branch vacua.
The Higgs branch is defined as follows. We can first solve the F-term conditions $\tilde{A}_jA_j = A_j\tilde{A}_j=0$ by setting either $A_j$ or $\tilde{A}_j$ to zero. With the positive FI parameters, it turns out that the scalar fields $A_j$ for $j\ge i$ and $\tilde{A}_j$ for $j<i-1$ should be non-vanishing. In particular, we need to be more careful with the fields $A_{i-1}$ and $\tilde{A}_{i-1}$. When $A_{i-1}=0$, the F-term condition $q_2\tilde{q}_1=0$ requires $\tilde{q}_1=0$ since $q_2$ must be non-zero. The Higgs branch in this case is therefore the space of solutions to
\begin{align}\label{linear-quiver-F-constraints}
	&B_i\tilde{A}_j - \tilde{A}_j B_{j+1} = 0 \quad {\rm for} \quad j=1,\cdots,i\!-\!2 \ ,\cr 
	&A_j B_j - B_{j+1} A_j = 0 \quad {\rm for} \quad j=i,\cdots, N\!-\!1 \ , \cr
	& B_{i-1} \tilde{A}_{i-1} - \tilde{A}_{i-1}B_i +q_1\tilde{q}_2 = 0 \ , \quad \tilde{A}_{i-1}q_2 = 0 
\end{align}
modulo the gauge transformations. We claim that this class of the Higgs branch vacua with $\n_{i-1}<\n_i$ describes the moduli space of vortices in the first chamber in the 3d theory. This Higgs branch has complex dimension
\begin{equation}
	{\rm dim}_{\mathbb{C}} \mathcal{M}_H = 2\n_i \ ,
\end{equation}
which agrees with the dimension of the vortex moduli space in the first chamber at $\n_{i-1}<\n_i$.

On the other hand if $\tilde{A}_{i-1}=0$, the F-term condition $q_1\tilde{q}_2=0$ requires that $\tilde{q}_2=0$.
The Higgs branch is then the solution space of
\begin{align}\label{linear-quiver-F-constraints-2}
	&B_j\tilde{A}_j - \tilde{A}_j B_{j+1} = 0 \quad {\rm for} \quad j=1,\cdots,i\!-\!2 \ ,\cr 
	&A_j B_j - B_{j+1} A_i = 0 \quad {\rm for} \quad j=i,\cdots, N\!-\!1 \ , \cr
	& A_{i-1}B_{i-1} - B_kA_{i-1} +q_2\tilde{q}_1 = 0 \ , \quad A_{i-1}q_1 = 0 
\end{align}
divided by the gauge transformations. We again claim that this class of the Higgs branch vacua with $\n_{i-1}>\n_i$ coincides with the vortex moduli space in the second chamber. The dimension of this space is
\begin{equation}
	{\rm dim}_{\mathbb{C}} \mathcal{M}_H = 2\n_{i-1} \ ,
\end{equation}
which also agrees with the dimension of the vortex moduli space in the second chamber at $\n_{i-1}>\n_i$.

Let us now focus on the first chamber with $\n_{i-1}<\n_i$. Turning on twisted masses $m_1,m_2,\epsilon$ for the flavor symmetry $U(1)_1\times U(1)_2\times U(1)_\epsilon$, the quantum mechanics has an isolated fixed point of the flavor symmetry where the diagonal elements of the complex scalars in the vectormultiplets take the values
\begin{equation}\label{phi-fp-abelian}
	\phi_{j,I} =  \left\{ \begin{array}{ll} -m_1 -I\epsilon+\frac{i-j}{2}\epsilon & \quad {\rm if} \quad j =1,\cdots, i-1  \\
	 -m_2 - I\epsilon-\frac{i-j-1}{2}\epsilon  & \quad {\rm if} \quad j = i\cdots, N \end{array} \right. \ .
\end{equation}
Each supersymmetric vacuum defines a vortex state $|\vec\n\ra$ in the Hilbert space of the 3d abelian quiver theory in an $\Omega$-background.

The equivariant weight of the tangent space to the fixed point can be obtained using equivariant index theorem. Its inverse can be expressed as the residue of
\begin{eqnarray}
	&&\frac{\prod_{I\neq J}^{\n_j}(\phi_{j,I}\!-\!\phi_{j,J})}{\prod_{I, J}^{\n_j}(\phi_{j,I}\!-\!\phi_{j,J}\!+\!\epsilon)}\cdot
	 \prod_{j=1}^{i-1} \prod_{I=1}^{\n_j}\prod_{J=1}^{\n_{j+1}} \frac{(\phi_{j+1,J}\!-\!\phi_{j,I}\!-\!\frac{\epsilon}{2})}{(\phi_{j,I}\!-\!\phi_{j+1,J}\!-\!\frac{\epsilon}{2})} \quad \\
	&& \times \prod_{j=i}^{N-1} \prod_{I=1}^{\n_j}\prod_{J=1}^{\n_{j+1}}\frac{(\phi_{j,I}\!-\!\phi_{j+1,J}\!-\!\frac{\epsilon}{2})}{(\phi_{j+1,J}\!-\!\phi_{j,I}\!-\!\frac{\epsilon}{2})}\cdot
     \prod_{I=1}^{\n_{i\!-\!1}}\frac{(-\!\phi_{i\!-\!1,I}\!-\!m_2)}{(\phi_{i\!-\!1,I}\!+\!m_1\!+\!\frac{\epsilon}{2})}
     \cdot\prod_{I=1}^{\n_i}\frac{1}{(\phi_{i,I}\!+\!m_2\!+\!\frac{\epsilon}{2})(\!-\!\phi_{i,I}\!-\!m_1)} \ , \quad \nonumber
\end{eqnarray}
at supersymmetric vacuum of \eqref{phi-fp-abelian}. Plugging \eqref{phi-fp-abelian} into this formula, we find the equivariant weight
\begin{equation}
	\omega_{\vec\n} = \prod_{j=1}^{i-1}\prod_{\ell=1}^{\n_j-\n_{j\!-\!1}}\ell\epsilon \cdot \prod_{\ell=1}^{\n_{i}-\n_{i\!-\!1}}(m_{12}-(\ell-\tfrac{1}{2})\epsilon) \cdot \prod_{j=i+1}^N\prod_{\ell=1}^{\n_{j\!-\!1}-\n_{j}}\ell\epsilon \ .
	\label{abelian-vortex-partition-function}
\end{equation}
If we identify the flavor mass parameter as $m_{12} = -2m -\frac{N-2i+2}{2}\epsilon$, we see that the quantum mechanics result perfectly agrees with the equivariant weight of the fixed point in the first chamber computed using the moduli matrix approach above in~(\ref{linear-weights}).

\subsubsection{Interface}
We can engineer the monopole operators in the 1d vortex quantum mechanics, as in Section \ref{sec-adhm}, by coupling the 1d boundary conditions introduced above to extra degrees of freedom localized at an interface at $t=0$. The interface interpolates between a pair of vortex quantum mechanics with $\otimes_{i=1}^N U(\n_i)$ and $\otimes_{i=1}^N U(\n'_i)$ gauge groups living on half-lines $t<0$ and $t>0$, respectively.
In the following we will consider vortices on the $i$-th vacuum with $\n_{i-1}<\n_i$.

Let us first discuss boundary conditions at $t=0$.
Boundary conditions should be chosen to be consistent with the Higgs vacuum. So we choose Neumann-type boundary conditions for the chiral fields that belong to the Higgs branch, whereas we choose Dirichlet-type boundary conditions for all other chiral multiplets vanishing in the Higgs branch. The vector multiplets will have Neumann-type boundary conditions.

For our case, the chiral multiplets with scalar fields $B_j,q_1,q_2,\tilde{q}_1$ and $A_j$ for $j\ge i$ and $\tilde{A}_j$ for $j<i$ will have Neumann b.c. and they induce $\mathcal{N}=(0,2)$ chiral multiplets at $t=0$. The remaining chiral multiplets obeying Dirichlet b.c. give rise to the boundary $\mathcal{N}=(0,2)$ Fermi multiplets with the superpotentials,
\begin{align}
	J_{\Lambda_j} &= B_j\tilde{A}_j-\tilde{A}_jB_{j+1} \ , \quad 
	J_{\tilde\Lambda_j} = A_j B_j - B_{j+1} A_j \ , \cr
	J_{\Lambda_{i-1}} & = B_{i-1}\tilde{A}_{i-1}-\tilde{A}_{i-1}B_i + q_1\tilde{q}_2 \ , \quad J_{\Psi_1} = \tilde{A}_{i-1}	q_2 \ ,
\end{align}
and $E_{\Lambda_j}=E_{\tilde{\Lambda}_j}=E_{\Psi_1} = 0$.  $\Lambda_j, \tilde{\Lambda}_j,\Psi_1$ are the Fermi multiplets induced from Dirichlet b.c. of the 1d chiral multiplets whose lowest components are $A_j$ with $j< i$ and $\tilde{A}_j$ with $j\ge i$ and $\tilde{q}_1$ respectively.

We propose that the monopole operator acting on vortex states is realized by an interface with the above boundary conditions coupled to the $\mathcal{N}=(0,2)$ matrix degrees of freedom as follows.
The interface contains the extra chiral and Fermi multiplets in the representations of the gauge groups as
\begin{align}
  \upsilon_j \ & : \ {\rm chiral \ multiplet \ in \ } ({\bf \bar{\n}'}_j,{\bf\n}_j) \ , \quad
  \gamma_j \  : \ {\rm Fermi \ multiplet \ in \ } ({\bf\bar{\n}}'_j,{\bf\n}_j) \ , 
\end{align}
for all $j$, and
\begin{equation}
  \left\{ 
  \begin{array}{lll}
  \tilde\upsilon_j \ : \ {\rm chiral \ in \ } ({\bf\n'}_{j+1},{\bf\bar\n}_{j}) \ , & 
  \tilde\gamma_j \ : \ {\rm Fermi \ in \ } ({\bf\bar{\n}'}_{j+1},{\bf\n}_{j}) & \ \ j<i \\
  \tilde\upsilon_j \ : \ {\rm chiral \ in \ } ({\bf\n'}_{j},{\bf\bar\n}_{j+1}) \ ,  &
  \tilde\gamma_j \ : \ {\rm Fermi \ in \ } ({\bf\bar{\n}'}_{j},{\bf\n}_{j+1}) & \ \ j\ge i
  \end{array}
  \right. \ ,
\end{equation}
and lastly
\begin{align}
	\rho \ & : \ {\rm chiral \ multiplet \ in \ } ({\bf\bar{\n}}_{i-1},{\bf1}) \ , \quad
  	\eta_1 \ : \ {\rm Fermi \ multiplet \ in \ } ({\bf1},{\bf\n}_{i-1})  \ , \cr
  	\eta_2 \ & : \ {\rm Fermi \ multiplet \ in \ } ({\bf1},{\bf\n}_i) \ , \quad
  	\tilde\eta \  : \ {\rm Fermi \ multiplet \ in \ } ({\bf\bar{\n}'}_k,{\bf1}) \ . \ 
\end{align}

The 0d matrix fields couple to the boundary conditions through the superpotentials of the extra boundary Fermi multiplets given by
\begin{align}
	& E_{\gamma_j} = \upsilon_j B_j' - B_j \upsilon_j \quad \ j=1,\cdots,N \ , \quad J_{\gamma_j} = \left\{ 
	\begin{array}{cl}
		\tilde{\upsilon}_{j-1}\tilde{A}_{j-1}-\tilde{A}_j' \tilde{\upsilon}_j & \quad j < i \\
		\tilde{\upsilon}_jA_j + \tilde{\upsilon}_{j-1}\tilde{A}_{j-1} & \quad j =i \\
		\tilde{\upsilon}_jA_j - A_{j-1}' \tilde\upsilon_{j-1} & \quad j>i 
	\end{array}\right.
	\ , \\
	&E_{\tilde\gamma_j} = \left\{ 
	\begin{array}{cl}
	\upsilon_j\tilde{A}'_j - \tilde{A}_j\upsilon_{j+1} & \quad j <i \\
	\upsilon_{j+1}A_j' - A_j\upsilon_j & \quad j\ge i
	\end{array}
	\right. \ ,
	\qquad
	J_{\tilde{\gamma}_j} = \left\{
	\begin{array}{cl}
	-\tilde{\upsilon}_jB_j + B_{j+1}'\tilde{\upsilon}_j & \quad j < i \\
	-\tilde{\upsilon}_jB_j + B_{j+1}'\tilde{\upsilon}_j + q'_2\rho & \quad j = i\!-\!1 \\
	-\tilde{\upsilon}_j B_{j+1} + B_j' \tilde{\upsilon}_j & \quad j\ge i 
	\end{array}
	\right. \nonumber
\end{align}
and 
\begin{align}
	\begin{array}{lll}
	E_{\eta_1} = \upsilon_{i-1}q_1'-q_1 \ , &\ E_{\eta_2} = \upsilon_{i}q_2'-q_2 \ , &\ E_{\tilde\eta} = \tilde{q}_2' - \tilde{q}_2\upsilon_k \ ,\\
	J_{\eta_1} = -\tilde{q}_2'\tilde{\upsilon}_{i-1} \ , &\ J_{\eta_2} = \rho\tilde{A}_{i-1} \ , &\ J_{\tilde{\eta}} = -\tilde{\upsilon}_{i-1}q_1  \ .
	\end{array}
\end{align}

In addition, the superpotentials of the Fermi multiplets from the Dirichlet b.c. need to be modified as follows:
\begin{eqnarray}
	&&J_{\Lambda_j} = B_i\tilde{A}_j - \tilde{A}_jB_{j+1} \ , \quad E_{\Lambda_j} = -\upsilon_{j+1} \tilde{\upsilon}_j \ , \nonumber \\
	&&J_{\Lambda'_j} =  B_j' \tilde{A}_j' - \tilde{A}'_jB_{j+1}' \ , \quad E_{\Lambda'_j} = \tilde\upsilon_j \upsilon_{j} \ ,
\end{eqnarray}
for $j<i\!-\!1$, and
\begin{eqnarray}
	&&J_{\Lambda_{i-1}} = B_{i-1}\tilde{A}_{i-1} - \tilde{A}_{i-1}B_{i} +q_1\tilde{q}_2\ , \quad E_{\Lambda_{i-1}} = -\upsilon_{i} \tilde{\upsilon}_{i-1} \ , \nonumber \\
	&&J_{\Lambda'_{i-1}} =  B_{i-1}' \tilde{A}_{i-1}' - \tilde{A}'_{i-1}B_{i}'+q_1'\tilde{q}'_2 \ , \quad E_{\Lambda'_{i-1}} = \tilde\upsilon_{i-1} \upsilon_{i-1} \ ,
\end{eqnarray}
and
\begin{eqnarray}
	&&J_{\tilde{\Lambda}_j} = A_j B_j - B_{j+1}A_j \ , \quad E_{\tilde{\Lambda}_j} = \upsilon_j \tilde{\upsilon}_j \ , \nonumber \\
	&&J_{\tilde{\Lambda}'_j} = A_j' B_j' - B_{j+1}'A'_j \ , \quad E_{\tilde{\Lambda}'_j} = -\tilde\upsilon_j \upsilon_{j+1} \ ,
\end{eqnarray}
for $j\ge i$, and
lastly
\begin{equation}
	J_{\Psi_1} = \tilde{A}_{i-1}q_2 \ , \quad E_{\Psi_1} = \rho \ , \quad J_{\Psi'_1} = \tilde{A}'_{i-1}q'_2 \ , \quad E_{\Psi_1} = -\rho\upsilon_{i-1} \ .
\end{equation}
The above superpotentials are chosen in order that all charges of the extra supermultiplets are uniquely fixed and also the supersymmetric condition $\sum_a E_a\cdot J_a=0$ is satisfied~\footnote{We note that there is another choice of superpotentials including the extra chiral field $\rho$ which gives the same, but opposite for $\rho$, charge assignments and the same partition function. The choice is that we turn on only one superpotential term for $\rho$ such as $J_{\Psi_1} = \tilde{A}_{i-1}q_2-\rho$. However, if we restrict to the case where the chiral field $\rho$ does not develop extra branch of moduli space, for being consistent with the physics at the interface, the current choice is preferred.}.

We remark here that the vortex quantum mechanics in the first vacuum at $i=1$ can be considered as the handsaw quiver theory in Figure~\ref{fig:handsaw} with $\rho_1=\rho_2=1$ and $\rho_j=0$ for $j>2$, and $q_j=\tilde{q}_j=0$ for $j\ge2$. So we expect that the interface in this section and the interface for the handsaw quiver given in section~\ref{sec:interface-triangular} after truncating the some fields appropriately will be the same.
Indeed one can check that our interface for the abelian quiver theory at $i=1$ coincides with the interface of the handsaw quiver theory by setting $q_j=\tilde{q}_j=\eta_j=\tilde\eta_j=0$ for $j>2$. So two interface constructions are compatible.

The localized partition function with the interface is given by
\begin{equation}
	Z_{\vec{\n}'\times\vec{\n}} = Z^{\rm 1-loop}_{t>0} Z^{\rm 1-loop}_{t=0}Z^{\rm 1-loop}_{t>0} \ ,
\end{equation}
where $Z^{\rm 1-loop}_{t>0} = 1/\omega_{\vec{\n}'}$ and $Z^{\rm 1-loop}_{t<0} = 1/\omega_{\vec\n}$ are the 1-loop contributions from the 1d bulk fields.
The contribution from the extra fields at $t=0$ is given by
\begin{eqnarray}
	Z^{\rm 1-loop}_{t=0} &=& \prod_{j=1}^{N}\prod_{I=1}^{\n_j}\prod_{J=1}^{\n'_j}\frac{(\phi_{j,I}\!-\!\phi_{j,J}'\!+\!\epsilon)}{(\phi_{j,I}-\phi_{j,J}')} \!
	\cdot\!\prod_{j=1}^{i-1}\prod_{I=1}^{\n_j}\prod_{J=1}^{\n'_{j+1}}\frac{(\phi_{j,I}\!-\!\phi'_{j\!+\!1,J}\!-\!\frac{\epsilon}{2})}{(\phi'_{j\!+\!1,J}\!-\!\phi_{j,I}\!-\!\frac{\epsilon}{2})} \!
	\cdot\! \prod_{j=i}^{N\!-\!1} \prod_{I=1}^{\n'_j}\prod_{J=1}^{\n_{j+1}} \frac{(\phi_{j\!+\!1,J}\!-\!\phi_{j,I}'\!-\!\frac{\epsilon}{2})}{(\phi_{j,I}'\!-\!\phi_{j\!+\!1,J}\!-\!\frac{\epsilon}{2})} \nonumber \\
	&& \times \prod_{I=1}^{\n_{i-1}}\frac{(\phi_{i-1,I}+m_1+\frac{\epsilon}{2})}{(-\phi_{i-1,I}-m_2)} \cdot \prod_{I=1}^{\n_i}(\phi_{i,I}+m_2+\frac{\epsilon}{2})\prod_{I=1}^{\n_i'}(-\phi_{i,I}'-m_1) \ ,
\end{eqnarray}
where the complex scalars $\phi_{i,I}$ and $\phi_{i,I}'$ take values at the supersymmetric vacua (\ref{phi-fp-abelian}).
By plugging the fixed point values of $\phi_{j,I}$ and $\phi'_{j,I}$ into this contribution, we find
\begin{equation}
	Z^{\rm 1-loop}_{t=0} = \prod_{j=1}^{i-1}\prod_{\ell=\n_{j\!-\!1}\!+\!1}^{\n_j}(\n_j'\!-\!\ell\!+\!1)\epsilon \cdot \prod_{j=i}^{N}\prod_{\ell = \n_{j\!+\!1}\!+\!1}^{\n_j}(\n_j'\!-\!\ell\!+\!1)\epsilon\cdot\prod_{\ell=n_{i\!-\!1}\!+\!1}^{\n_i'}(m_{12}\!-\!(\n_i'\!-\!\ell\!+\!\tfrac{1}{2})\epsilon) \ .
\end{equation}

The partition function of the identity interface at $\vec{\n}' = \vec{\n}$ correctly yields the overlap of the vortex state as
\be
	\la \vec{\n}|\vec{\n}\ra = Z_{\vec{\n}\times \vec{\n}} = 1/\omega_{\vec{\n}} \ .
\ee
On the other hand, the partition function with $\vec{\n}' = \vec{\n}+\bm\delta_i$ computes the correlation functions of the monopole operators as
\be
	\la \vec{\n}+\bm\delta_i|v^+_i|\vec{\n} \ra = \la\vec{\n}|v^-_i|\vec{\n}+\bm\delta_i\ra = Z_{\vec{\n}'\times \vec{n}} \ .
\ee

Using the correlation function, we find the actions of the monopole operators:
\begin{align} \label{abelian-ADHM}
	v_{j}^+|\vec\n\rangle &= | \vec\n +\bm\delta_{j}\rangle 
	\left\{ \begin{array}{ll} (\n_j-\n_{j-1}+1)\epsilon & \ \ j < i\\
	(\n_i-\n_{i+1}+1)\epsilon(m_{12}-(\n_i-\n_{i-1}+\frac{1}{2})\epsilon) \ \ & \ \ j=i \\
	(\n_{j}-\n_{j+1}+1)\epsilon \ & \ \ j>i\end{array}\right. \ ,\cr
	v_{j}^-|\vec\n\rangle &= |\vec\n -\bm\delta_{j}\rangle  
	\left\{ \begin{array}{ll} (\n_{j+1}-\n_j+1)\epsilon & \ \ j<i-1 \\
	(m_{12}-(\n_i-\n_{i-1}-\frac{1}{2})\epsilon) & \ \ j = i-1 \\
	1 & \ \ j=i \\
	(\n_{j-1}-\n_j+1)\epsilon \ \ & \ \ j>i\end{array}\right. \ .
\end{align}
The results show perfect agreement with the monopole actions in (\ref{abelian-Vplus}) and (\ref{abelian-Vminus}).
This strongly supports that the interface in the vortex quantum mechanics constructed in this section realizes the action of monopole operators on vortex states in the abelian quiver theory.

\subsection*{Acknowledgements}

We are grateful to many friends and colleagues for stimulating discussions on the subjects of this paper, in particular  David Ben-Zvi, Alexander Braverman, Kevin Costello, Michael Finkelberg, Sergei Gukov, Lotte Hollands, Joel Kamnitzer, Gregory Moore, Hiraku Nakajima, Daniel Park, Ben Webster, and Alex Weekes. This paper also benefited greatly from interaction at the workshop ``Symplectic Duality and Gauge Theory'' at the Perimeter Institute, supported in part by the John Templeton Foundation, and at the Aspen Center for Physics Program ``Boundaries and Defects in Quantum Field Theories,'' supported by NSF grant PHY-1066293.

The work of M.B. was supported by ERC Starting Grant no. 306260 `Dualities in Supersymmetric Gauge Theories, String Theory and Conformal Field Theories'.
The work of T.D. was supported in part by ERC Starting Grant no. 335739 ``Quantum fields and knot homologies,'' funded by the European Research Council under the European Union's Seventh Framework Programme, and by the Perimeter Institute for Theoretical Physics.
The work of D.G. was supported by the Perimeter Institute for Theoretical Physics.
The work of J.H. was supported in part by the Visiting Graduate Fellowship Program at the Perimeter Institute for Theoretical Physics.
The work of H.K. was supported in part by NSF grant PHY-1067976 and by the Perimeter Institute for Theoretical Physics.
Research at the Perimeter Institute is supported by the Government of Canada through Industry Canada and by the Province of Ontario through the Ministry of Economic Development \& Innovation.

\bibliographystyle{JHEP}
\bibliography{vortex}

\def\cprime{$'$} \def\cprime{$'$}
\providecommand{\href}[2]{#2}\begingroup\raggedright\begin{thebibliography}{10}

\bibitem{BDG-Coulomb}
M.~Bullimore, T.~Dimofte, and D.~Gaiotto, {\it {The Coulomb Branch of 3d
  $\mathcal{N}=4$ Theories}},  \href{http://arxiv.org/abs/1503.04817}{{\tt
  arXiv:1503.04817}}.

\bibitem{Nak-Coulomb}
H.~Nakajima, {\it {Towards a mathematical definition of Coulomb branches of
  $3$-dimensional $\mathcal N=4$ gauge theories, I}},
  \href{http://arxiv.org/abs/1503.03676}{{\tt arXiv:1503.03676}}.

\bibitem{BFN-II}
A.~Braverman, M.~Finkelberg, and H.~Nakajima, {\it {Towards a mathematical
  definition of Coulomb branches of $3$-dimensional $\mathcal N=4$ gauge
  theories, II}},  \href{http://arxiv.org/abs/1601.03586}{{\tt
  arXiv:1601.03586}}.

\bibitem{BDGH}
M.~Bullimore, T.~Dimofte, D.~Gaiotto, and J.~Hilburn, {\it Boundaries, mirror
  symmetry, and symplectic duality in 3d $\mathcal{N}=4$ gauge theory},
  \href{http://arxiv.org/abs/1603.08382v1}{{\tt arXiv:1603.08382v1}}.

\bibitem{Shadchin-2d}
S.~Shadchin, {\it On f-term contribution to effective action},  {\em JHEP} {\bf
  08} (Jan, 2007) 052, [\href{http://arxiv.org/abs/hep-th/0611278v1}{{\tt
  hep-th/0611278v1}}].

\bibitem{DGH}
T.~Dimofte, S.~Gukov, and L.~Hollands, {\it Vortex counting and lagrangian
  3-manifolds},  {\em Lett. Math. Phys.} {\bf 98} (2011) 225--287,
  [\href{http://arxiv.org/abs/1006.0977v1}{{\tt arXiv:1006.0977v1}}].

\bibitem{HananyWitten}
A.~Hanany and E.~Witten, {\it Type iib superstrings, bps monopoles, and
  three-dimensional gauge dynamics},  {\em Nucl. Phys.} {\bf B492} (1997)
  152--190, [\href{http://arxiv.org/abs/hep-th/9611230v3}{{\tt
  hep-th/9611230v3}}].

\bibitem{HananyHori}
A.~Hanany and K.~Hori, {\it Branes and n=2 theories in two dimensions},  {\em
  Nucl. Phys.} {\bf B513} (1998) 119--174,
  [\href{http://arxiv.org/abs/hep-th/9707192v2}{{\tt hep-th/9707192v2}}].

\bibitem{HananyTong-branes}
A.~Hanany and D.~Tong, {\it {Vortices, instantons and branes}},  {\em JHEP}
  {\bf 07} (2003) 037, [\href{http://arxiv.org/abs/hep-th/0306150}{{\tt
  hep-th/0306150}}].

\bibitem{Nakajima-handsaw}
H.~Nakajima, {\it Handsaw quiver varieties and finite w-algebras},
  \href{http://arxiv.org/abs/1107.5073v3}{{\tt arXiv:1107.5073v3}}.

\bibitem{Abrikosov}
A.~A. Abrikosov, {\it {On the Magnetic properties of superconductors of the
  second group}},  {\em Sov. Phys. JETP} {\bf 5} (1957) 1174--1182. [Zh. Eksp.
  Teor. Fiz.32,1442(1957)].

\bibitem{NielsenOlesen}
H.~B. Nielsen and P.~Olesen, {\it {Vortex Line Models for Dual Strings}},  {\em
  Nucl. Phys.} {\bf B61} (1973) 45--61.

\bibitem{Taubes-LG}
C.~H. Taubes, {\it {Arbitrary N-Vortex Solutions to the First Order
  Landau-Ginzburg Equations}},  {\em Commun. Math. Phys.} {\bf 72} (1980)
  277--292.

\bibitem{JaffeTaubes}
A.~M. Jaffe and C.~H. Taubes, {\em {Vortices and Monopoles: Structure of Static
  Gauge Theories}}.
\newblock Boston, USA: Birkh{\"a}user 287 pp. (Progress In Physics, 2), 1980.

\bibitem{Tong-TASI}
D.~Tong, {\it {TASI lectures on solitons: Instantons, monopoles, vortices and
  kinks}},  in {\em {Theoretical Advanced Study Institute in Elementary
  Particle Physics: Many Dimensions of String Theory (TASI 2005) Boulder,
  Colorado, June 5-July 1, 2005}}, 2005.
\newblock \href{http://arxiv.org/abs/hep-th/0509216}{{\tt hep-th/0509216}}.

\bibitem{MorrisonPlesser}
D.~R. Morrison and M.~R. Plesser, {\it {Summing the instantons: Quantum
  cohomology and mirror symmetry in toric varieties}},  {\em Nucl. Phys.} {\bf
  B440} (1995) 279--354, [\href{http://arxiv.org/abs/hep-th/9412236}{{\tt
  hep-th/9412236}}].

\bibitem{Vafa-MS}
C.~Vafa, {\it {Topological mirrors and quantum rings}},  in {\em {Mirror
  symmetry I}} (S.~Yau, ed.), pp.~97--120, 1991.
\newblock \href{http://arxiv.org/abs/hep-th/9111017}{{\tt hep-th/9111017}}.

\bibitem{Givental-MS}
A.~Givental, {\it A mirror theorem for toric complete intersections},  in {\em
  Topological Field Theory, Primitive Forms and Related Topics} (M.~Kashiwara,
  A.~Matsuo, K.~Saito, and I.~Satake, eds.), pp.~141--175.
\newblock Birkh{\"a}user Boston, Boston, MA, 1998.

\bibitem{HoriVafa}
K.~Hori and C.~Vafa, {\it {Mirror symmetry}},
  \href{http://arxiv.org/abs/hep-th/0002222}{{\tt hep-th/0002222}}.

\bibitem{GiventalKim}
A.~Givental and B.-s. Kim, {\it {Quantum cohomology of flag manifolds and Toda
  lattices}},  {\em Commun. Math. Phys.} {\bf 168} (1995) 609--642,
  [\href{http://arxiv.org/abs/hep-th/9312096}{{\tt hep-th/9312096}}].

\bibitem{GiventalLee}
A.~Givental and Y.-P. Lee, {\it {Quantum K-Theory on Flag Manifolds,
  Finite-Difference Toda Lattices and Quantum Groups}},  {\em Invent. Math.}
  {\bf 151} (2003) 193--219, [\href{http://arxiv.org/abs/math/0108105}{{\tt
  math/0108105}}].

\bibitem{CoatesGivental}
T.~Coates and A.~Givental, {\it Quantum riemann?roch, lefschetz and serre},
  {\em Ann. Math.} {\bf 165} (2007) 15--53,
  [\href{http://arxiv.org/abs/math/0110142}{{\tt math/0110142}}].

\bibitem{CV-tt*}
S.~Cecotti and C.~Vafa, {\it {Topological antitopological fusion}},  in {\em
  {In *Trieste 1991, Proceedings, High energy physics and cosmology, vol. 2*
  682-784}}, 1991.

\bibitem{IntegrableHierarchies}
M.~Aganagic, R.~Dijkgraaf, A.~Klemm, M.~Marino, and C.~Vafa, {\it {Topological
  strings and integrable hierarchies}},  {\em Commun. Math. Phys.} {\bf 261}
  (2006) 451--516, [\href{http://arxiv.org/abs/hep-th/0312085}{{\tt
  hep-th/0312085}}].

\bibitem{AGGTV}
L.~F. Alday, D.~Gaiotto, S.~Gukov, Y.~Tachikawa, and H.~Verlinde, {\it Loop and
  surface operators in n=2 gauge theory and liouville modular geometry},  {\em
  JHEP} {\bf 1001} (2010) 113, [\href{http://arxiv.org/abs/0909.0945v2}{{\tt
  arXiv:0909.0945v2}}].

\bibitem{DGG}
T.~Dimofte, D.~Gaiotto, and S.~Gukov, {\it Gauge theories labelled by
  three-manifolds},  {\em Comm. Math. Phys.} {\bf 325} (2014) 367--419,
  [\href{http://arxiv.org/abs/1108.4389v1}{{\tt arXiv:1108.4389v1}}].

\bibitem{Nek-SW}
N.~A. Nekrasov, {\it Seiberg-witten prepotential from instanton counting},
  {\em Adv. Theor. Math. Phys.} {\bf 7} (Dec, 2004) 831--864,
  [\href{http://arxiv.org/abs/hep-th/0206161v1}{{\tt hep-th/0206161v1}}].

\bibitem{LNS-SW}
A.~Losev, N.~Nekrasov, and S.~Shatashvili, {\it Testing seiberg-witten
  solution},  {\em Strings, branes and dualities (Carg{\`e}se, 1997), NATO Adv.
  Sci. Inst. Ser. C Math. Phys. Sci.} {\bf 520} (1999) 359--372,
  [\href{http://arxiv.org/abs/hep-th/9801061v1}{{\tt hep-th/9801061v1}}].

\bibitem{Moore:1998et}
G.~W. Moore, N.~Nekrasov, and S.~Shatashvili, {\it {D particle bound states and
  generalized instantons}},  {\em Commun. Math. Phys.} {\bf 209} (2000) 77--95,
  [\href{http://arxiv.org/abs/hep-th/9803265}{{\tt hep-th/9803265}}].

\bibitem{Moore:1997dj}
G.~W. Moore, N.~Nekrasov, and S.~Shatashvili, {\it {Integrating over Higgs
  branches}},  {\em Commun. Math. Phys.} {\bf 209} (2000) 97--121,
  [\href{http://arxiv.org/abs/hep-th/9712241}{{\tt hep-th/9712241}}].

\bibitem{NShatashvili}
N.~A. Nekrasov and S.~L. Shatashvili, {\it Quantization of integrable systems
  and four dimensional gauge theories},  {\em 16th International Congress on
  Mathematical Physics, Prague, August 2009 (World Scientific 2010)} (2009)
  265--289, [\href{http://arxiv.org/abs/0908.4052v1}{{\tt arXiv:0908.4052v1}}].

\bibitem{DG-RMQ}
T.~Dimofte and S.~Gukov, {\it Refined, motivic, and quantum},  {\em Lett. Math.
  Phys.} {\bf 91} (Jan, 2010) 1, [\href{http://arxiv.org/abs/0904.1420v1}{{\tt
  arXiv:0904.1420v1}}].

\bibitem{DGOT}
N.~Drukker, J.~Gomis, T.~Okuda, and J.~Teschner, {\it Gauge theory loop
  operators and liouville theory},  {\em JHEP} {\bf 1002} (2010) 057,
  [\href{http://arxiv.org/abs/0909.1105v2}{{\tt arXiv:0909.1105v2}}].

\bibitem{GMNIII}
D.~Gaiotto, G.~W. Moore, and A.~Neitzke, {\it Framed bps states},  {\em Adv.
  Theor. Math. Phys.} {\bf 17} (2013) 241--397,
  [\href{http://arxiv.org/abs/1006.0146v1}{{\tt arXiv:1006.0146v1}}].

\bibitem{RW}
L.~Rozansky and E.~Witten, {\it Hyper-kahler geometry and invariants of
  three-manifolds},  {\em Selecta Math.} {\bf 3} (1997) 401--458,
  [\href{http://arxiv.org/abs/hep-th/9612216v3}{{\tt hep-th/9612216v3}}].

\bibitem{BT-twists}
M.~Blau and G.~Thompson, {\it Aspects of $n_{T}\geq 2$ topological gauge
  theories and d-branes},  {\em Nucl. Phys.} {\bf B492} (1997) 545--590,
  [\href{http://arxiv.org/abs/hep-th/9612143v1}{{\tt hep-th/9612143v1}}].

\bibitem{Yagi-quantization}
J.~Yagi, {\it Omega-deformation and quantization},  {\em JHEP} {\bf 1408}
  (2014) 112, [\href{http://arxiv.org/abs/1405.6714v3}{{\tt
  arXiv:1405.6714v3}}].

\bibitem{Brav-W}
A.~Braverman, {\it Instanton counting via affine lie algebras i: Equivariant
  j-functions of (affine) flag manifolds and whittaker vectors},  {\em CRM
  Proc. Lecture Notes} {\bf 38} (2004) 113--132,
  [\href{http://arxiv.org/abs/math/0401409v2}{{\tt math/0401409v2}}]. Amer.
  Math. Soc., Providence, RI.

\bibitem{BFFR-W}
A.~Braverman, B.~Feigin, M.~Finkelberg, and L.~Rybnikov, {\it A finite analog
  of the agt relation i: finite w-algebras and quasimaps' spaces},  {\em Comm.
  Math. Phys.} {\bf 308} (2011), no.~2 457--478,
  [\href{http://arxiv.org/abs/1008.3655v3}{{\tt arXiv:1008.3655v3}}].

\bibitem{Kostant-Whittaker}
B.~Kostant, {\it {On Whittaker vectors and representation theory}},  {\em
  Invent. Math.} {\bf 48} (1978) 101--184.

\bibitem{GW-Sduality}
D.~Gaiotto and E.~Witten, {\it {S-Duality of Boundary Conditions In N=4 Super
  Yang-Mills Theory}},  {\em Adv. Theor. Math. Phys.} {\bf 13} (2009), no.~3
  721--896, [\href{http://arxiv.org/abs/0807.3720}{{\tt arXiv:0807.3720}}].

\bibitem{Tjin-W}
T.~Tjin, {\it Finite w-algebras},  {\em Phys. Lett.} {\bf B292} (1992) 60--66,
  [\href{http://arxiv.org/abs/hep-th/9203077v1}{{\tt hep-th/9203077v1}}].

\bibitem{dBT-W}
J.~de~Boer and T.~Tjin, {\it Representation theory of finite w algebras},  {\em
  Comm. Math. Phys.} {\bf 158} (1993)
  [\href{http://arxiv.org/abs/hep-th/9211109v1}{{\tt hep-th/9211109v1}}].

\bibitem{Losev-W}
I.~Losev, {\it Finite w-algebras},
  \href{http://arxiv.org/abs/1003.5811v1}{{\tt arXiv:1003.5811v1}}.

\bibitem{AGT}
L.~F. Alday, D.~Gaiotto, and Y.~Tachikawa, {\it Liouville correlation functions
  from four-dimensional gauge theories},  {\em Lett. Math. Phys.} {\bf 91}
  (2010), no.~2 167--197, [\href{http://arxiv.org/abs/0906.3219v1}{{\tt
  arXiv:0906.3219v1}}].

\bibitem{BFN-instW}
A.~Braverman, M.~Finkelberg, and H.~Nakajima, {\it {Instanton moduli spaces and
  W-algebras}},  \href{http://arxiv.org/abs/1406.2381}{{\tt arXiv:1406.2381}}.

\bibitem{SchiffmannVasserot}
O.~Schiffmann and E.~Vasserot, {\it Cherednik algebras, w algebras and the
  equivariant cohomology of the moduli space of instantons on $a^2$},
  \href{http://arxiv.org/abs/1202.2756v2}{{\tt arXiv:1202.2756v2}}.

\bibitem{MaulikOkounkov}
D.~Maulik and A.~Okounkov, {\it Quantum groups and quantum cohomology},
  \href{http://arxiv.org/abs/1211.1287v1}{{\tt arXiv:1211.1287v1}}.

\bibitem{Tachikawa-instanton}
Y.~Tachikawa, {\it {A review on instanton counting and W-algebras}},  in {\em
  New Dualities of Supersymmetric Gauge Theories} (J.~Teschner, ed.),
  pp.~79--120.
\newblock 2016.
\newblock \href{http://arxiv.org/abs/1412.7121}{{\tt arXiv:1412.7121}}.

\bibitem{Gaiotto-states}
D.~Gaiotto, {\it {Asymptotically free $\mathcal{N} = 2$ theories and irregular
  conformal blocks}},  {\em J. Phys. Conf. Ser.} {\bf 462} (2013), no.~1
  012014, [\href{http://arxiv.org/abs/0908.0307}{{\tt arXiv:0908.0307}}].

\bibitem{MMM-AGT}
A.~Marshakov, A.~Mironov, and A.~Morozov, {\it {On non-conformal limit of the
  AGT relations}},  {\em Phys. Lett.} {\bf B682} (2009) 125--129,
  [\href{http://arxiv.org/abs/0909.2052}{{\tt arXiv:0909.2052}}].

\bibitem{Taki-AGT}
M.~Taki, {\it {On AGT Conjecture for Pure Super Yang-Mills and W-algebra}},
  {\em JHEP} {\bf 05} (2011) 038, [\href{http://arxiv.org/abs/0912.4789}{{\tt
  arXiv:0912.4789}}].

\bibitem{Alday:2010vg}
L.~F. Alday and Y.~Tachikawa, {\it {Affine SL(2) conformal blocks from 4d gauge
  theories}},  {\em Lett. Math. Phys.} {\bf 94} (2010) 87--114,
  [\href{http://arxiv.org/abs/1005.4469}{{\tt arXiv:1005.4469}}].

\bibitem{KT-chainsaw}
H.~Kanno and Y.~Tachikawa, {\it Instanton counting with a surface operator and
  the chain-saw quiver},  {\em JHEP} {\bf 06} (2011) 119,
  [\href{http://arxiv.org/abs/1105.0357v2}{{\tt arXiv:1105.0357v2}}].

\bibitem{GaiottoKim}
D.~Gaiotto and H.-C. Kim, {\it {Duality walls and defects in 5d N=1 theories}},
   \href{http://arxiv.org/abs/1506.03871}{{\tt arXiv:1506.03871}}.

\bibitem{BPW-I}
T.~Braden, N.~Proudfoot, and B.~Webster, {\it Quantizations of conical
  symplectic resolutions i: local and global structure},
  \href{http://arxiv.org/abs/1208.3863v3}{{\tt arXiv:1208.3863v3}}.

\bibitem{BLPW-II}
T.~Braden, A.~Licata, N.~Proudfoot, and B.~Webster, {\it Quantizations of
  conical symplectic resolutions ii: category $\mathcal o$ and symplectic
  duality},  \href{http://arxiv.org/abs/1407.0964v1}{{\tt arXiv:1407.0964v1}}.

\bibitem{GomisAssel}
B.~Assel and J.~Gomis, {\it {Mirror Symmetry And Loop Operators}},  {\em JHEP}
  {\bf 11} (2015) 055, [\href{http://arxiv.org/abs/1506.01718}{{\tt
  arXiv:1506.01718}}].

\bibitem{Witten-path}
E.~Witten, {\it {A New Look At The Path Integral Of Quantum Mechanics}},
  \href{http://arxiv.org/abs/1009.6032}{{\tt arXiv:1009.6032}}.

\bibitem{Witten-Morse}
E.~Witten, {\it {Supersymmetry and Morse theory}},  {\em J. Diff. Geom.} {\bf
  17} (1982), no.~4 661--692.

\bibitem{ClossetCremonesi}
C.~Closset and S.~Cremonesi, {\it {Comments on $ \mathcal{N} $ = (2, 2)
  supersymmetry on two-manifolds}},  {\em JHEP} {\bf 07} (2014) 075,
  [\href{http://arxiv.org/abs/1404.2636}{{\tt arXiv:1404.2636}}].

\bibitem{AtiyahBott}
M.~F. Atiyah and R.~Bott, {\it {The Moment map and equivariant cohomology}},
  {\em Topology} {\bf 23} (1984) 1--28.

\bibitem{Eto-matrix}
M.~Eto, Y.~Isozumi, M.~Nitta, K.~Ohashi, and N.~Sakai, {\it {Solitons in the
  Higgs phase: The Moduli matrix approach}},  {\em J. Phys.} {\bf A39} (2006)
  R315--R392, [\href{http://arxiv.org/abs/hep-th/0602170}{{\tt
  hep-th/0602170}}].

\bibitem{Eto-moduli}
M.~Eto, Y.~Isozumi, M.~Nitta, K.~Ohashi, and N.~Sakai, {\it {Moduli space of
  non-Abelian vortices}},  {\em Phys. Rev. Lett.} {\bf 96} (2006) 161601,
  [\href{http://arxiv.org/abs/hep-th/0511088}{{\tt hep-th/0511088}}].

\bibitem{Groth-class}
A.~Grothendieck, {\it Sur la classification des fibr\'es holomorphes sur la
  sph\`ere de riemann},  {\em Amer. J. Math.} {\bf 79} (1957) 121--138.

\bibitem{Kapustin-Witten}
A.~Kapustin and E.~Witten, {\it Electric-magnetic duality and the geometric
  langlands program},  {\em Comm. Num. Th. and Phys.} {\bf 1} (2007) 1--236,
  [\href{http://arxiv.org/abs/hep-th/0604151v3}{{\tt hep-th/0604151v3}}].

\bibitem{Lusztig-qg}
G.~Lusztig, {\em Introduction to quantum groups}.
\newblock Birkh\"auser Boston, Inc., Boston, MA, 1993.
\newblock Volume 10 of Progress in Mathematics.

\bibitem{CPS-weight}
E.~Cline, B.~Parshall, and L.~Scott, {\it {Finite-dimensional algebras and
  highest weight categories}},  {\em J. Reine Agnew. Math.} {\bf 391} (1988)
  85--99.

\bibitem{BFN-III}
A.~Braverman, M.~Finkelberg, and H.~Nakajima, {\it {Coulomb branches of $3d$
  $\mathcal N=4$ quiver gauge theories and slices in the affine Grassmannian
  (with appendices by Alexander Braverman, Michael Finkelberg, Joel Kamnitzer,
  Ryosuke Kodera, Hiraku Nakajima, Ben Webster, and Alex Weekes)}},
  \href{http://arxiv.org/abs/1604.03625}{{\tt arXiv:1604.03625}}.

\bibitem{ClossetCremonesiPark}
C.~Closset, S.~Cremonesi, and D.~S. Park, {\it {The equivariant A-twist and
  gauged linear sigma models on the two-sphere}},  {\em JHEP} {\bf 06} (2015)
  076, [\href{http://arxiv.org/abs/1504.06308}{{\tt arXiv:1504.06308}}].

\bibitem{NS-curved}
N.~A. Nekrasov and S.~L. Shatashvili, {\it {Bethe/Gauge correspondence on
  curved spaces}},  {\em JHEP} {\bf 01} (2015) 100,
  [\href{http://arxiv.org/abs/1405.6046}{{\tt arXiv:1405.6046}}].

\bibitem{CV-class}
S.~Cecotti and C.~Vafa, {\it {On classification of N=2 supersymmetric
  theories}},  {\em Commun. Math. Phys.} {\bf 158} (1993) 569--644,
  [\href{http://arxiv.org/abs/hep-th/9211097}{{\tt hep-th/9211097}}].

\bibitem{GMW-infrared}
D.~Gaiotto, G.~W. Moore, and E.~Witten, {\it {Algebra of the Infrared: String
  Field Theoretic Structures in Massive ${\cal N}=(2,2)$ Field Theory In Two
  Dimensions}},  \href{http://arxiv.org/abs/1506.04087}{{\tt
  arXiv:1506.04087}}.

\bibitem{2010arXiv1009.0676F}
M.~{Finkelberg} and L.~{Rybnikov}, {\it {Quantization of Drinfeld Zastava in
  type A}},  {\em ArXiv e-prints} (Sept., 2010)
  [\href{http://arxiv.org/abs/1009.0676}{{\tt arXiv:1009.0676}}].

\end{thebibliography}\endgroup

\end{document}